\newif{\ifarxiv}
\newif{\ifdraft}
\newif{\ifremarks}

\arxivtrue  

\ifremarks
\newcommand{\remarktb}[1]{{\renewcommand{\bfdefault}{b}\color[RGB]{0,150,0}{\textbf{#1}}}}
\newcommand{\remarkfc}[1]{{\renewcommand{\bfdefault}{b}\color[RGB]{0,0,150}{\textbf{#1}}}}
\newcommand{\remarkpv}[1]{{\renewcommand{\bfdefault}{b}\color[RGB]{150,0,0}{\textbf{#1}}}}
\fi
\providecommand{\remarktb}[1]{\ignorespaces}
\providecommand{\remarkfc}[1]{\ignorespaces}
\providecommand{\remarkpv}[1]{\ignorespaces}

\newcommand\beqa{\begin{eqnarray}}
\newcommand\eeqa{\end{eqnarray}}

\documentclass[12pt,a4paper]{article}
\ifdraft\else\fi
\usepackage{siunitx} 
\usepackage{amsmath,amssymb,mathtools}
\usepackage{longtable}
\usepackage{pdflscape} 
\usepackage{rotating}
\usepackage{array}
\newcolumntype{L}[1]{>{\raggedright\let\newline\\\arraybackslash\hspace{0pt}}m{#1}}

\usepackage[nosort]{cite}
\usepackage[bulletsep]{collref} 
\ifdraft\usepackage{showkeys}\fi 
\usepackage[bookmarks=true]{hyperref} 
\usepackage{graphicx}
\usepackage[usenames,dvipsnames,svgnames,table]{xcolor}
\usepackage{float}
\usepackage[normalem]{ulem}
\usepackage{booktabs} 
\usepackage{calc}
\usepackage{listings}
\usepackage{graphbox}
\usepackage[mathbold,autobold,greekcaps,greeklower]{mathfixs}
\providecommand{\mathbold}{\mathbf}
\usepackage[theorems,breakable]{tcolorbox}

\usepackage{pgffor} 
\newcounter{FootnoteCounter}
\newcounter{Init}

\NewDocumentEnvironment{myinset}{ O{} }%
{%
\begin{tcolorbox}[breakable, width=\textwidth, colback=white,
boxrule=0.4pt, before skip=4mm]
\setcounter{Init}{\value{footnote}}
\setcounter{FootnoteCounter}{\value{Init}}
\def\footnote##1{%
  \stepcounter{footnote}%
  \stepcounter{FootnoteCounter}%
  \footnotemark[\arabic{FootnoteCounter}]%
  \def\temp{##1}%
  \expandafter\expandafter\expandafter\global\expandafter%
  \let\csname F\arabic{FootnoteCounter}\endcsname\temp}%
\footnotesize
}%
{%
\end{tcolorbox}%
\endlist%
\ifdim\arabic{Init}pt<\arabic{FootnoteCounter}pt%
\stepcounter{Init}%
\foreach \f in {\arabic{Init},...,\arabic{FootnoteCounter}}%
{\footnotetext[\f]{\csname F\f\endcsname}}%
\fi%
}

\definecolor{mathgreen}{RGB}{0,90,39}
\definecolor{mypurple}{rgb}{0.6,0,0.8}

\usepackage{lmodern}
\usepackage[T1]{fontenc}
\usepackage{microtype}

\usepackage[left=2.3cm,right=2.3cm,top=2.8cm,bottom=2.8cm]{geometry}

\allowdisplaybreaks
\usepackage{enumitem}

\newcommand{\beq}{\begin{equation}}
\newcommand{\eeq}{\end{equation}}
\newcommand{\nn}{\nonumber}

\newenvironment{myeqnarray}{\arraycolsep0pt\begin{eqnarray}}{\end{eqnarray}\ignorespacesafterend}
\newenvironment{myeqnarray*}{\arraycolsep0pt\begin{eqnarray*}}{\end{eqnarray*}\ignorespacesafterend}

\def\[{\begin{equation}}
\def\]{\end{equation}}
\def\<{\begin{myeqnarray}}
\def\>{\end{myeqnarray}}

\numberwithin{equation}{section}

\def\etal.{et\penalty50\ al.}
\usepackage{xspace}
\newcommand*{\eg}{e.\,g.\@\xspace}
\newcommand*{\ie}{i.\,e.\@\xspace}

\usepackage[font=small,labelfont=bf,margin=0.05\textwidth]{caption}

\providecommand{\hypersetup}[1]{}
\providecommand{\hyperref}[2][]{#2}
\providecommand{\texorpdfstring}[2]{#1}
\providecommand{\pdfbookmark}[3][]{}

\hypersetup{plainpages=false}
\hypersetup{pdfpagemode=UseOutlines}
\hypersetup{bookmarksnumbered=true}
\hypersetup{bookmarksopen=true}
\hypersetup{pdfstartview=FitH}
\hypersetup{colorlinks=true}
\hypersetup{citecolor=[rgb]{0 .3 0}}
\hypersetup{urlcolor=[rgb]{.35 0 0}}
\hypersetup{linkcolor=[rgb]{0 0 .5}}
\hypersetup{citebordercolor={.6 .9 .6}}
\hypersetup{urlbordercolor={.7 .8 1}}
\hypersetup{linkbordercolor={1 .7 .7}}
\hypersetup{pdfborder={0 0 .5}}

\newcommand{\namedref}[2]{\hyperref[#2]{#1~\ref*{#2}}}
\newcommand{\secref}[1]{\namedref{Section}{#1}}
\newcommand{\appref}[1]{\namedref{Appendix}{#1}}
\newcommand{\tabref}[1]{\namedref{Table}{#1}}
\newcommand{\figref}[1]{\namedref{Figure}{#1}}

\makeatletter
\def\mr@ignsp#1 {\ifx\:#1\@empty\else #1\expandafter\mr@ignsp\fi}%
\newcommand{\multiref}[1]{\begingroup
\xdef\mr@no@sparg{\expandafter\mr@ignsp#1 \: }%
\def\mr@comma{}%
\@for\mr@refs:=\mr@no@sparg\do{\mr@comma\def\mr@comma{,}\ref{\mr@refs}}%
\endgroup}
\makeatother
\renewcommand{\eqref}[1]{(\multiref{#1})}

\usepackage{fixltx2e}
\MakeRobust{\eqref}

\makeatletter
\let\@myabstract\@empty
\let\@keywords\@empty
\let\@subject\@empty
\providecommand{\affiliation}[1]{\gdef\@affiliation{#1}}
\providecommand{\myabstract}[1]{\gdef\@myabstract{#1}}
\providecommand{\keywords}[1]{\gdef\@keywords{#1}}
\providecommand{\subject}[1]{\gdef\@subject{#1}}
\def\thetitle{\@title}
\def\theauthor{\@author}
\def\theaffiliation{\@affiliation}
\def\theabstract{\@myabstract}
\def\thesubject{\@subject}
\def\thedate{\@date}
\def\thekeywords{\@keywords}
\makeatother
\AtBeginDocument{
\hypersetup{pdftitle={\thetitle}}%
\hypersetup{pdfauthor={\theauthor}}%
\hypersetup{pdfsubject={\thesubject}}%
\hypersetup{pdfkeywords={\thekeywords}}%
}

\newcommand{\order}[1]{\mathcal{O}(#1)}

\newcommand{\superN}{\mathcal{N}}

\newcommand{\Integers}{\mathbb{Z}}

\newcommand{\dd}{d}
\newcommand{\Nc}{N\subrm{c}}
\newcommand{\Csphere}{{}^\bullet\kern-1.2pt C}
\newcommand{\Ctorus}{{}^\circ\kern-1.2pt C}

\newcommand{\oct}{\mathbb{O}}
\newcommand{\gen}{\mathcal{C}}

\newcommand{\arccosh}{\operatorname{arccosh}}
\newcommand{\tr}{\operatorname{tr}}

\newcommand{\op}[1]{\mathcal{#1}}

\DeclareMathOperator*{\disc}{disc}
\newcommand{\rotpos}{\mathbin{\circlearrowleft}}
\newcommand{\rotneg}{\mathbin{\circlearrowright}}

\DeclareMathOperator{\im}{Im}
\DeclareMathOperator{\Li}{Li}

\newcommand{\subrm}[1]{_{\text{#1}}}
\newcommand{\alg}[1]{\mathfrak{#1}}
\newcommand{\grp}[1]{\mathrm{#1}}

\RequirePackage[extdef]{delimset}

\usepackage[mode=buildnew]{standalone}

\title{Octagons II:\texorpdfstring{\\}{ }Strong Coupling}

\author{%
T. Bargheer\texorpdfstring{$^{a,b}$}{},
F. Coronado\texorpdfstring{$^{c,d}$}{},
P. Vieira\texorpdfstring{$^{c,d}$}{}}

\myabstract{The octagon function is the fundamental building block
yielding correlation functions of four large BPS operators in
$\superN=4$ super Yang--Mills theory at any
value of the 't~Hooft coupling and at any genus order. Here we compute
the octagon at strong coupling, and discuss various interesting
limits and implications, both at the planar and non-planar level.}

\subject{Mathematical Physics}

\keywords{4d gauge theory, integrability, correlation functions,
planar limit, non-planar corrections, hexagonalization, Riemann
surface, moduli space, worldsheet}

\begin{document}

\pdfbookmark[1]{Title Page}{title}

\thispagestyle{empty}
\setcounter{page}{0}
\renewcommand{\thefootnote}{\fnsymbol{footnote}}
\setcounter{footnote}{0}

\mbox{}
\vfill

\begin{center}

{\Large\textbf{\mathversion{bold}\thetitle}\par}

\vspace{1cm}

\textsc{\theauthor}

\bigskip

\begingroup
\footnotesize\itshape

$^{a}$Institut f\"ur Theoretische Physik, Leibniz Universit\"at Hannover,\\
Appelstra{\ss}e 2, 30167 Hannover, Germany

\medskip

$^{b}$DESY Theory Group, DESY Hamburg,\\
Notkestra\ss e 85, D-22603 Hamburg, Germany\\

\medskip

$^{c}$Perimeter Institute for Theoretical Physics,\\
31 Caroline St N Waterloo, Ontario N2L 2Y5, Canada

\medskip

$^{d}$Instituto de F\'isica Te\'orica, UNESP - Univ. Estadual Paulista,\\
ICTP South American Institute for Fundamental Research,\\
Rua Dr. Bento Teobaldo Ferraz 271, 01140-070, S\~ao Paulo, SP, Brasil

\endgroup

\bigskip

\newcommand{\email}[1]{\href{mailto:#1}{#1}}

{\small\ttfamily
\email{till.bargheer@desy.de},
\email{fcoronado@perimeterinstitute.ca},
\email{pedrogvieira@gmail.com}}
\par

\vspace{1cm}

\textbf{Abstract}\vspace{5mm}

\begin{minipage}{12cm}
\theabstract
\end{minipage}

\end{center}

\vfill
\vfill

\newpage
\renewcommand{\thefootnote}{\arabic{footnote}}

\pdfbookmark[1]{\contentsname}{contents}
\tableofcontents

\section{Introduction}

\begin{figure}
\centering
\includegraphics[scale=0.5]{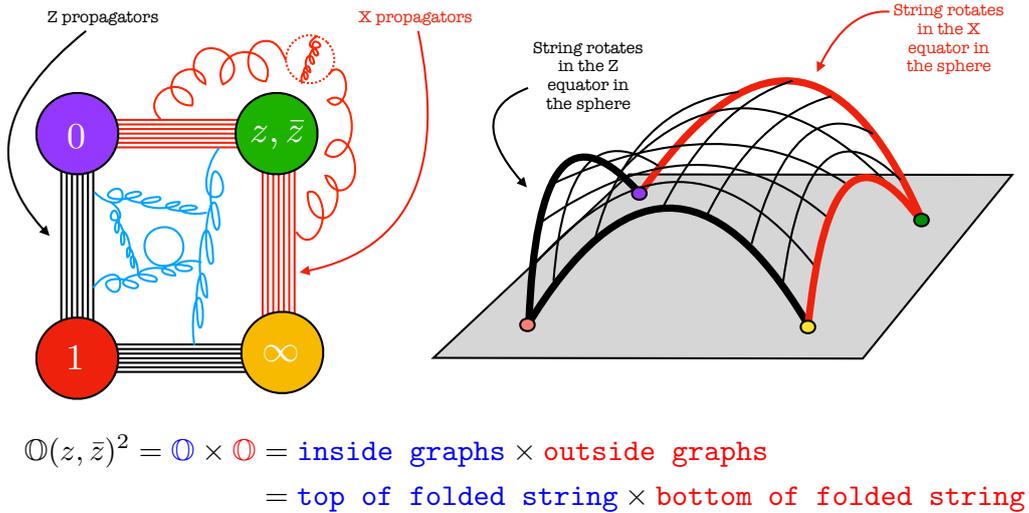}
\vspace{-3cm}
\caption{We work here with the so-called \textit{simplest} correlator
introduced in~\cite{Coronado:2018ypq}. Two operators are BPS primaries
made out of only $Z$ or only $X$ fields respectively while the other
two are BPS descendants composed of both $\bar Z$ and $\bar X$ so that
at tree level there is a single square frame diagram describing this
correlator. At loop level, for large operators, the inside and outside
decouple. In string theory language the correlator is described by a
folded string ending on some spinning
geodesics~\cite{Bargheer:2019kxb}; the top and bottom folds decoupling
is the string counterpart of the inside/outside gauge theory
decoupling. }
\label{fig:recall}
\end{figure}

The octagon function $\mathbb{O}(z,\bar z | \lambda)$ introduced
in~\cite{Coronado:2018ypq} provides a finite 't~Hooft coupling
representation for four-point correlation functions of large BPS
operators, as recalled in~\figref{fig:recall}. The octagon is also the
fundamental building block for these correlators beyond the planar
limit \cite{Bargheer:2019kxb}.
In~\cite{Coronado:2018cxj} the octagon was bootstrapped, providing an
all-loop weak-coupling perturbative expansion, and
in~\cite{Kostov:2019stn} a beautiful finite-coupling representation
in terms of an infinite-dimensional Pfaffian was provided. In this small
note, we study the octagon at strong coupling.
Our study is split in two parts: The derivation of the strong-coupling
result and its analysis.

\begin{figure}[tb]
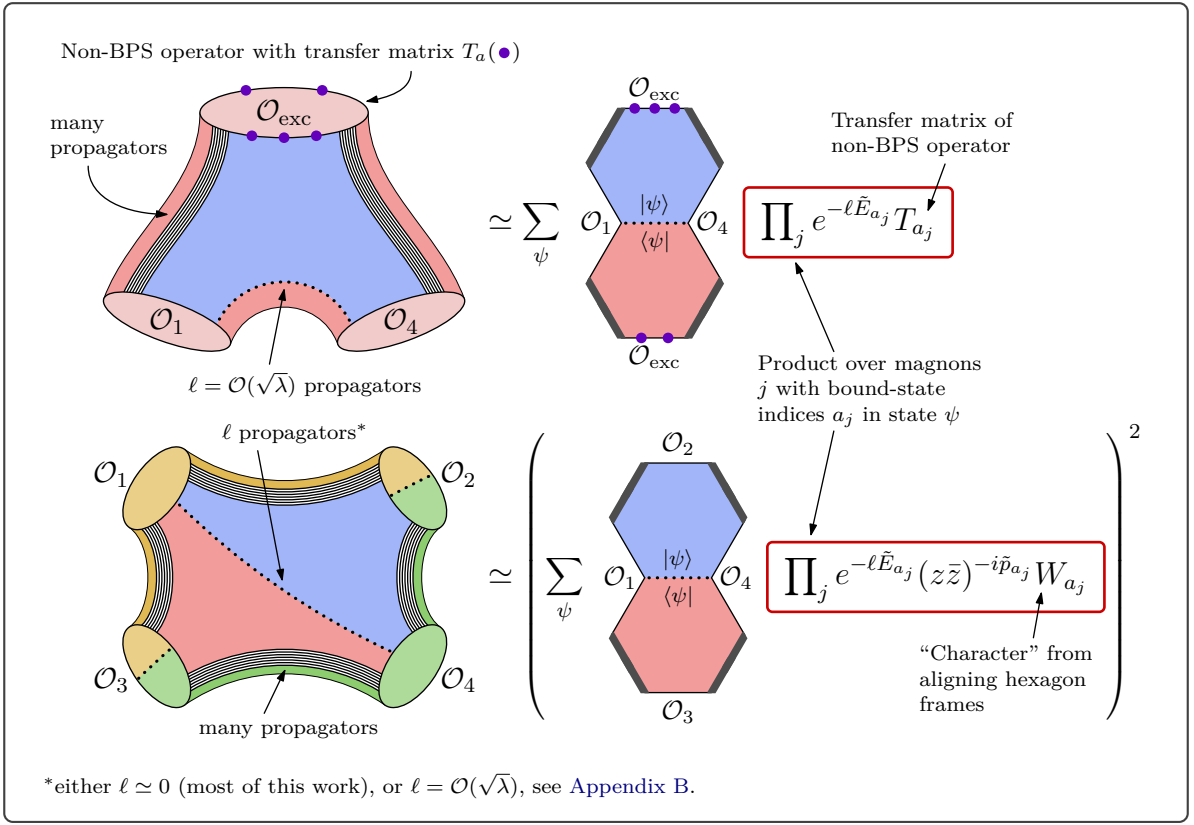

\centering
\begin{tcolorbox}[width=-0.8cm+\textwidth, top=-3pt, colback=white, boxrule=0.8pt]
\begin{align*}
\includegraphics[align=c,vshift=0.05cm,rmargin=-0.6cm]{Fig3ptWSHexcut}
&\simeq
\sum_\psi
\mspace{8mu}
\includegraphics[align=c]{Fig2hex3pt}
\mspace{7mu}
\tcboxmath[
    colframe=red!80!black,
    colback=white,
    size=small,
    boxrule=1.0pt
    ]{\prod\nolimits_j e^{-\ell\tilde{E}_{a_j}}T_{a_j}}
\includegraphics[smash=tc,vshift=0.25cm,hshift=-0.1cm]{FigLabelTransferMatrix}
\\[1ex]
\includegraphics[align=c,vshift=-0.05cm]{Fig4ptWSHexcut}
&\simeq
\brk*{
\sum_\psi
\mspace{8mu}
\includegraphics[align=c]{Fig2hex4pt}
\mspace{7mu}
\tcboxmath[
    colframe=red!80!black,
    colback=white,
    size=small,
    boxrule=1.0pt
    ]{
      \prod\nolimits_j
      \includegraphics[smash=tc,vshift=0.65cm,hshift=0.5cm]{FigLabelMagnonProduct}
      e^{-\ell\tilde{E}_{a_j}}
      (z\bar z)^{-i\tilde{p}_{a_j}}
      W_{a_j}
      \includegraphics[smash=bc,vshift=-0.3cm,hshift=-0.7cm]{FigLabelCharacter}
    }
}^2
\end{align*}
{\scriptsize $^*$either $\ell\simeq0$ (most of this work), or $\ell=\order{\sqrt{\lambda}}$, see~\appref{sec:withBridge}.}
\end{tcolorbox}
\caption{\emph{Top:} Wrapping correction in the ``bottom'' channel of a
three-point function between two BPS and one non-BPS operator; these were re-summed
in~\cite{Jiang:2016ulr}. \emph{Bottom:} Virtual corrections for the
``simplest'' four-point function ($=$ octagon$^2$). We see that they exhibit
strong similarities: Both sums run over the same states $\psi$, only
the weights (red boxes) are different. The upper weight is a product
of transfer matrix eigenvalues associated
to the non-BPS operator; the lower weight is a product of
characters associated to the cross-ratios of the four-point function.
Under a simple replacement, we can thus re-use the clustering
analysis of~\cite{Jiang:2016ulr} to derive the strong-coupling octagon
representation, see the final expression~\eqref{eq:prediction} below. }
\label{fig:PantsVs4p}
\end{figure}

The octagon is obtained by gluing two hexagon form factors together
along one common edge. At weak coupling, each edge along which two
hexagons are glued together becomes a ``bridge'' of planar
propagator contractions between physical operators participating in
the correlator. The number of propagators in a bridge is called the
``bridge length'', and it is a measure of the distance between the two
adjacent hexagons. When the bridge length is asymptotically large, the
two hexagons decouple. Beyond the asymptotic regime, one has to sum over
a complete basis of virtual excitations (mirror magnons) that
propagate across the bridge between the two hexagons. These
excitations capture the finite-size effects within the correlator, and
computing their sum is difficult in general, especially when the
mirror magnons can propagate across multiple bridges. In our case, the
octagon is framed by asymptotically large bridges, and hence the
mirror magnons are confined to a single bridge that splits the octagon
into two hexagons. It turns out that this setup is very similar to the
case of a three-point function between two BPS operators and one
non-BPS operator, as illustrated in~\figref{fig:PantsVs4p}.
This case was considered by Komatsu,
Kostov, Serban, and Jiang~\cite{Jiang:2016ulr}, and we can follow
their ``clustering'' analysis almost verbatim.

In the three-point function case, the bridges between the non-BPS
operator and the two BPS operators are taken to be large, such that
the mirror magnons are confined to the single bridge that connects the
two BPS operators to each other. The sum over mirror magnons is
weighted by the transfer matrix of the non-BPS operator, which
accounts for the interaction between the mirror magnons and the
physical magnons on the non-BPS operator. If the third operator were
also BPS, the correlator would be protected and the sum over mirror
magnons would be trivial. The non-trivial weight breaks supersymmetry
and thus leads to a non-trivial result. The authors
of~\cite{Jiang:2016ulr} have shown how to evaluate this
``bottom-wrapping'' non-trivial sum at strong coupling.

In the octagon case, the mirror magnons also live on a bridge
connecting two BPS operators. This time, the opposite sides of the two
hexagons do not connect to the same non-BPS operator, but to two
\emph{different} BPS operators. In order to perform the sum over
mirror magnons, the two hexagons have to be brought to the same frame
by a $\grp{PSU}(2,2|4)$ transformation that maps the two different BPS
operators onto each other. This change of frame induces a non-trivial
character-like Boltzmann weight into the sum over mirror magnons.
Again, this weight breaks supersymmetry and leads to a non-trivial
result.

We thus see that the two cases are very similar at a technical level,
and that is why we can recycle the analysis of~\cite{Jiang:2016ulr}
rather efficiently. We simply have to spot and replace the transfer
matrices of the three-point case by the character weights of the
octagon.
An important part of the analysis in~\cite{Jiang:2016ulr} was that
the energy of mirror particles is constrained to be small, of order
$1/\sqrt{\lambda}$, because they are multiplied by the length of
their supporting bridge which is of order $\sqrt{\lambda}$. Here,
$\lambda$ is the (large) 't~Hooft coupling. In the octagon case, the
mirror particles are weighted by their mirror momenta multiplied by
(logarithms of) space-time cross ratios. The latter are naturally of
order $1$, and the mirror momenta is also of order $1$ precisely when the
mirror energy is of order $1/\sqrt{\lambda}$. Hence the kinematical
regime in the octagon case is exactly the same as in the three-point
function case.

Having realized this, the derivation exercise becomes rather
straightforward and is presented in~\secref{Derivation}. The reader
might want to skip directly to the final result,
equation~\eqref{eq:prediction} below. We observe a nice exponentiation as
\begin{equation}
\mathbb{O}(z,\bar z | \lambda)
\simeq
e^{-\sqrt{\lambda} \,\mathbb{A}(z,\bar z)}
\label{Oexp}
\end{equation}
As explained in~\cite{Bargheer:2019kxb}, $\mathbb{A}$ should be the minimal
area of a string that ends on four BMN geodesics in AdS \emph{and}
rotates in the sphere, as sketched in~\figref{fig:recall}. The fact
that the string moves in both AdS and the sphere makes it quite
non-trivial to compute this minimal area from the string sigma model.%
\footnote{In~\appref{sec:minimal-areas-ending}, based on discussions with
Martin Kruzcenski, we comment on how the simpler problem of minimal
areas in AdS ending on geodesics -- without any sphere -- can often be
solved.}
Still, the form of $\mathbb{A}$ clearly indicates that it should be
possible to develop some Y-system like technology as
in~\cite{Alday:2009dv,Alday:2010vh,Janik:2011bd,Kazama:2011cp,Kazama:2012is,Caetano:2012ac}
to directly compute this minimal area at
strong coupling, starting from the string sigma model. It would be very
interesting to study this problem.

In~\secref{sec:Analysis}, we analyze this result. We note that the area
$\mathbb{A}$ is real and positive in the Euclidean
regime where
\begin{equation}
z,\bar z = e^{-\varphi\pm i \phi}
\label{eq:crossRatios}
\end{equation}
with $\varphi$ and $\phi$ both real, and we explore what happens as we
analytically continue the cross-ratios to various other interesting
kinematical regimes, such as the Lorentzian regime and various OPE-like
limits. We also make contact with~\cite{Bargheer:2019kxb}, and explore the
consequences of these results for the full non-planar expansion of the
correlator of four large BPS operators.
We conclude with some speculations and open problems
in~\secref{Conclusions}.

\section{Derivation}
\label{Derivation}

As explained in the introduction, we can recycle the results
of~\cite{Jiang:2016ulr} to obtain the expression for the octagon
function $\oct$ at strong coupling. To see how this comes about, let
us briefly review the construction of the octagon.

\begin{myinset}
{\bf Brief Review.} The octagon $\oct_\ell$
is obtained by fusing two hexagon operators along a common mirror
edge,
%
%
\begin{equation}
\includegraphics[align=c]{FigOctagon}
=\sum_{\psi}\,
\includegraphics[align=c]{FigTwoHexagons}
\label{eq:oct=2hex}
\end{equation}
%
%
The fusion amounts to summing and integrating over a complete basis of
states $\psi$ on the common mirror edge. Such a basis is given by all
$n$-magnon states, where each magnon is completely characterized by a
rapidity $u$, a bound-state index $a$,%
\footnote{A magnon with bound-state index $a$ is composed of $a$
fundamental magnons.}
and indices $A$, $\dot A$ for the $a$'th antisymmetric representations of
the two (left and right) $\alg{su}(2|2)$ algebras~\cite{Basso:2015zoa,Fleury:2016ykk}:
\begin{equation}
\oct_\ell=\int\brk[s]{\dd\psi}\,
\mu_\psi\,e^{-\tilde E_\psi\ell}
\braket{\mathcal{H}_1}{\psi}
\braket{\psi}{\mathcal{H}_2}
\,,\qquad
\int\brk[s]{\dd\psi}=
\sum_{n=0}^\infty
\frac{1}{n!}
\sum_{\substack{a_i=1\\i=1..n}}^\infty
\sum_{A_i,\,\dot A_i}
\int \dd u_1\dots \dd u_n
\,.
\end{equation}

Each state in the sum is weighted by a measure factor $\mu_\psi$ for
its creation, and a factor $e^{-\tilde E_\psi\ell}$ for its propagation
across $\ell$ propagators, where $\tilde E_\psi$ is the (mirror) energy of $\psi$.
In this work, we mostly consider $\ell=0$, and hence $e^{-\tilde E_\psi\ell}=1$.%
\footnote{Our strong-coupling result remains correct for $l\neq0$, as
long as $\ell\ll \sqrt{\lambda}$. For $\ell\sim{\sqrt{\lambda}}$,
the result gets modified, but is still calculable, see~\appref{sec:withBridge}.}
Since we only study correlators of BPS operators, and since all
outer mirror edges of the octagon will be occupied by a large number
of propagators, there will be no (mirror) magnons on any of the outer
edges of the two hexagons.

The evaluation of a hexagon amplitude requires a choice of
``frame'', which is defined by the spacetime positions and
R-symmetry orientations of the three BPS ``vacuum'' operators that
attach to the three physical edges of the hexagon.%
\footnote{Three half-BPS vacuum operators are preserved by a common
(diagonal) $\alg{su}(2|2)$ algebra~\cite{Drukker:2009sf}, which fixes
the hexagon amplitude to a large extent\cite{Basso:2015zoa}.}
The two hexagons defining the octagon share two external operators
($1$ and $2$ in~\eqref{eq:oct=2hex}),
but the third BPS operators attaching to the two hexagons are
generically different (operators $3$ and $4$ in~\eqref{eq:oct=2hex}).
In order to consistently perform the sum over
mirror states, the frames have to be aligned by a finite
$\grp{PSU}(2,2|4)$ transformation $g$ that maps one of these two different operators
onto the other~\cite{Fleury:2016ykk}: Choosing the first hexagon to be canonical
$\mathcal{H}_1=\mathcal{\hat H}$, the second hexagon is related to the
first by conjugation with $g$: $\mathcal{H}_2=g\mathcal{\hat
H}g^{-1}$. The transformation $g$ is composed of a dilatation, a
Lorentz rotation, and similar transformations in internal R-symmetry
space. It is diagonal in the multi-magnon state basis, and hence
amounts to a weight factor $\mathcal{W}_\psi$ in the sum over mirror states. The octagon
function $\oct_{\ell=0}$ therefore becomes
\begin{equation}
\oct\equiv\oct_{\ell=0}
=\int\brk[s]{\dd\psi}\,
\braket{\mathcal{\hat H}}{\psi}
\mu_\psi
\mathcal{W}_\psi
\braket{\psi}{\mathcal{\hat H}}
\,,\qquad
\mathcal{W}_\psi=
\bra{\psi}g\ket{\psi}=
e^{-i\tilde p_\psi\log(z\bar z)}
\,e^{iL_\psi\phi}
\,e^{iR_\psi\theta}
\,e^{iJ_\psi\varphi}
\,.
\label{eq:octagonexpression}
\end{equation}
Here, $\tilde p_\psi$ is the mirror momentum with $2i\tilde p_\psi=D_\psi-J_\psi$,
and $D_\psi$, $L_\psi$, $R_\psi$ as well as $J_\psi$ are the charges
of the state $\psi$ under the dilatation, Lorentz rotation, and the
corresponding rotations in R-symmetry space that make up the
transformation $g$, see~\cite{Fleury:2016ykk}.
For the case relevant to us, they take the values%
\footnote{The octagon is most clearly isolated in correlators of the
type considered in~\cite{Coronado:2018ypq,Bargheer:2019kxb}, where two
of the operators are BPS primaries $\op{O}_1=\tr\brk{X^{2k}}$, $\op{O}_4=\tr\brk{Z^{2k}}$, and
two operators are BPS descendants
$\op{O}_2=\op{O}_3=\tr\brk{\bar{Z}^k\bar{X}^k}+\text{(permutations)}$.
See~\appref{sec:operators} for more details and how this implies $\alpha=\bar\alpha=1$.}
\begin{equation}
\theta=0
\,,\qquad
e^{-\varphi+i\phi}=z
\,,\qquad
e^{-\varphi-i\phi}=\bar z
\,,
\label{eq:alphafix}
\end{equation}
where $z$ and $\bar z$ as usual parametrize the spacetime cross ratios:
\begin{equation}
z\bar z=\frac{x_{12}^2x_{34}^2}{x_{13}^2x_{24}^2}
\,,\qquad
(1-z)(1-\bar z)=\frac{x_{14}^2x_{23}^2}{x_{13}^2x_{24}^2}
\,.
\label{eq:zzbar}
\end{equation}
The sum over magnon flavors $A_i$, $\dot A_i$ can be performed
and gives~\cite{Coronado:2018ypq}
\begin{gather}
\oct=
\int\brk[s]{\dd\psi}'\,
(z\bar z)^{-i\tilde p_\psi}
W_{\brk[c]{a_i}}
\mu_\psi
\prod_{i<j}P_{a_i,a_j}(u_i,u_j)
\,,\qquad
\int\brk[s]{\dd\psi}'=
\sum_{n=0}^\infty
\frac{1}{n!}
\sum_{\substack{a_i=1\\i=1..n}}^\infty
\int \dd u_1\dots \dd u_n
\,,
\label{eq:flavorsummed}
\end{gather}
where $P_{ab}(u,v)$ is a function of the rapidities $u$, $v$, and
bound-state indices $a$, $b$, see (A.8) in~\cite{Coronado:2018ypq}.
The weight factor $(z\bar z)^{-i\tilde p_\psi}W_{\brk[c]{a_i}}$
contains the ``character'' $W_{\brk[c]{a_i}}$.
With~\eqref{eq:alphafix}, it takes the form
\begin{equation}
W_{\brk[c]{a_i}}=\prod_{j=1}^nW_{a_j}
\,,\qquad
W_{a}=
-4\sinh\brk*{\frac{\varphi}{2}+\frac{i\phi}{2}}
\sinh\brk*{\frac{\varphi}{2}-\frac{i\phi}{2}}
\frac{\sin(a\phi)}{\sin\phi}
\,.
\label{eq:character}
\end{equation}
{\bf Identification.} The formula~\eqref{eq:flavorsummed} for the complete octagon sum
closely resembles the ``bottom wrapping''
expression of~\cite{Jiang:2016ulr} (see eqs.~(5.2) and~(5.3) there) that sums all
mirror excitations on a mirror edge between two BPS operators in a
three-point function with another non-BPS operator, see~\figref{fig:PantsVs4p}. In that
latter case, each state $\psi$ in the mirror state sum is weighted by the
product $\prod_jT_{a_j}(u_j)$ over constituent magnons $j$ of $\psi$,
with $a_j$ and $u_j$ being the $j$'th magnon's bound-state index and
rapidity, and $T_a$ being the transfer matrix eigenvalue
of the third (non-BPS) operator. Instead of this product over
transfer matrix eigenvalues, our sum~\eqref{eq:flavorsummed} is weighted by the
factor $(z\bar z)^{-i\tilde p_\psi}W_{\brk[c]{a_i}}$  that originates in the
misalignment of the two ``opposite'' BPS operators of our four-point
function. Importantly, this weight also factorizes into a product over
the $n$ constituent magnons of $\psi$:
\begin{equation}
(z\bar z)^{-i\tilde p_\psi}W_{\brk[c]{a_i}}
=\prod_{j=1}^n(z\bar z)^{-i\tilde{p}_{a_j}\!(u_j)}W_{a_j}
\,.
\end{equation}
This factorization is due to the additivity of the mirror momentum
$\tilde{p}$ and the factorization of the
character~\eqref{eq:character}.
We therefore recover the expression~(5.3) of~\cite{Jiang:2016ulr} for
the bottom wrapping sum by identifying the transfer matrix eigenvalue
$T_a(u)$ in that expression as $T_a(u)=(z\bar z)^{-i\tilde p_a(u)}W_a$.%
\footnote{The bottom wrapping expression~(5.3)
in~\cite{Jiang:2016ulr} is written as a sum over occupation numbers $n_a$ of
bound-state indices $a$. Since all factors in the integrand
of~\eqref{eq:flavorsummed} and in~\cite{Jiang:2016ulr} are symmetric under permutations of magnons
with identical bound-state indices, it is easy to
convert back and forth between a sum over the total number $n$ of
constituent magnons and sums over occupation numbers $n_a$. In
particular, the weight $(z\bar z)^{-i\tilde p_\psi}W_{\brk[c]{a_i}}$
can equally be written as
$(z\bar z)^{-i\tilde p_\psi}W_{a_i}
=\prod_{a=1}^\infty
\prod_{j_a=1}^{n_a}
(z\bar z)^{-i\tilde p_a(u^a_{j_a})}W_a$,
where $u^a_1,\dots,u^a_{n_a}$ are the rapidities of the $n_a$ magnons
with bound-state index $a$, and $\tilde p_a(u)$ is the mirror momentum of the
$a$'th bound state.}
This straightforward identification at the level of individual magnons
makes the ``clustering'' analysis of~\cite{Jiang:2016ulr} also
applicable to our case, and lets us directly re-use their final
re-summed result (eqs.~(5.49) and~(5.43) there)!
To state the final result, we only need the strong-coupling expression
of the mirror bound-state momentum
\begin{equation}
\tilde p_a(u)=u-g \brk*{\frac{1}{x^{\brk[s]{+a}}}+\frac{1}{x^{\brk[s]{-a}}}}
\,,
\label{eq:mirrormomentum}
\end{equation}
with the Zhukowsky variable $x(u)$ and the shorthand $x^{\brk[s]{\pm
a}}$ being defined via
\begin{equation}
\frac{u}{g}=x+\frac{1}{x}
\qquad \Rightarrow \qquad
x(u)=\frac{u+\sqrt{u-2g}\sqrt{u+2g}}{2g}
\,,\qquad
x^{\brk[s]{\pm a}}(u)\equiv x\brk{u\pm a\,i/2}
\,.
\end{equation}
At strong coupling,~\eqref{eq:mirrormomentum} becomes
\begin{equation}
\tilde p_a(u)\simeq \frac{a\,u}{2 \sqrt{4g^2-u^2}}
\qquad \text{for} \qquad
g\to\infty
\,.
\end{equation}
With $\log(z\bar z)=-2\varphi$, we thus find at strong coupling
\begin{equation}
T_a(u)=(z\bar z)^{-i\tilde p_a(u)}W_a
\simeq
\exp\brk3{a\varphi\frac{iu}{\sqrt{4g^2-u^2}}}W_a
\,.
\end{equation}

\end{myinset}

In summary, we can immediately recycle the
results of~\cite{Jiang:2016ulr} with appropriate replacements of the
non-BPS transfer matrices there by the characters produced by the two
differently aligned hexagons in the four-point function. As explained
above, a simple comparison of
the starting point in~\cite{Jiang:2016ulr} with the octagon infinite sum representation
indicates that we should take
\begin{equation}
T_a=
\overbrace{e^{a \varphi \frac{i u}{\sqrt{4g^2-u^2}}}}^{(z\bar{z})^{\,\texttt{mirror momentum}}}
\,\underbrace{X\,\sin(a\phi)}_{\texttt{character}}
\,,\qquad
X\equiv
\overbrace{
-\frac{
4\sinh\left(\frac{\varphi}{2}+\frac{i\phi}{2}\right)\sinh\left(\frac{\varphi}{2}-\frac{i\phi}{2}\right)
}{
\sin(\phi)
}}^{\mathclap{a\texttt{-independent part of character }X=\frac{-2i(1-z)(1-\bar z)}{z-\bar z} }}
\label{eq:Ta}
\end{equation}
and plug it into the final strong-coupling expression (5.49) and
(5.43) in~\cite{Jiang:2016ulr}. This gives
\begin{equation}
\log \mathbb{O}  \simeq  \int\limits_{-2g}^{+2g} \frac{du}{2\pi}
\sum_{n=1}^{\infty} \frac{1}{n} \sum_{\{n_a\}} (-1)^{(K-1)} (K-1)!
\prod_{a} \frac{(T_a)^{n_a} }{n_a!}   \,,
\label{eq:startingPoint}
\end{equation}
where the sum over the positive mode numbers $n_a$ is constrained as
$\sum_a a\, {n_a}=n$, and where $K$ is defined as $K\equiv \sum_a n_a$.
What follows is a straightforward simplification of this expression.
The reader might want to jump to the final simplified
result~\eqref{eq:prediction}.

\begin{myinset}
{\bf Simplification.} The goal is to factorize the integrand in a way that we can decouple
the sum over mode numbers $n_a$ into independent sums which we can then
perform. For instance, to factorize the factorial factor we simply
write $(K-1)!=\int dt e^{-t} t^{K-1} $ so that $K=\sum_a n_a$ appears
in exponents and thus breaks apart into a product of factors for the
various mode numbers. Next we want to cancel the $1/n$ factor
in~\eqref{eq:startingPoint} which is easy by a simple integration by
parts.
For that we note that the rapidity $u$ dependence only arises through
the exponential factor in $T_a$; using the definition of $n$ we can
take it out of the $n$ sum completely and write
\begin{align}
\frac{1}{n}
\int_{-2g}^{2g}du\,e^{i n \varphi u/\sqrt{4g^2-u^2}}
& = \frac{2g}{n} \int_{-\infty}^{\infty} d\theta\,
\frac{d \tanh(\theta)}{d\theta} e^{i n \varphi \sinh(\theta)}
&&
\Big|\, u=2g \tanh(\theta)
\nn\\
&= -2g i \varphi \int_{-\infty}^{\infty} d\theta\,
\sinh(\theta) e^{i n \varphi \sinh(\theta)}
&&
\Big|\, \text{integration by parts}
\label{eq:IBP}
\\
&= 2g \varphi \int_{-\infty}^{\infty} d\theta\,
\cosh(\theta) e^{-n \varphi \cosh(\theta)}
&&
\Big|\, \theta\to \theta+i \pi/2 \text{ shift}
\label{eq:shift}
\end{align}
Here we assumed
$\text{Re}(\varphi)>0$, otherwise we should pick an opposite shift in
\eqref{eq:shift}.
We obtain the desired factorization
\begin{equation}
\log\oct\simeq
-\frac{g}{\pi}\,\varphi
\int\limits_{-\infty}^\infty d\theta\,\cosh(\theta)
\int\limits_0^\infty\frac{dt}{t}e^{-t}
\sum_{n=1}^{\infty} \sum_{\{n_a\}} \prod_{a}
\frac{\brk!{-X t \sin(a\phi)\,e^{-a \varphi \cosh\theta}}^{n_a}}{n_a!}
\,,
\end{equation}
where $n=\sum a\,n_a$.
Adding and subtracting a $n=0$ term
(corresponding to all $n_a=0$) we get a final factorization into
\emph{unconstrained} mode numbers as
\begin{equation}
\log \mathbb{O}\simeq -
\frac{g}{\pi}\,\varphi
\int\limits_{-\infty}^\infty d\theta\, \cosh(\theta)
\int\limits_0^\infty \frac{dt}{t}e^{-t}
\brk[s]*{
\prod_{a=1}^{\infty}\sum_{n_a=0}^\infty
\frac{\brk!{-X t \sin(a\phi)\,e^{-a \varphi \cosh\theta}}^{n_a}}{n_a!}-1
} \,,
\end{equation}
We can now perform the sum over the mode numbers $n_a$,
\begin{equation}
\log\mathbb{O}\simeq
-\frac{g}{\pi}\,\varphi
\int\limits_{-\infty}^\infty d\theta\, \cosh(\theta)
\int\limits_0^\infty \frac{dt}{t}e^{-t}
\brk[s]*{
\exp\brk*{-\sum\limits_{a=1}^\infty t X  \sin(a\phi) e^{-a \varphi
\cosh(\theta)}}  -1
}\,,
\end{equation}
The sum over $a$ can also be done, leading to
\begin{equation}
\log\mathbb{O}\simeq
- \frac{g}{\pi}\,\varphi
\int\limits_{-\infty}^\infty d\theta\, \cosh(\theta)
\int\limits_0^\infty \frac{dt}{t}e^{-t}
\brk[s]*{ e^{-t Y(\theta)}-1} \,,
\end{equation}
where $Y(\theta)$ is given by~\eqref{eq:Yfunction} below. Finally, the
integral over $t$ yields $-\log(1+Y)$ and so we
obtain~\eqref{eq:prediction} once we recall the relation to the 't Hooft
coupling $g=\sqrt{\lambda}/4\pi$.
\end{myinset}

\paragraph{Result.}

\begin{figure}[t]
\includegraphics{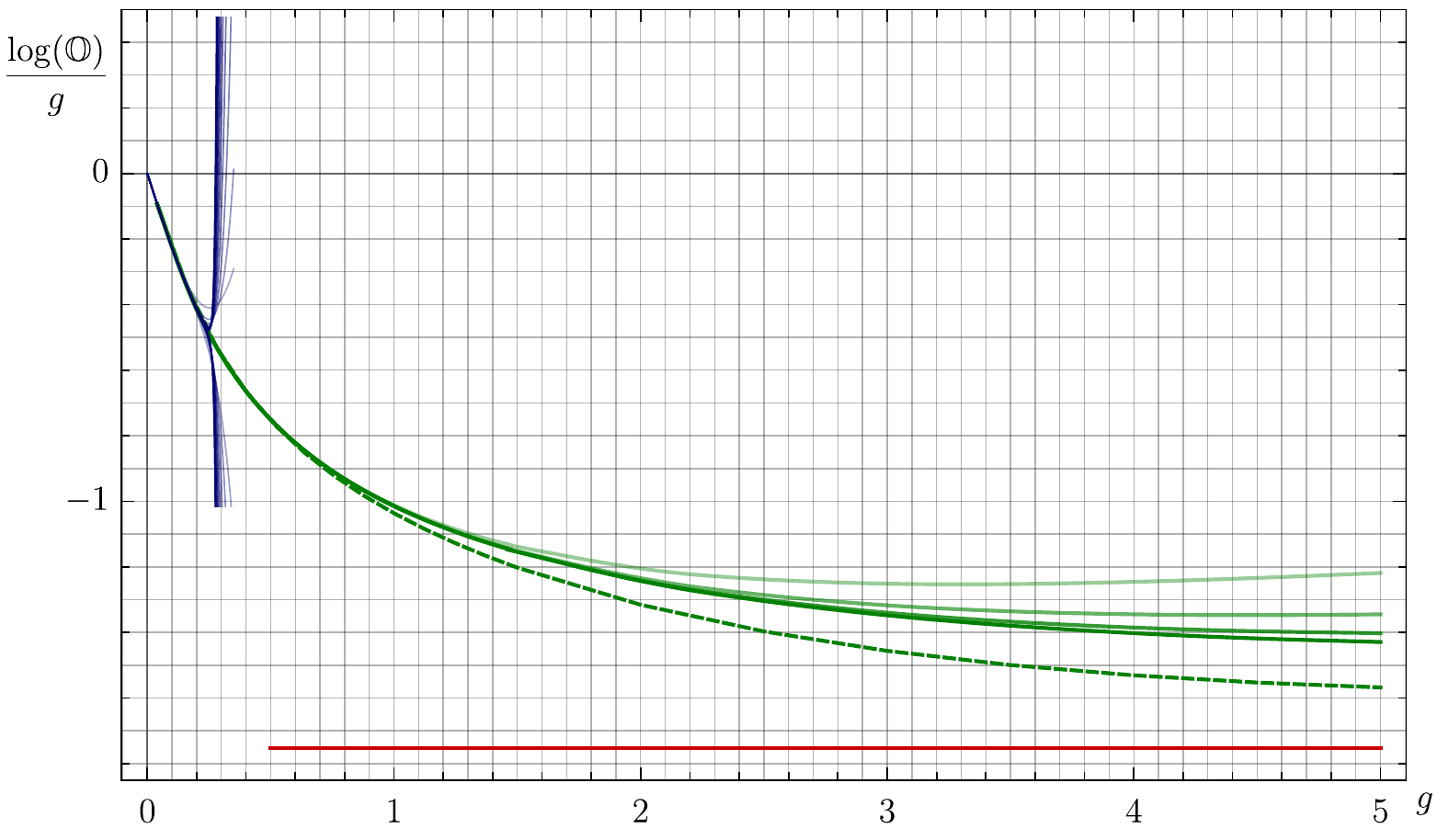}
\caption{Comparison of various data for the ratio $\log(\oct)/g$ as a
function of the coupling $g$. The thin blue lines that diverge near
the radius of convergence at $g=1/4$ are the perturbative
results of~\cite{Coronado:2018ypq}, from two loops all the way to $20$
loops. The solid green lines
are the numerical evaluation of the
determinant representation of~\cite{Kostov:2019stn}. The latter nicely agrees with
the perturbative representation, and continues it beyond its radius of
convergence. To evaluate the determinant, we truncated the
semi-infinite matrix to sizes $N=10,15,20,25$ with darker lines
corresponding to larger sizes; clearly it becomes more and more
important to consider larger matrices at strong coupling. The dashed
green line represents an extrapolation of these results towards
infinite matrices (using the results from $N=2$ to $N=25$ and a simple
fit $a+b/N$). Finally, the red horizontal line is the strong-coupling
prediction in~\protect\eqref{eq:prediction}. In this plot, we use
Euclidean cross-ratios $\varphi=1/10$, $\phi=\pi/3$. }
\label{fig:comparison}
\end{figure}

We thus find
\begin{equation}
\log\mathbb{O} \simeq  \frac{\sqrt{\lambda}}{2\pi }
\int\limits_{-\infty}^\infty \frac{d\theta}{2\pi} \,\varphi
\cosh(\theta)  \log(1+Y(\theta)) \,,
\label{eq:prediction}
\end{equation}
where
\begin{equation}
Y(\theta)= - \frac{\sin\big(\frac{\phi}{2}+i
\frac{\varphi}{2}\big)\sin\big(\frac{\phi}{2}-i
\frac{\varphi}{2}\big)}{\sin\big(\frac{\phi}{2}+i \frac{\varphi}{2}
\cosh(\theta)\big)\sin\big(\frac{\phi}{2}-i \frac{\varphi}{2}
\cosh(\theta)\big)} \,.
\label{eq:Yfunction}
\end{equation}
We derived this result from the octagon expression as two fused
hexagons. It would be very nice to derive
it from the string world-sheet \textit{a la}~\cite{Alday:2009dv}. The
TBA-like result~\eqref{eq:prediction} is very reminiscent of the type of
expressions coming out in those papers from exploring the classical
string integrability. Note in particular that
\begin{equation}
1+Y=\frac{
\sinh\brk*{\frac{\varphi}{2}-\frac{\varphi}{2}\cosh(\theta)}
\sinh\brk*{\frac{\varphi}{2}+\frac{\varphi}{2}\cosh(\theta)}
}{
\sinh\brk*{\frac{i\phi  }{2}-\frac{\varphi}{2}\cosh(\theta)}
\sinh\brk*{\frac{i\phi  }{2}+\frac{\varphi}{2}\cosh(\theta)}
}
\label{1pY}
\end{equation}
takes a very nice factorized form, which allows us to split the
putative area in~\eqref{eq:prediction}; perhaps one contribution will
come from the AdS part and another from the sphere. We can also
include a finite internal bridge length in the octagon
(see~\appref{sec:withBridge}), and try to reproduce that more
general result from the string world-sheet.

In~\figref{fig:comparison}, we compare the strong-coupling result~\eqref{eq:prediction} to
the finite-coupling representation of the octagon recently worked out
in~\cite{Kostov:2019stn}. It looks consistent, but it would be very nice to work out
the one-loop prefactor in~\eqref{Oexp}, and to improve the
determinant evaluation at strong coupling to perform a more
conclusive comparison. Furthermore, if we could compute
the one-loop prefactor from the octagon representation, it would
provide us with yet another powerful data point to reproduce
from the string sigma model.

The result~\eqref{eq:prediction} was derived for real $\phi \in [0,2\pi]$ and
real $\varphi>0$ (see the shift~\eqref{eq:shift}). This translates into $\bar z=z^*$ and $|z|,|\bar z|<1$. Of
course, we can (and will) move away and study any range of parameters
-- both real and complex -- but we need to carefully analytically
continue the result starting from this safe starting point. This is
particularly obvious even if we remain in the fully Euclidean region
where $\phi$ and $\varphi$ are both real. The $Y$-function in~\eqref{eq:Yfunction}
is \emph{even} under $\varphi \to -\varphi$, but
because of the $\varphi$ outside the log in the area~\eqref{eq:prediction},
the integrand is odd, and thus it seems
that the full result is \emph{odd}. That is wrong. The full
area is \emph{even}, nicely realizing the $z\leftrightarrow 1/{\bar z}$ symmetry of the
octagon. To see that, however, is a bit non-trivial. It turns out that
as we rotate from $\varphi >0 $ to $\varphi <0$, infinitely many
singularities hit the integration contour, which therefore needs to be
re-arranged non-trivially. This
produces an additional minus sign required to convert the \emph{naive
odd} guess into the \emph{correct even} result. This contour
re-arrangement illustrates very nicely the kind of
manipulations involved in more general analytic continuations, and is presented
in detail in \appref{sec:EvenAp}.

\section{Analysis}
\label{sec:Analysis}

\begin{figure}
\centering
\includegraphics[align=c]{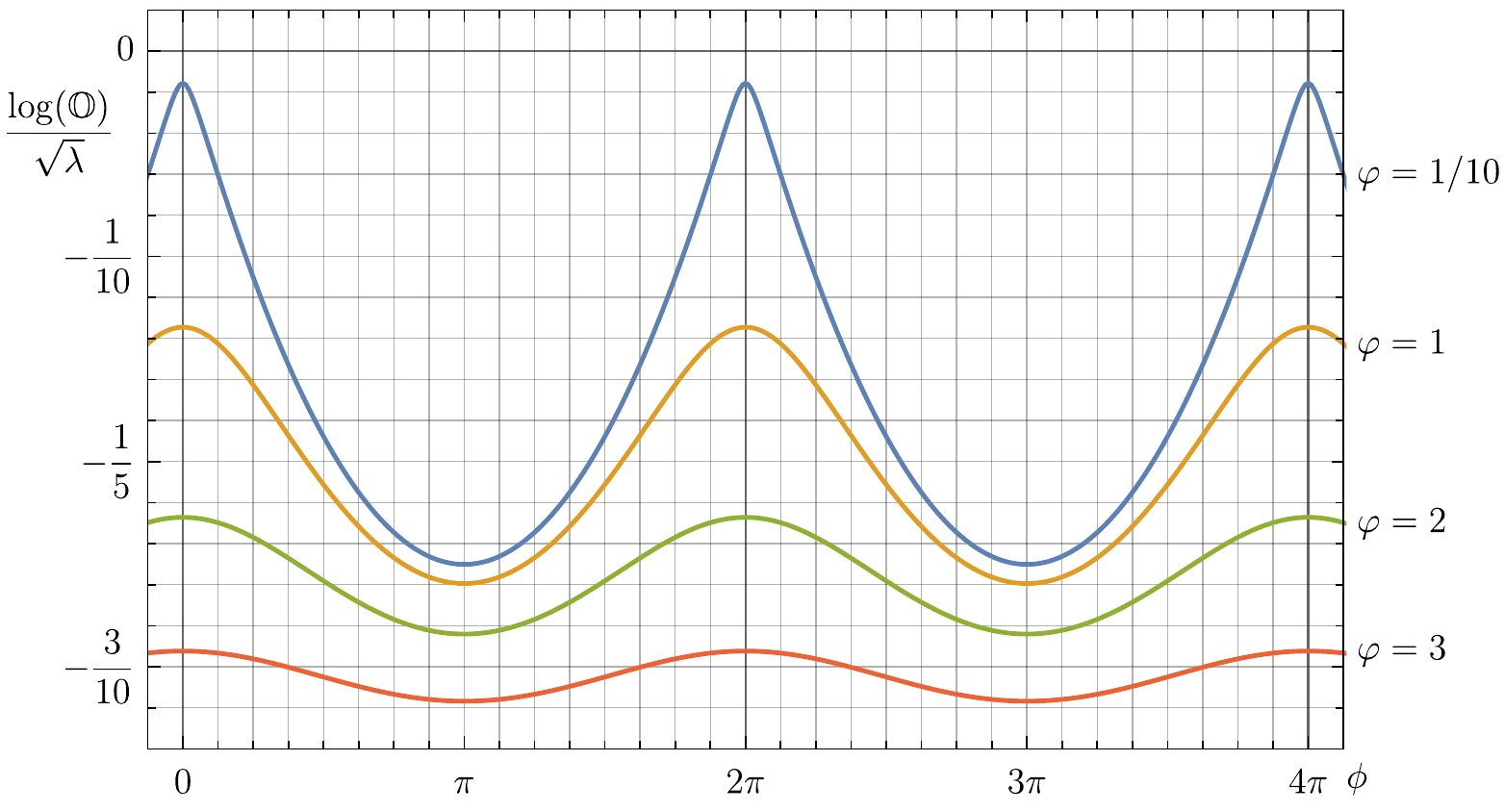}
\caption{In the Euclidean regime, the area is single-valued on the
physical sheet, \ie it is periodic in the angle $\phi$. As a function
of this angle, we see that it develops kinks as $\varphi \to 0$, and
approaches a constant as $\varphi \to \infty$.}
\label{fig:plotEuclidean}
\end{figure}

In the Euclidean regime, the two cross-ratio variables $z$ and $\bar z$ are complex conjugate to
each other, and therefore~$\varphi$ and~$\phi$ in~\eqref{eq:crossRatios}
are real. Then the area
\begin{equation}
\log\mathbb{O}
\simeq
\frac{\sqrt{\lambda}}{2\pi}
\int\limits_{-\infty}^\infty
\frac{d\theta}{2\pi}
\,\varphi\cosh(\theta)
\log\brk[s]*{1-\frac{
\sin\big(\frac{\phi}{2}+i\frac{\varphi}{2}\big)\sin\big(\frac{\phi}{2}-i\frac{\varphi}{2}\big)
}{
\sin\big(\frac{\phi}{2}+i\frac{\varphi}{2}\cosh(\theta)\big)\sin\big(\frac{\phi}{2}-i\frac{\varphi}{2}\cosh(\theta)\big)
}}
\,,
\label{eq:predictionA}
\end{equation}
is manifestly real. The logarithm is negative and  $\varphi$ is positive, thus the full integrated right hand side is negative. It is also
manifestly periodic in
$\phi$, leading to a single-valued expression in the Euclidean regime,
as expected (see \figref{fig:plotEuclidean}).
Since it is multiplied by a large
string tension $\sqrt{\lambda}$, we see that
the octagon is exponentially small in the Euclidean regime, and thus
\begin{equation}
\mathbb{O} \to 0
\quad\text{when}\quad
\lambda \to \infty
\,.
\label{eq:octtozero}
\end{equation}
This has nice implications for the
double-scaling limit of the ``simplest'' correlator of~\cite{Coronado:2018ypq},%
\footnote{See~\secref{sec:operators} for a brief review.}
where the charges of the external operators scale as
$\sqrt{\Nc}$. This limit was
analyzed in~\cite{Bargheer:2019kxb}. Since all configurations that
involve the octagon $\oct$ vanish due to~\eqref{eq:octtozero},
the full non-planar
correlator in the Euclidean regime reduces to a sum of BMN-like
configurations, in which the string world-sheet degenerates into various
point-like geodesics. Such configurations first show up at genus one.
All in all, these BMN configurations re-sum into the explicit
expression~(4.3) in~\cite{Bargheer:2019kxb}.

The octagon expression~\eqref{eq:predictionA} also exhibits a rich
behavior in various interesting kinematical regimes, as we will now
explore. To study these limits, it is useful to derive two
mathematical formulae for our expression. The first is
\begin{equation}
\log\mathbb{O}
\simeq
-\frac{\sqrt{\lambda}}{4\pi^{3/2}}
\sqrt{-\log(z\bar z)}
\brk*{
 \Li_{3/2}(1)
+\Li_{3/2}(z \bar z)
-\Li_{3/2}(z)
-\Li_{3/2}(\bar z)
}
\,,
\label{eq:Limit2}
\end{equation}
it is valid in the Euclidean sheet with $|z \bar z| \le 1$, and for
kinematics such that the small rapidity~$\theta\ll 1$ region dominates
the integral. The second expression reads
\begin{equation}
\log \mathbb{O}
\simeq
-\frac{\sqrt{\lambda}}{4\pi^2}
\,\frac{\log(z/\bar z)}{2i}
\brk*{2\pi-\frac{\log(z/\bar z)}{2i}}
\,,
\label{eq:Limit1}
\end{equation}
and it holds whenever the kinematics are such that it is the large
$\theta\gg 1$ region which controls the integral. It turns out that
in all OPE limits discussed below, it is indeed either the small or the large rapidity
regions which control the octagon behavior, so these expressions will
be all we need in the next section. In the following, we sketch the
derivations of these two expressions:
%
\begin{myinset}
\textbf{\eqref{eq:Limit2} derivation}:
The kinematical limit in~\eqref{eq:Limit2} is dominated by small
$\theta$. In order to analyze limits of $z$ or $\bar{z}$ individually,
we write the $Y$-function as:
\begin{equation}\label{eq:Yzzb}
Y = -\frac{\sinh\left(\frac{1}{2}\log z\right)\,\sinh\left(\frac{1}{2}\log \bar{z}\right)}{\sinh\left(\frac{1}{2}\log z + \frac{1}{2}\sinh^{2}\left(\frac{\theta}{2}\right)\,\log(z\bar{z})\right)\,\sinh\left(\frac{1}{2}\log \bar{z} + \frac{1}{2}\sinh^{2}\left(\frac{\theta}{2}\right)\,\log(z\bar{z})\right)} \,.
\end{equation}
We see that when $\theta$ is real and
$z$, $\bar z$ are either real or complex conjugate, then $-1\leq Y\leq
0$. Moreover, we recognize that when a cross ratio $z$
or $\bar{z}$ approaches $0$ or $1$, then the
$Y$--function~\eqref{eq:Yzzb} approaches $-1$ only in the region of very
small rapidity $\theta$. Therefore, it is this region which dominates the
integral in~\eqref{eq:prediction}, so that the
integrand
\begin{equation}
\cosh(\theta)\log\left(1+Y\right)=
\cosh(\theta)\log\frac{
\brk!{1-e^{\log(z\bar{z})\sinh^{2}(\theta/2)}}
\brk!{1-z\bar{z}\,e^{\log(z\bar{z})\sinh^{2}(\theta/2)}}
}{
\brk!{1-z\,e^{\log(z\bar{z})\sinh^{2}(\theta/2)}}
\brk!{1-\bar{z}\,e^{\log(z\bar{z})\sinh^{2}(\theta/2)}}
}
\label{integrandZ}
\end{equation}
can be dramatically simplified by expanding $\sinh(\theta/2)
\simeq \theta/2$ and $\cosh(\theta/2)\simeq 1$, and using
\begin{equation}\label{eq:limit0}
\int\limits_{-\infty}^{\infty}\,d\theta\,\log\left(1-x\exp\brk{- y\, \theta^{2}}\right) \,=\,-\sqrt{\pi/y}\,\Li_{3/2}(x) \,
\end{equation}
to establish~\eqref{eq:Limit2}. From a thermodynamic Bethe ansatz
context, these limits resemble non-relativistic limits, where particles
have small rapidities.
\end{myinset}
and
\begin{myinset}
\textbf{\eqref{eq:Limit1} derivation}:
The kinematical limit in~\eqref{eq:Limit1} is dominated by large
$\theta$. It is the relevant limit, for example, when $\log(z\bar z)
\to 0 $ so we see that all exponents in~\eqref{integrandZ} can be
effectively set to zero \textit{unless}
$\theta$ is huge, of order $\log\log(z\bar z)$. Furthermore, the full
expression is multiplied by $\varphi = -\log(z\bar z)/2$ and thus
vanishes unless $\theta$ is huge indeed. So we can freely replace all
hyperbolic functions by large~$\theta$ single exponential terms. Once
that is done, it suffices to use
\begin{equation}
\int\limits_{-\infty}^{\infty}\,d\theta\,{e^{\theta}}\log\left(1-x\exp\brk{-
y\, e^{\theta}}\right)
=
-\frac{1}{y}\Li_2(x)
\end{equation}
to evaluate all resulting integrals and thus get that the area is
proportional to $\text{Li}_2(1)+\text{Li}_2(z\bar
z)-\text{Li}_2(z)-\text{Li}_2(\bar z)$. Expanding for $\log(z\bar z)
\to 0 $ (\ie for $z\to 1/\bar z$) does simplify this expression
into~\eqref{eq:Limit1}. As indicated in~\eqref{eq:Limit1}, there are
other limits that are dominated by large~$\theta$. One such limit is
$\log(z/\bar z)\to-\infty$. In this case, expanding the $\Li_2$
expression again leads
to the same right hand side of~\eqref{eq:Limit1}.
The result blows up, and we trust its divergent part.
To find the finite part, a more careful analysis is
needed, as discussed in \secref{sec:puzzles}.
From a thermodynamic
Bethe ansatz point of view, these manipulations go by the name of
high-temperature analysis, dominated by very energetic particles with
large rapidities.
\end{myinset}

\subsection{OPE Limits}

\begin{figure}[t]
\centering
\includestandalone[width=\textwidth]{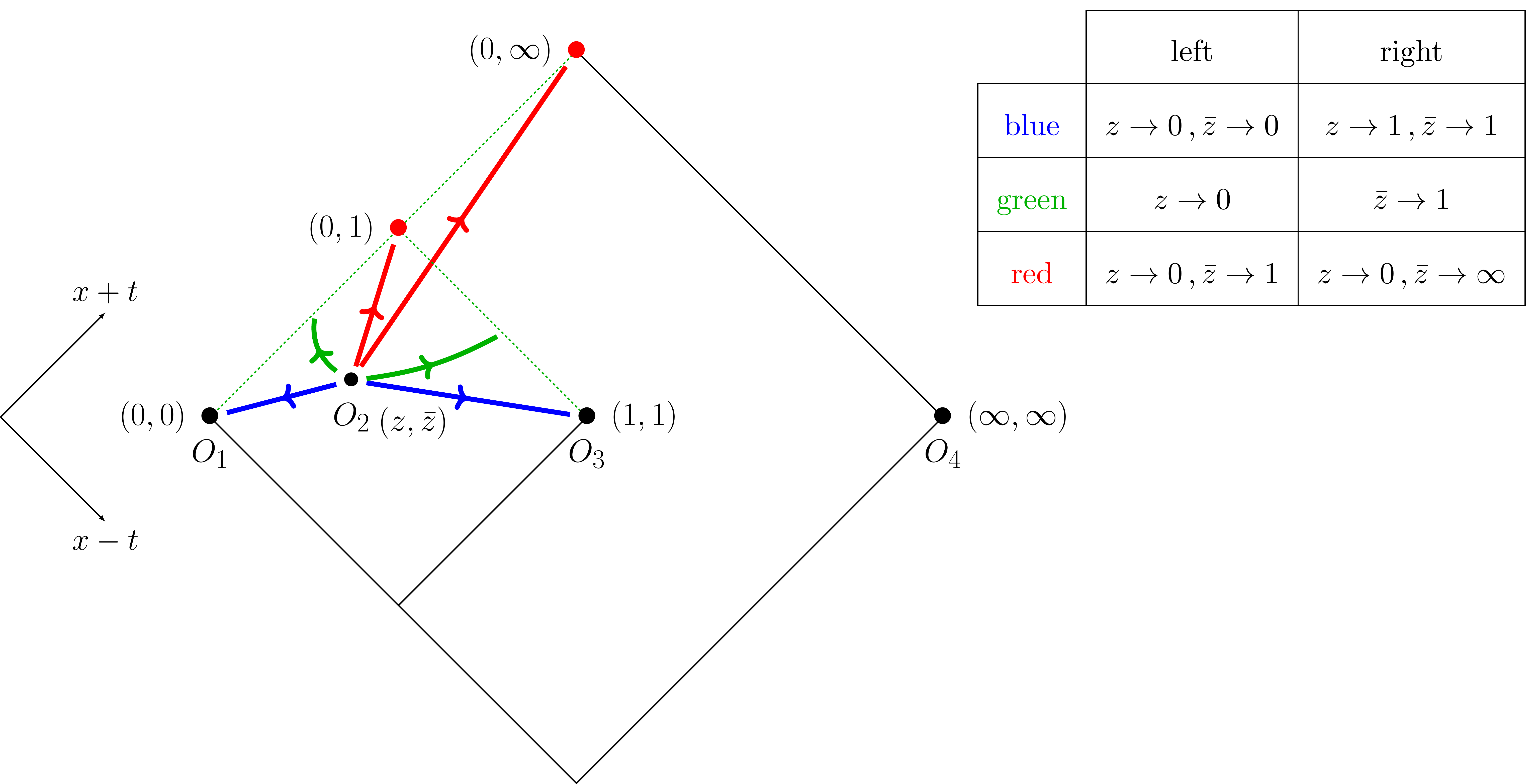}
\caption{Poincare patch in light-cone coordinates with operators
$O_{2}$ at $(z,\bar{z})$ and $O_{1},O_{3},O_{4}$ in canonical
positions. OPE limits are obtained when $O_{2}$ approaches any of the
corners or edges of the dashed squares. We indicate in blue the
Euclidean or space-like OPE's, in green the single light-like OPE's and
in red the double light-like OPE's.}
\label{fig:SteinmannTimelike}
\end{figure}

\begin{figure}[t]
\centering
\includegraphics[scale=1]{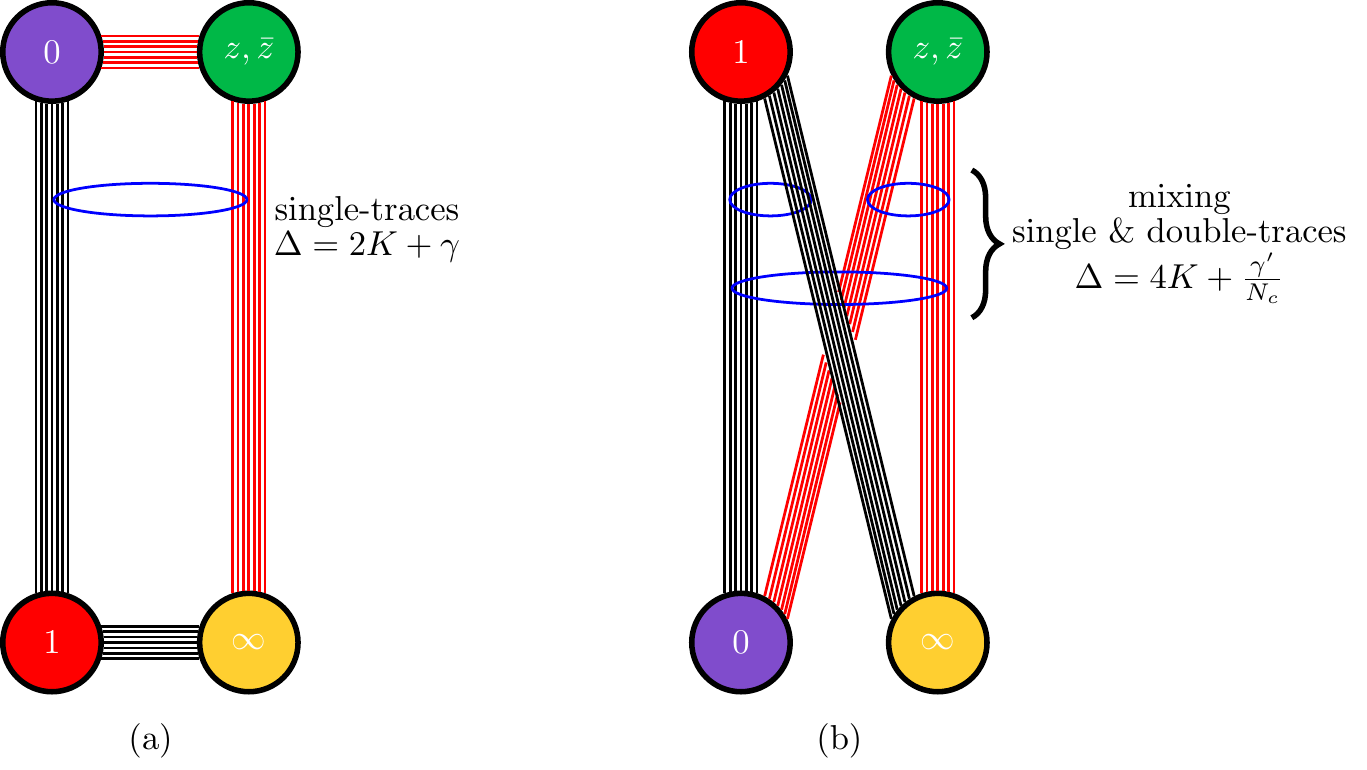}
\caption{(a) The OPE $z,\bar z \to \infty$ or $z, \bar z \to 0$ is
controlled by large single trace operators. (b) The other OPE limit
connecting diagonals of the square, that is $z,\bar z \to 1$ is
dominated by double traces. The area will indeed behave strikingly
differently in both limits.}
\label{fig:twoOPEs}
\end{figure}

Armed with~\eqref{eq:Limit2} and~\eqref{eq:Limit1}, we can now
straightforwardly explore all the various interesting OPE limits
summarized in \figref{fig:SteinmannTimelike}. They can be
Euclidean OPE's (if two points approach each other), light-like OPE's
(if two points become null separated) or double-light-like OPE limits
(if the four points approach the cusps of a null square). For each of
these limits, it matters whether adjacent operators (at points $x_i$,
$x_{i+1}$) or diagonally opposite operators (at points $x_i$,
$x_{i+2}$) collide (or become null separated).
This makes a big difference, since
in our correlator, operators at points $x_i$ and $x_{i+1}$ are connected by an
edge of the large R-charge frame (see \figref{fig:recall}), whereas
operators at $x_i$ and $x_{i+2}$ are located at
non-neighboring cusps of the square. This important difference is
illustrated in an example in \figref{fig:twoOPEs}. Hence all in all,
there are six different interesting limits that we can take, as summarized in
\figref{fig:SteinmannTimelike}.%
\footnote{All these limits can
be approached from the Euclidean regime; genuinely Lorentzian
configurations will be considered in the
next subsection.}

We find that the area $\mathbb{A}\equiv-(\log\oct)/\sqrt{\lambda}$ in the
various limits is approximately given by the expressions in~\tabref{tab:OPElimits}.
\begin{table}[t]
\centering
\begin{tabular}{L{1.5cm} L{6cm} L{5cm}}
\toprule
& $z\to 0$
& $z\to 1$
\\ \midrule
$\bar z\to 0$
& {\color{Blue} Euclidean OPE} $\color{red}{\star}$\newline
  $4\pi^{3/2}\mathbb{A} \simeq \zeta(3/2)\sqrt{-\log(z\bar z)}$
& Equivalent to $z\to 0, \bar z\to 1$
\\ \midrule
$\bar z$ \text{ finite}
& {\color{ForestGreen} Light-like OPE} $\color{red}{\star}$\newline
  ${4\pi^{3/2}}\mathbb{A}\simeq \sqrt{\log \tfrac{1}{z}}
  \big(\zeta\big(\frac{3}{2}\big)-\textrm{Li}_{\frac{3}{2}}(\bar{z}) \big)$
& {\color{ForestGreen} Light-like OPE} $\color{blue}{\ast}$\newline
  $2\pi\mathbb{A} \simeq \sqrt{1-z}\sqrt{-\log\bar{z}}$
\\ \midrule
$\bar z\to 1$
& {\color{Red} Double Light-like OPE}\newline
  $2\pi \mathbb{A}\simeq  \sqrt{1-\bar{z}}\sqrt{-\log\,z}$
& {\color{Blue} Euclidean OPE} $\color{blue}{\ast}$\newline
  $2\pi\mathbb{A} \simeq  \sqrt{1-z}\sqrt{1-\bar{z}}$
\\ \midrule
$\bar z\to \infty$
& {\color{red} Double Light-like OPE} $\color{red}{\star}$\newline
 $16\pi^2\mathbb{A}\simeq   (\log(-z)+\log(-1/\bar z))^2$
& Equivalent to $z\to 1, \bar z\to 0$
\\ \bottomrule
\end{tabular}
\caption{The area $\mathbb{A}\equiv-(\log\mathbb{O})/\sqrt{\lambda}$
in various OPE limits.}
\label{tab:OPElimits}
\end{table}
A few comments are in order:
\begin{itemize}
\item Some areas are very large (identified with
$\color{red}{\star}$), some are very small (identified with
$\color{blue}{\ast}$), and some can be either. The large areas are the
ones where the points colliding or becoming null separated are
neighbors in the square, while the vanishingly small area are those
where the points colliding or becoming null separated are
non-neighboring cusps in the square. This is in nice agreement with
the intuition that the first is a single-trace OPE channel while the
latter ought to be dominated by double-trace operators, see
\figref{fig:twoOPEs}. In particular, when the area is very large, we can
extract the effective spin and dimension of the exchanged operators, since
those dominate the OPE saddle point. When the
area vanishes, instead, the prefactor multiplying the classical area
would become important; it would be very interesting to study this
prefactor, at least in the limits considered here. The technology to
probe this fascinating limit should be very close to that recently
developed in \cite{Basso:2019diw}%
\footnote{We thank Benjamin Basso for enlightening discussions on the physics of this OPE channel
and its relation to (the singularities of) extremal three-point
functions way before their paper was published.}
\item Some limits arise as particular cases of~\eqref{eq:Limit2}; others
of~\eqref{eq:Limit1}. For example, the first line is a specialization
of~\eqref{eq:Limit2} while the last line follows
from~\eqref{eq:Limit1}.
\item Most limits commute with each other. For example, from the
second line we can further specialize to the elements in the first and
third lines. One exception is the limit where $z\to 0$ and $\bar z \to
\infty$ with their product held fixed. This leads to the fourth line,
which is not a trivial expansion of the second line at large $\bar z$.
In this case, the order of limits matters. This last limit,
with $z\to 0$ and $\bar z\to \infty$, was
very important in bootstrapping the octagon to all loops
in~\cite{Coronado:2018cxj}, and will be the subject of a more detailed
discussion in the next section.
\item The Euclidean OPE limit {$z, \bar z \to 1$ corresponds to
$\varphi \to 0$ and $\phi\to 0$}. We can reach it starting with $z\to
1$, \ie under the conditions of~\eqref{eq:Limit2}, and then taking the
limit $\bar z\to1$, leading to the expression in \tabref{tab:OPElimits}, see also~\eqref{eq:LeadingZ1} in
\appref{sec:LargeExpansionSection}.
We can also first take $\varphi \to 0$ (\ie $\bar z\to 1/z$) and
then send $z\to 1$ to obtain a specialization of the result in the
table, namely $2\pi\mathbb{A} \simeq 1-z$. To obtain this result, we
note that when $\varphi \to 0$, we are under
the conditions of~\eqref{eq:Limit1} even before taking any limit on
$\phi$. We hence obtain, when $\varphi \to 0$,
\begin{equation}
\log\mathbb{O}
\simeq
-\frac{\sqrt{\lambda}}{4\pi^2}
\,\widetilde\phi\,(2\pi-\widetilde\phi)
\,.
\label{highT}
\end{equation}
Here $\widetilde \phi = \phi\!\mod 2\pi$, since the derivation
implicitly assumes that $\phi=-i/2\log(z/\bar z)=-i\log(z)$ is defined
to be between $0$ and $2\pi$ on the principal sheet. Outside this
range, the result should be continued periodically to match with the manifestly
periodic function~\eqref{eq:predictionA} in the Euclidean regime, see
\figref{fig:plotEuclidean}. When $\phi\to0$, we recover indeed
$2\pi\mathbb{A}\propto1-z$, which is consistent with the above.
\item Another very important limit for the weak-coupling bootstrap
of~\cite{Coronado:2018cxj} was the observation of Steinmann-like
relations: The expansion around $z=\bar z=1$ generates only single logarithms of
$z-1$ and $\bar z-1$ in perturbation theory. As such,
the double discontinuity, around $z=1$ say, vanishes. We cannot
see this effect at strong coupling, since it would require knowing the one-loop
pre-factor to our classical result.
\end{itemize}

\subsection{The Null Octagon (Results and Speculations)}
\label{sec:puzzles}

In this section, we would like to open a longish parenthesis and study
the very interesting limit where the four points approach the cusps
of a null square. More precisely, we want to focus on the cyclic order
where the points which are becoming consecutively null are those which
are also connected by geodesics. In terms of cross ratios, this corresponds to the limit
\begin{equation}
z\to 0^{-} \,,\qquad \bar z\to -\infty  \quad \text{with fixed }z\bar{z}\label{DLC}
\,.
\end{equation}
This limit, also dubbed as double light-like
limit, was studied in detail for small operators in
perturbation theory in~\cite{Alday:2013cwa}. There, it was highlighted
that this limit is controlled by the exchange of large-spin operators
of leading twist. In the analysis of~\cite{Alday:2013cwa}, there is a single lowest-twist
family (of twist $2$), which simplified the analysis and allowed for
all-loop re-summations of the four-point functions (and associated OPE
data) in this limit. In our case, we are dealing with large operators, so
that even at leading twist there is a huge degeneracy involved. Some
progress has been made towards taming
this degeneracy (see \eg~\cite{Alday:2016mxe}), but for generic large
operators, this problem remains
unsolved. It would be very interesting to clean this up and settle many
of the interesting issues raised in the discussion that follows.

This double light-like limit was identified as very important in
bootstrapping the octagon in perturbation theory
in~\cite{Coronado:2018cxj}. In this work, it was noted that the
correlator admits a beautiful exponentiated result in this limit, as
\begin{equation}
\log \mathbb{O} =
-\frac{(\log(-z)+\log(-1/\bar z))^2}{8\pi^2}\Gamma(\lambda)
+\frac{1}{8} C(\lambda)
+\frac{\lambda}{16\pi^2}\log(z\bar z)^2
\,.
\label{perturbation}
\end{equation}
The functions $\Gamma$ and $C$ were explicitly evaluated to twenty-four
loops, and a general algorithm for finding them to arbitrarily high loop
order was provided. We will come back to these terms in a moment.

For now, let us stress that the last term in~\eqref{perturbation} is
very weird, as it is one loop exact. To our knowledge, it is the only
instance where the 't~Hooft coupling appears explicitly in a physical
observable.%
\footnote{Unless we count the magnon dispersion relation
as a physical quantity. Strictly speaking, we measure anomalous
dimensions and not the magnon dispersion, and hence it is not a direct
observable.}
It seems like we could define the coupling non-perturbatively as the
coefficient of $\log(z\bar z)^2$ in the double light-like limit. In
perturbation theory, this is indeed the case. However, extra care is
needed if we want to go to finite coupling. We would like to claim
that we cannot take~\eqref{perturbation} as the correct double light-like limit at
finite coupling. \emph{If} it were correct, it would lead to a
correlator proportional to the exponential of~$\lambda$ at strong coupling,
which is much larger than the classical string tension
$\sqrt{\lambda}$. It would thus contradict the most basic AdS/CFT
dictionary. Indeed, above we argued that the octagon
exponentiates with a nice exponent proportional to $\sqrt{\lambda}$, so
there is indeed no sign of such a weird term at strong coupling. We
conclude that we face a clear example of an order of limits
issue. In perturbation theory, even if we take $\log(z/\bar z)$ and
$\log(z\bar z)$ to infinity, the products $\Gamma \log^2(z/\bar z)$
and $\lambda \log^2(z\bar z)$ are always very small, since the coupling
is the smallest parameter. Hence it is dangerous to extrapolate these
expressions to finite or strong coupling, where these products would become
very large.

Instead, we conjecture that in the double light-like limit~\eqref{DLC}
and at finite coupling, we have
\begin{equation}
\log \mathbb{O}
\simeq
-\frac{(\log(-z)+\log(-1/\bar z))^2}{8\pi^2}\Gamma(\lambda)
+\frac{1}{8} C(\lambda)
+\log\log (\bar z/z) f(z \bar z, \lambda)
+h(z \bar z, \lambda)
\,.
\label{perturbation2}
\end{equation}
Strong evidence for this proposal comes from the strong-coupling
analysis performed here. We observe precisely this structure, with
\begin{equation}
\Gamma \simeq
\frac{\sqrt{\lambda}}{2}
\,, \qquad
C\simeq -2\sqrt{\lambda}
\,, \qquad
f\simeq \frac{\sqrt{\lambda}}{16\pi^2} \log(z \bar z)^2
\,.
\label{eq:OurLightCs}
\end{equation}
The subleading contribution $h$ is considerably more
complicated. It
vanishes if $z=1/\bar z$, and admits a simple series expansion away
from this limit, see \appref{app:DoubleLight}.
Inspired by the type of expressions observed for small operators (see
below), it is tempting to interpret the function $f$ as a sort of
recoil contribution, and conjecture that $f\propto\Gamma(\lambda)\log(z\bar z)^2$ in the strict light-like limit
-- \ie when the coupling is \textit{not} the smallest parameter.

In a beautiful recent work~\cite{Belitsky:2019fan}, Belitsky and Korchemsky addressed the
double light-like limit of the octagon in the further simplifying
\emph{diagonal limit}
\begin{equation}
z=1/\bar z
\,.
\label{eq:diagonal}
\end{equation}
This is indeed a beautiful simplifying limit, which projects out the
last subtle term, leading to
\begin{equation}
\log \mathbb{O}(z,1/z) =
-\frac{\log(-z)^2}{2\pi^2} \Gamma(\lambda)
+\frac{1}{8} C(\lambda)
\qquad \text{for }
z\to 0
\,.
\label{perturbation3}
\end{equation}
The authors of~\cite{Belitsky:2019fan} found that the octagon
determinant representation of~\cite{Kostov:2019stn,Kostov:2019auq}
simplifies enormously in this
diagonal light-like limit, and took advantage of this
simplification to derive analytic expressions for both $\Gamma$ and
$C$ as
\begin{equation}
\Gamma_{BK}(\lambda)=\log\cosh\frac{\sqrt{\lambda}}{2}
\,, \qquad
C_{BK}(\lambda)=-\log\frac{\sinh\sqrt{\lambda}}{\sqrt{\lambda}}
\,.
\end{equation}
They observed that these re-summed expressions have a rather remarkable
property: Their strong-coupling expansions truncate at one-loop
order! Namely,
\begin{equation}
\Gamma_{BK}(\lambda)=\frac{\sqrt{\lambda}}{2}-\log2-\sum_{n=1}^{\infty}\frac{(-1)^{n}}{n}\,e^{-n \sqrt{\lambda}}
\,, \quad
C_{BK}(\lambda)=-\sqrt{\lambda}+\frac{1}{2}\log\lambda+\log2+\sum_{n=1}^{\infty}\frac{e^{-2n\sqrt{\lambda}}}{n}
\,.
\label{BKStrong}
\end{equation}
It would be fascinating to understand why these series truncate. Note also that $C(\lambda)$
has an amusing logarithmic term, which seems to indicate an
interesting prefactor structure of the octagon in this double
light-like limit as
\begin{equation}
\mathbb{O} \simeq
e^{-\frac{\sqrt{\lambda}}{4\pi^{2}}\log(-z)^{2}
-\frac{\sqrt{\lambda}}{8}} \lambda^{1/16} \times O(1)
\,,
\label{prefactorExpression}
\end{equation}
which would also be fascinating to understand. Now, the attentive
reader probably noticed that the classical piece in $\Gamma$
in~\eqref{BKStrong} perfectly agrees with our strong coupling
evaluation,
while the classical part in $C$ is off by a factor of two.

What is this factor of two? One option is a glitch in our computation.%
\footnote{We carefully checked our computation, and all factors appear
to be correct. At strong coupling, the
limit when $z\to 1/\bar z$ is a simple high-temperature limit of the
free-energy type formula for the area, for which we have good control over
$C(\lambda)$. Also, it is hard to see how a potentially missing factor of two
in $C(\lambda)$ of~\cite{Belitsky:2019fan} would come about,
since we did check that the expression for $C(\lambda)$
in~\cite{Belitsky:2019fan} does agree with the perturbative results
of~\cite{Coronado:2018cxj} up to $24$ loops.}
The other option is
physics. Perhaps both results are perfectly correct, and the
disagreement is simply because we are again taking different limits.
In this work, we first go to strong coupling $\lambda\to\infty$, and then take the double
light-like limit~\eqref{DLC}. In~\cite{Belitsky:2019fan} on the other hand, the light-like limit is
taken first. So our result could be a purely classical/minimal area
result, while the prediction from~\cite{Belitsky:2019fan} appears to be a
more strict double light-like limit.

There was a similar setup in the context of null
polygonal Wilson loops, where such limits were similarly subtle.
In~\cite{Basso:2014jfa}, it was understood that in the collinear limit
of scattering amplitudes (corresponding to an OPE limit of Null Wilson
loops~\cite{Alday:2010ku}), an additional enhanced contribution from
nearly massless scalars modifies the classical minimal area result
by an additional constant term, and also produces a funny power of
$\lambda$ in the prefactor. Could the mismatch we are observing,
together with the interesting prefactor
in~\eqref{prefactorExpression}, have a similar origin? Perhaps in the
double light-like limit additional massless modes come into play, the
naive expansion around the BMN vacuum needs to be reorganized, and a
more careful analysis is needed along those lines? We are currently
analyzing this possibility.

It is very likely that this potential order of limits problem and the order of
limits issue related to the function $f$ introduced above are not
mathematically unrelated. So another source of clarification would
come from repeating the double light-like OPE limit analysis
of~\cite{Belitsky:2019fan} without the diagonal
restriction~\eqref{eq:diagonal} to see if the picture above -- including
the interesting function $f$ -- is indeed realized. Expanding around the
diagonal limit, that is for $\log(z\bar z)^2\ll 1$ should hopefully
not be that hard, and would be very illuminating.

Finally, we could compare the discussion above with the double
light-like predictions for small operators from Alday and Bissi
in~\cite{Alday:2013cwa,Alday:2016mxe} which we alluded to above. We
use the notation of~\cite{Belitsky:2019fan}, equation~(1.10) therein,
from where we extract the correlator in the null limit as
\begin{equation}
G_4 \simeq
H(g) \int\limits_0^\infty
\brk*{\prod_{j=1}^2 2 K_0(2 \sqrt{y_j})dy_j}
e^{-S}
\,, \quad
S= \tfrac{1}{2} \Gamma_\text{cusp}
\log\brk!{ -\tfrac{z}{y_1} }\log\brk!{-\tfrac{1}{\bar zy_2}}
-\tfrac{1}{2}\Gamma_v \log\brk!{ \tfrac{z}{\bar z y_1 y_2}}
\,,
\end{equation}
where ${K}_0$ is the modified Bessel function of the second
kind, and the contribution shown explicitly is what is interpreted as the recoil contribution.
In the null limit, we can estimate the integral by saddle point,%
%
\footnote{The leading saddle location is at
$(y_1,y_2)\simeq\Gamma_\text{cusp}^2\brk!{\log(-z)^2,\log(-1/{\bar{z}})^2}/4$.}
so that finally
\begin{align}
\log (G_4) \simeq&
-\tfrac{1}{2}\Gamma_\text{cusp} \log(-z)\log(-1/\bar z)
+\tfrac{1}{2} \gamma \log\brk*{{z}/{\bar{z}}}
\nn \\
&{}+\Gamma_{\text{cusp}} \log \brk*{-{1}/{\bar{z}}}
\log\brk!{\Gamma_{\text{cusp}} \log (-{1}/{\bar{z}}) }
+\Gamma_{\text{cusp}} \log (-z) \log\brk!{\Gamma_{\text{cusp}} \log (-z)}
\nn \\
&{}+\text{constant}
\,,
\end{align}
where $\gamma=\Gamma_v-\log(4e^2)\Gamma_\text{cusp}$. This expression
is indeed quite similar to the expressions above.
However, the $\log\log(-z)$ term here is dressed by a term $\log(-z)$
that is linearly divergent in
the double light-like limit, while for us it was multiplied by a finite
factor $\log(z\bar z)^2$, which is held fixed in the
limit~\eqref{DLC}. The second term in the first line is also absent for us.
Perhaps this is related to the absence of recoil in our correlator,
with its huge R-charge frame. It would be fascinating to investigate this
further, \ie to analyze the octagon from a light-cone bootstrap
perspective.

In sum, we have a very rich behavior of the correlation function
given by the octagon in the null limit.%
\footnote{We thank Grisha Korchemsky and Andrei Belitsky for useful
correspondence on these matters.}
We can approach it in three different ways:
\begin{itemize}
\item Weak coupling: We first expand around $\lambda=0$, and then
expand in the light-cone limit~\eqref{DLC}. This was originally
studied in~\cite{Coronado:2018cxj} to all loop orders in perturbation
theory.
\item Null Limit First: We first take the light-cone
limit~\eqref{DLC}, keeping $z \bar z$ fixed. In its generic form, this
limit was not analyzed yet. In the diagonal limit, where $z\bar z=1$,
it was recently studied in~\cite{Belitsky:2019fan}.
\item Strong coupling: We first take $\lambda \to \infty$, and then
expand in the light-cone limit~\eqref{DLC}. This is what we studied in
this section.
\end{itemize}
All three limits lead to beautiful exponentiation of the correlator, as
just reviewed. The leading term in the exponent seems to be universal,
and independent on how we approach the null limit. It is governed by
the function $\Gamma_\text{cusp}$. The subleading terms, however, seem
sensitive to the order of limits. In particular, in the diagonal limit
(which as of now is the only one we can compare), the second and third
limits differ in a mild way, by a factor of two in the constant
subleading term. We suspect this $2$ to be a smoking gun for some yet
to be unveiled interesting physics.%
\footnote{The extrapolation from weak coupling differs from the
strong-coupling result in an even more drastic fashion in the
non-diagonal limit. But in that case, we have no reason to expect the
difference to be explained by some interesting physics, since the two
limits are genuinely very different. In the weak-coupling limit,
$\lambda^n \log(z/\bar z)$ is very small; in the second and third
limits, this combination is very large.}

\subsection{Lorentzian Continuations}

Lorentzian correlators can be obtained as analytic continuations of
the Euclidean correlator. This follows from a Wick rotation of the
time-coordinates of the operators, whose effect on the cross ratios is
to take them from being complex conjugate $\bar{z}=z^{*}$ in the
original Euclidean configuration to real and independent values in the
final Lorentzian configuration. If we start in the Euclidean cylinder
$\mathbb{R}\times S^3$, then at the end of the Wick rotation we are in
the Lorentzian cylinder, where $\mathbb{R}$ is a time. We will only use
a circle subspace of the full three-sphere, so we will be working in an
$\mathbb{R}\times S^1$ subspace. Then each operator insertion is
parametrized by an angle $\phi_j$ on the circle and a time $t_j-i
\epsilon_j$, where the order of the imaginary epsilons dictates the
order of the operator insertions, see \eg
\cite{Cornalba:2006xk,Hartman:2015lfa,talk:Simmons-Duffin:TASI:6.2019}.
The physical cross-ratios then take the
nice form
\begin{equation}
z=\frac{\sin \psi^+_{12} \,\sin \psi^+_{34}}{\sin \psi^+_{13}\,\sin \psi^+_{24}}
\qquad \Longleftrightarrow \qquad
1-z=\frac{\sin \psi^+_{14} \,\sin \psi^+_{23}}{\sin \psi^+_{13}\,\sin \psi^+_{24}}
\,,
\label{zExpressions}
\end{equation}
where $\psi^+_{ij}=\psi^+_{i}-\psi^+_{j}$, with $\psi^+_j =
(\phi_j+t_j)/2$, and with a similar expression for $\bar z$ in terms of the
other light-cone direction $\psi^+_j\to \psi^-_j = (\phi_j-t_j)/2$.

\begin{figure}[t]
\centering
\includegraphics[trim=0 300 0 0,clip,width=0.9\textwidth]{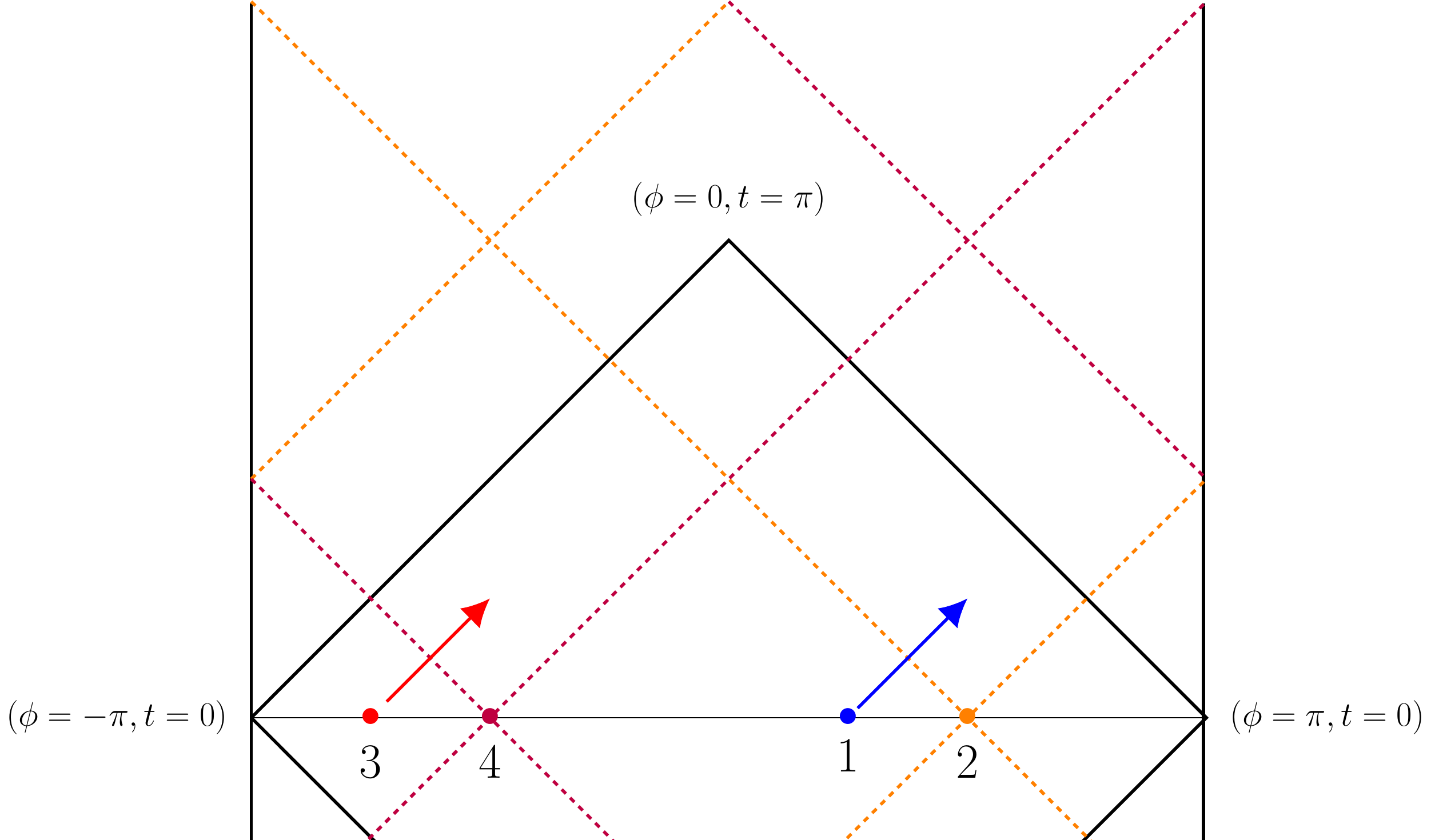}
\\
\caption{Operators $O_{4}$ and $O_{1}$ being lifted in Lorentzian time
$t$, entering the light-cones of $O_{1}$ and $O_{3}$ as we perform the
Wick rotation. As reference, we depict a Poincare patch as a solid
diamond, specifying its corners in global coordinates $t$ and $\phi$
of $R_{t}\times S_{\phi}$. The lines $\phi=-\pi$ and $\phi=\pi$ must
be identified.}
\label{fig:LiftingTwo}
\end{figure}

\begin{figure}[t]
\centering
\includegraphics[width=\textwidth,trim=35 15 60 10,clip]{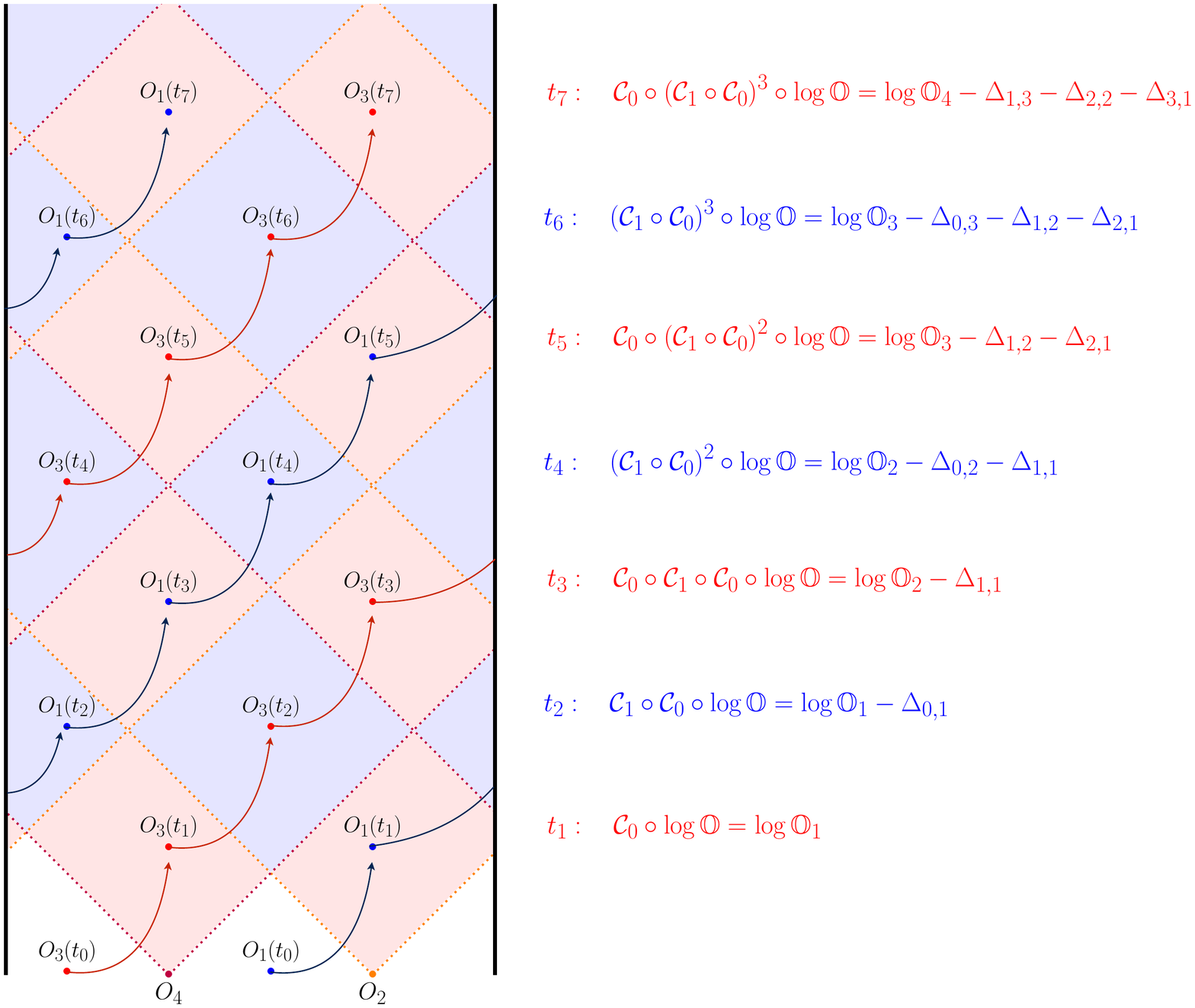}%
\caption{Left panel: Lorentzian configurations with operators $O_{3}$
and $O_{1}$ inside the light-cones of $O_{4}$ and $O_{2}$, at
different Lorentzian times $t_{j}$ in $\mathbb{R}_{t}\times S^{1}$.
Right panel: the correspondent monodromies $\mathcal{C}_{x}$,
counter-clockwise around the branch points $x=0$ or $1$, and their
effect on the Euclidean correlator $\log\mathbb{O}$. See
\eqref{eq:logOn} and \eqref{eq:deltapq} for notation used on the
right-hand side.}
\label{fig:monodromiesRegions}
\end{figure}

Consider the setup where operators $O_1$ and $O_3$ are lifted in
Lorentzian time and enter the light-cone(s) of the other pair of
operators $O_2$ and $O_4$, see figure \figref{fig:LiftingTwo}. As
indicated in the figure, we move up the Lorentzian cylinder along the
light-cone directions $\psi^+$, so that nothing relevant happens along
the $\psi^-$ directions; we can thus ignore the $\bar z$ cross-ratio
altogether. We do not touch operators $O_2$ and $O_4$, and we move
$O_1$ and $O_3$ simultaneously, so that their distance is always
space-like; then we can also ignore the denominator factors
in~\eqref{zExpressions} which always stay finite, basically untouched.
All the fun is in the four sine factors in the numerators
in~\eqref{zExpressions}. To figure out what happens there, we need the
$i\epsilon$'s. Since we are moving $O_1$ and $O_3$ up, and are looking
for a time-ordered configuration at the very end, it suffices to take
$\epsilon_1=\epsilon_3>0=\epsilon_2=\epsilon_4$. Then the numerators
never vanish. Instead, each of them picks an $e^{i\pi}$ half-monodromy
as $O_{1}$ or $O_3$ crosses a light-cone of $O_{2}$ or $O_4$. Two such
half-monodromies can then combine into full monodromies. For example,
when $O_1$ crosses the light-cone of $O_2$, and $O_3$ crosses the
light-cone of $O_4$, then each factor in $z$ picks a half-monodromy, so
that $z$ acquires a full monodromy around zero, as illustrated in
\figref{fig:monodromiesRegions}. If we continue moving up the Lorentzian
cylinder, two more light-cones are crossed, and then it will be $1-z$
numerator factors which become relevant, and we now end up picking an
extra monodromy, this time around $z=1$, see also
\figref{fig:monodromiesRegions}.

This describes the analytic continuations required to move up the
Lorentzian cylinder along the light-like helices, where $z$ changes and
picks monodromies, whereas $\bar z$ does not. What about other paths?
They all give the same of course. For example, paths using the other
light-cone helices would generate $\bar z$ monodromies instead, with
counter clock-wise orientation. Since we start in the Euclidean
correlator, which is single-valued, we can always trade those
monodromies for regular clock-wise oriented $z$ monodromies, see
\eg~\cite{Cornalba:2006xm}.
Similarly, if we were to move the points
vertically, we would generate both types of monodromies, which we could
again relate to purely $z$ monodromies using single-valuedness, so that
all is nice and consistent.

What about moving other points up the Lorentzian cylinder? That would
of course be different, but trivially related to the previous case,
since we can reach these other cases by simply relabeling the points.
For instance, with the exchange $O_{3}\leftrightarrow O_{4}$, we have
$z\to z/(z-1)$, and the corresponding time-ordered correlator is
obtained by the monodromies
\begin{multline}
\mathbb{O}^2
\xrightarrow{\substack{ O_{1}/O_4 \text{ cross}\\\text{LCs of }O_{2}/O_3}}
\mathcal{C}_0 \circ \mathbb{O}^2
\xrightarrow{\substack{O_{1}/O_4 \text{ cross}\\ \text{LCs of }O_{3}/O_2}}
\mathcal{C}_\infty \circ \mathcal{C}_0 \circ \mathbb{O}^2
\xrightarrow{\substack{O_{1}/O_4 \text{ cross}\\ \text{LCs of }O_{2}/O_3}}
\mathcal{C}_0 \circ \mathcal{C}_\infty \circ\mathcal{C}_0 \circ \mathbb{O}^2 \dots
\\[0.8ex]
=\frac{
\bra{0}
O_{4}(t-i \epsilon,\vec{x}_{4})
O_{1}(t-i\epsilon,\vec{x}_{1})
O_{3}(0,\vec{x}_{3})
O_{2}(0,\vec{x}_{2})
\ket{0}
}{
\texttt{Tree Level}
}
\end{multline}
once we move $O_4$ and $O_1$ up the cylinder. In practice, it is useful
to replace the infinity monodromy by counterclockwise monodromies
around $1$ and $0$, as $\mathcal{C}_\infty=\bar{\mathcal{C}}_1\circ
\bar{\mathcal{C}}_0$.

\begin{figure}[tb]
\centering
\includegraphics[width=\textwidth]{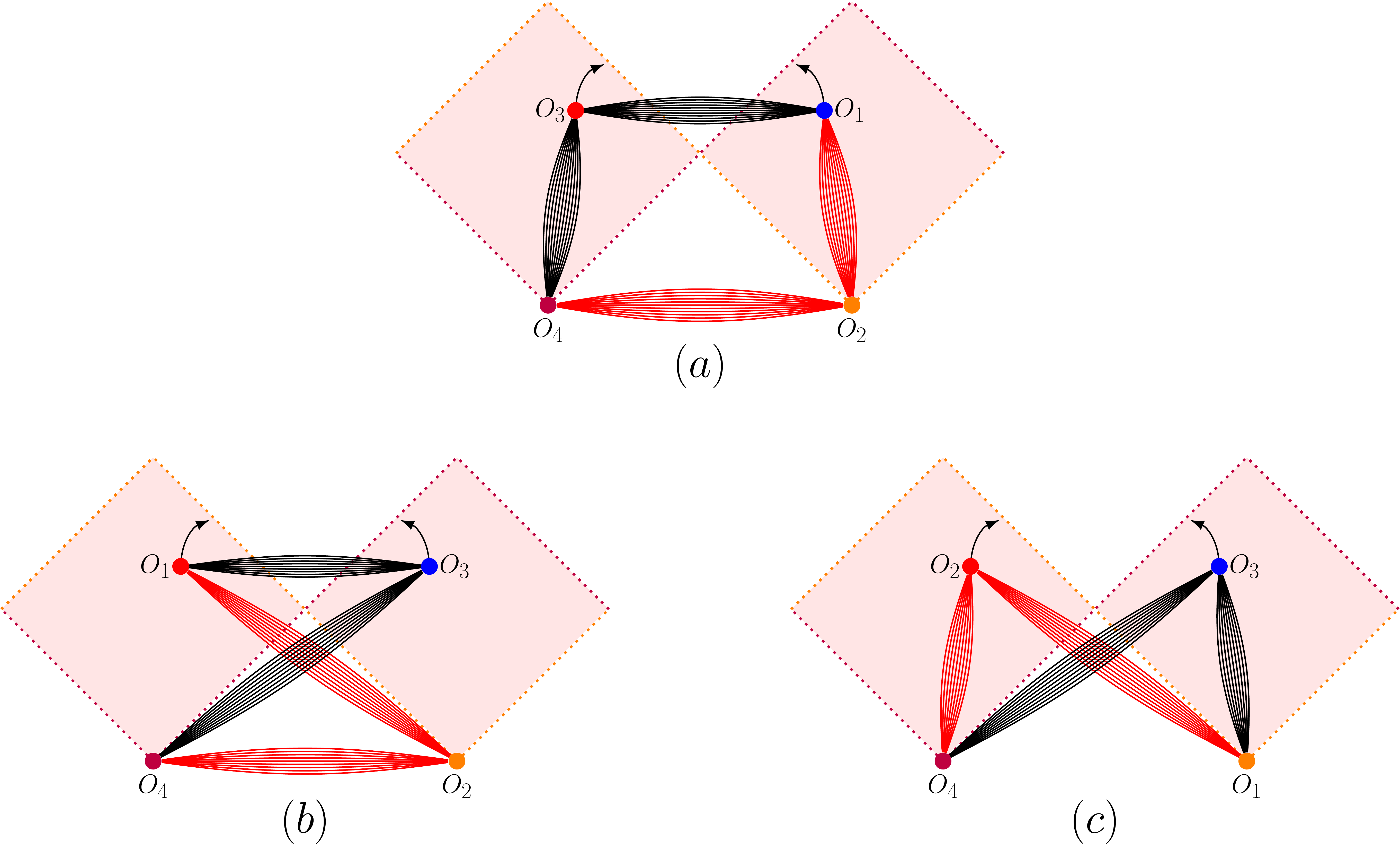}
\caption{The large R-charges in our setup admit three qualitatively
different Regge limits, which are reached as follows:
\mbox{\emph{(a)} $z,\bar z\to 1$} after $z \rotpos 0$,
\mbox{\emph{(b)} $z,\bar z\to 0$} after $z \rotpos 1 $,
\mbox{\emph{(c)} $z,\bar z\to 0$} after $z \rotpos \infty $.}
\label{fig:Regge}
\end{figure}

This concludes the analysis of the analytic continuation paths, that
is of the kinematics. For the dynamics, we would like to know how the
octagon transforms under all such monodromies as we analytically
continue it from its Euclidean representation~\eqref{eq:predictionA}
into Lorentzian kinematics. There are two important contributions. On
the one hand, we have the variables $\phi$ and $\varphi$ which are
themselves \emph{not} periodic as we perform monodromies around
$z=0$. Instead, they transform as
$(\phi,\varphi) \to (\phi+\pi
n,\varphi-i \pi n)$ under $n$ monodromies around $z=0$. The
area~\eqref{eq:predictionA} is periodic in such shifts of $\phi$, but
not of $\varphi$, so from here we get an obvious contribution, which we
conveniently define as
\begin{equation}
\log\oct_n \equiv
\log\oct(\varphi-i\pi n,\phi+\pi n)  = \log\oct(\varphi-i\pi n,\phi)
\,,
\label{eq:logOn}
\end{equation}
This is however not the full story. As we analytically continue the
cross-ratios, the singularities of the integrand will move, and can cross the
contour of integration, at which point we produce further interesting
contributions. This does indeed happen, albeit not for the $z=0$
monodromies; it happens for the monodromies around $z=1$. Such extra
new terms are very simple and take the form
\begin{equation}
\Delta_{p,q}:=
\frac{\sqrt{\lambda}}{\pi}
\sqrt{\brk!{\log(z)+2\pi ip}\brk!{\log(\bar{z})+2\pi iq}}
\,.
\label{eq:deltapq}
\end{equation}
All octagon expressions in any Lorentzian region can be cast as simple
combinations of $\log\oct_n$ and $\Delta_{p,q}$, as summarized in the
example of \figref{fig:monodromiesRegions}. In concise formulae,
carefully derived in \appref{sec:AppendixContinuations},
\begin{align}
\left(\mathcal{C}_{1}\circ\mathcal{C}_{0}\right)^{n}\circ\log\mathbb{O}
\,&=\,
\log\mathbb{O}_{n} - \sum_{i=0}^{n-1}\Delta_{i,n-i}
\,, \nn\\
\mathcal{C}_{0}\circ\left(\mathcal{C}_{1}\circ\mathcal{C}_{0}\right)^{n}\circ\log\mathbb{O}
\,&=\,
\log\mathbb{O}_{n+1} - \sum_{i=0}^{n-1}\Delta_{i+1,n-i}
\,.
\label{eq:alloctcond}
\end{align}
As a simple application, we can consider the Regge limit depicted in
\figref{fig:Regge}.
As depicted there, since our configuration is carries
large R-charges, the behavior in this limit can be strikingly different,
depending on whether the dominating exchange is charged or chargeless.
Indeed, we see that the area can either blow up or vanish, depending
on the R-charge setup! One of these Regge configurations can be
reached by taking a monodromy around $z=0$ and then approaching
$z,\bar z=1$, and this limit is dominated by double-trace exchanges,
see \figref{fig:Regge}a. The other configurations are dominated by
single-trace exchange, they are
reached by taking $z$ around $1$ (or $\infty$) and then approaching
$z,\bar z=0$, as represented in \figref{fig:Regge}b(c).
For the area function $\log\oct$, we find
\begin{equation}
z,\bar z\to 0 \text{ after } z \rotpos 1 \text{ or } z \rotpos  \infty
\; : \qquad
\log \mathbb{O} \simeq -\frac{\sqrt{\lambda}}{\pi}\sqrt{\log(z)\log(\bar z)}
\,,
\label{eq:ReggeLimitA}
\end{equation}
in the channels dominated by single-traces, while
\begin{equation}
z,\bar z\to 1 \text{ after } z \rotpos 0
\; : \qquad
\log \mathbb{O} \simeq
-\frac{\sqrt{\lambda}}{2\sqrt{\pi}}(1-i) (\sqrt{1-z}+\sqrt{1-\bar z} - \sqrt{2-z-\bar z} )
\,.
\label{eq:ReggeLimitB}
\end{equation}
in the double-trace dominated channel. We could have
obtained~\eqref{eq:ReggeLimitA} directly from the Euclidean OPE
expression in the second row of~\tabref{tab:OPElimits} in
the previous section. For the second limit~\eqref{eq:ReggeLimitB}, we need to
be more careful, since we want to approach $z,\bar z=1$ with a fixed
angle. In this case, we can obtain the expression directly from
the expansion in \appref{sec:LargeExpansionSection}. With hindsight,
the attentive reader could argue that with these two shortcuts, we
could have sidelined all the subtle continuations and contour analysis
for this particular Regge limit. That is true. Our apologies to the
reader.

\section{Conclusions}
\label{Conclusions}

The main result of this short note is a compact representation for the
octagon at strong coupling. It takes the form
\begin{equation}
\mathbb{O} \simeq e^{-\sqrt{\lambda}\, \mathbb{A}(z,\bar z)} \,.
\end{equation}
It would be very interesting to compute the function $\mathbb{A}$ directly from
string theory. It should be a nice minimal area. It would be even more
interesting to compute the one-loop prefactor multiplying the
exponential, both from the integrability representation for the
octagon and from string theory. Together with the area, the prefactor
should provide strong insights about the finite-coupling nature of
this object. We also expect many interesting properties of the
correlator, such as the bulk point limit and various Steinmann
discontinuity properties to be manifest only once we tackle this
prefactor.

For physical kinematics (both Euclidean and Lorentzian),
the real part of $\mathbb{A}$ is positive. This is probably good, since
otherwise we would obtain an exponentially large octagon. As it
stands, we obtain an exponentially small one, as expected for a
tunneling process. A particular implication is that in the non-planar
limit explored in \cite{Bargheer:2019kxb}, the correlator becomes
simply given by the very simple expression~(4.3) therein.

On the other hand, we can easily take unphysical kinematics and obtain
an exponentially large correlator. For example, we can
take~\eqref{eq:Limit1} in the unphysical regime where $\phi=(1/2i)\log(z/\bar{z})$ is
real and very large to get a huge octagon. This is not surprising. It
happens commonly in such classical analyses, as in the high-energy
scattering of strings in flat space by Gross--Mende~\cite{Gross:1987kza} and
Gross--Manes~\cite{Gross:1989ge}, as recently highlighted by Sever and
Zhiboedov~\cite{Sever:2017ylk}. There, when $s/t$ is very large and~$t$ is negative,
we are in a physical regime, and the amplitude is exponentially small
indeed, but if $t$ is positive, then we get an exponentially large
result. In these complexified situations, where the octagon is very
large, we can again take advantage of the non-planar re-summations
in~\cite{Bargheer:2019kxb} to conclude that the correlator would now be
given by~(4.4) therein. It would be very interesting if there was some
universality in these large-area limits, akin to those recently
explored by~\cite{Caron-Huot:2016icg} and~\cite{Sever:2017ylk} in
the context of Regge theory.

It is rare to have access to a full-fledged four-point function at
strong coupling. As illustrated here, the result exhibits a plethora
of interesting limits and monodromies, as expected in a rich
strong-coupling CFT as the one under consideration. In
\appref{sec:Weak}, we
compare these properties with those observed in the perturbative
weak-coupling regime. It would be very nice to address these at finite
coupling, perhaps making use of the recent determinant
representation~\cite{Kostov:2019stn}. Perhaps the most intriguing limit
of all -- and thus the most interesting to address, given all the
puzzles of \secref{sec:puzzles} -- would be the double light-like limit.

\subsection*{Acknowledgments}

We thank
Benjamin Basso,
Andrei Belitsky,
Nathan Berkovits,
Jo{\~a}o Caetano,
Simon Caron-Huot,
Thiago Fleury,
Valentina Forini,
Vasco Gon\c{c}alves,
Andrea Guerrieri,
Grisha Korchemsky,
Ivan Kostov,
Juan Maldacena,
Jo{\~a}o Penedones,
Amit Sever,
Sasha Zhiboedov
and Shota Komatsu
for numerous enlightening discussions and suggestions.
We are specially grateful to Vasco Gon\c{c}alves for collaboration at
early stages of this work. Research at the Perimeter Institute is
supported in part by the Government of Canada through NSERC and by the
Province of Ontario through MRI. This work was additionally supported
by a grant from the Simons Foundation \#488661.

\appendix

\section{Octagon Correlators}
\label{sec:operators}

It was explained in~\cite{Coronado:2018ypq} and~\cite{Bargheer:2019kxb} how the octagon
$\oct_{\ell=0}$ with zero bridge length completely captures the
planar and non-planar loop corrections of a specific ``simplest'' correlator. We
will briefly review
(a slight variation of) the argument
in the following, and thereby
justify the identification $\alpha=\bar\alpha=1$
in~\eqref{eq:alphafix}. For more details, see Sections~4.1 and~4.2
in~\cite{Coronado:2018ypq}.

The cleanest isolation of the octagon (with zero internal bridge
length) occurs for a correlator of the four operators
\begin{align}
\op{O}_1&=\tr\brk{\bar{Z}^k\bar{X}^k}+\text{(permutations)}
\,,&
\op{O}_2&=\tr\brk{X^{2k}}
\,,\nn\\
\op{O}_3&=\tr\brk{Z^{2k}}
\,,&
\op{O}_4&=\tr\brk{\bar{Z}^k\bar{X}^k}+\text{(permutations)}
\label{eq:operators}
\end{align}
at large $k$ in the planar limit.
Here, $X$ and $Z$ are orthogonal complex scalars, \eg
$X=\phi_1+i\phi_2$ and $Z=\phi_5+i\phi_6$. The operators $\op{O}_2$
and $\op{O}_3$ are BPS superconformal primaries, and $\op{O}_1$
and $\op{O}_4$ are BPS descendants. At tree level, the
correlator $\avg{\op{O}_1\dots\op{O}_4}$ is given by a single
square-shaped Feynman diagram, where each edge of the square consists
of $k$ parallel propagators. In the large-$k$ limit, all loop
corrections are confined to the individual regions ``inside'' and ``outside''
the square tree-level graph.

In the hexagonalization prescription~\cite{Fleury:2016ykk}, all loop
corrections are captured by hexagon form factors. Two such hexagon
form factors fuse to an octagon, and there is one octagon inside and
one octagon outside the square tree-level graph. However, two of the physical
edges of each octagon touch the descendant operators $\op{O}_1$ and
$\op{O}_4$. These descendants carry a large number ${\sim}\,k$ of
(zero-momentum) magnons on top of a BPS primary ``vacuum'' (\eg
$\tr\brk{\bar Z}$ or $\tr\brk{\bar X}$). The presence of these physical magnons
complicate the computation of the octagon form factor.

This complication can be avoided by considering a slightly different
correlator of BPS primary operators
\begin{equation}
\op{O}(y):=\tr\brk[s]!{\brk{y\cdot\Phi}^{2k}}
\,,\qquad
y^2=0
\,,
\label{eq:BPSy}
\end{equation}
with (for example)
\begin{align}
y_1&=\frac{1}{\sqrt{2}}\brk{\beta_1,-i\beta_1,0,0,1,-i}
\,,&
y_2&=\frac{1}{\sqrt{2}}\brk{1,i,0,0,0,0}
\,,\nn\\
y_3&=\frac{1}{\sqrt{2}}\brk{0,0,0,0,1,i}
\,,&
y_4&=\frac{1}{\sqrt{2}}\brk{1,-i,0,0,\beta_4,-i\beta_4}
\,.
\label{eq:yi}
\end{align}
Here, $\Phi=\brk{\phi_1,\dots,\phi_6}$ are the six real scalars of
$\superN=4$ SYM, and $y_i$ are six-component complex null vectors that
parametrizes the internal polarizations of the operators.
Since all operators $\op{O}(y_i)$ are BPS primary ``vacua'', there
will be no magnons on any physical edge of the two octagons. The
polarizations $y_i$ are chosen such that all tree-level graphs are
still square-shaped:
\begin{equation}
\includegraphics[align=c]{FigBoxGraph}
\label{eq:skelgraph}
\end{equation}
but the bridge length $m$ (number of parallel propagators) may take
all values $0\leq m\leq 2k$ (and $2k-m$ accordingly). The
hexagonalization prescription requires to sum over all these skeleton
graphs with different $m$. For small (or
large) values of $m$, there will be interactions between the inside
and the outside of the graph (the front and back of the lower left
world-sheet in~\figref{fig:PantsVs4p}). In order to confine all
interactions to the individual inside and outside octagons, we have to
restrict to $m\sim k$, as is made sure by the choice of
operators~\eqref{eq:operators}. Moreover, the correlator
$\avg{\op{O}_1\dots\op{O}_4}$ can be extracted from
$\avg{\op{O}(y_1)\dots\op{O}(y_4)}$ by a simple projection: The
reduced correlators
\begin{equation}
\mathcal{G}:=
\brk*{x_{12}^2x_{34}^2}^{2k}
\avg!{\op{O}_1(x_1)\dots\op{O}_4(x_4)}
\,,\qquad
\mathcal{\tilde{G}}:=
\brk*{\frac{x_{12}^2x_{34}^2}{y_{12}^2y_{34}^2}}^{2k}
\avg!{\op{O}(y_1,x_1)\dots\op{O}(y_4,x_4)}
\end{equation}
depend only on the spacetime and internal cross
ratios~\eqref{eq:zzbar} and
\begin{equation}
\alpha\bar \alpha=\frac{y_{12}^2y_{34}^2}{y_{13}^2y_{24}^2}=\sigma
\,,\qquad
(1-\alpha)(1-\bar \alpha)=\frac{y_{14}^2y_{23}^2}{y_{13}^2y_{24}^2}=\tau
\,,
\label{eq:sigmatau}
\end{equation}
and by R-symmetry conservation,
$\mathcal{G}$ can be extracted from $\mathcal{\tilde{G}}$ via
\begin{equation}
\mathcal{G}(z,\bar z)=
\eval[s]!{\mathcal{\tilde{G}}(z,\bar z,\sigma,\tau)}_{\text{coefficient of }\tau^0\sigma^{-k}}
\,.
\end{equation}
The correlator $\mathcal{\tilde{G}}$ is a finite power series in
$\sigma$. From the hexagonalization point of view, there are two
sources for powers of $\sigma$: The octagons $\oct_{\ell=0}$ inside
and outside the square skeleton graph, and the
skeleton graph~\eqref{eq:skelgraph} itself, which is proportional to $(y_{12}^2y_{34}^2/\sigma)^{2k}\sigma^{m}$.
Each skeleton graph is weighted by the same function
$\oct_{\ell=0}^2$. The latter has a power expansion
$\sum_{i=-p}^pc_i\sigma^i$, with $p\leq L$ at $L$ loops.%
\footnote{At tree level, the correlator
$\avg{\op{O}_1\dots\op{O}_4}$ is exactly given by the skeleton graph~\eqref{eq:skelgraph}
with $m=k$. At higher loop orders, there will also be contributions
from skeleton graphs with $m$ deviating from $k$ by
(small) finite numbers.}
Together with
the final projection to the $\sigma^{-k}$ coefficient, each skeleton
graph thus picks a different term in the octagon expansion. At large
enough $k$, each term is picked exactly once, and thus the whole sum
amounts to the full octagon squared function $\oct_{\ell=0}^2$
evaluated at $\sigma=1$. The polarizations~\eqref{eq:yi} in addition
imply $\tau=0$. Together, this is equivalent to
$\alpha=\bar\alpha=1$, and therefore
\begin{equation}
\mathcal{G}=
\mathcal{G}^{\text{tree}}
\,\oct^2_{\ell=0}(\alpha=1,\bar\alpha=1)
\,.
\end{equation}
For general values of $\alpha$ and $\bar\alpha$, the
parameters (angles) of the transformation $g$ in the
octagon expression~\eqref{eq:octagonexpression} are
\begin{equation}
\phi=-\frac{i}{2}\log \frac{z}{\bar z}
\,,\qquad
\theta=-\frac{i}{2}\log \frac{\alpha}{\bar\alpha}
\,,\qquad
\varphi=\frac{1}{2}\log \frac{\alpha\bar\alpha}{z\bar z}
\,.
\end{equation}
In terms of these angles, the character takes the form
\begin{equation}
W_{\brk[c]{a_i}}=\frac{W^+_{\brk[c]{a_i}}+W^-_{\brk[c]{a_i}}}{2}
\,,\qquad
W_{\brk[c]{a_i}}^\pm=\prod_{j=1}^n2\brk!{\cos\phi-\cosh\brk{\varphi\pm i\theta}}\frac{\sin(a_j\phi)}{\sin\phi}
\,.
\label{eq:characterdef}
\end{equation}
For $\alpha=\bar\alpha=1$, the angles reduce
to~\eqref{eq:alphafix}. In this case,
\begin{equation}
2\brk!{\cos\phi-\cosh\brk{\varphi\pm i\theta}}
=-\frac{\brk{1-z}\brk{1-\bar z}}{\sqrt{z\bar z}}
=-4\sinh\brk*{\frac{\varphi}{2}+\frac{i\phi}{2}}\sinh \brk*{\frac{\varphi}{2}-\frac{i\phi}{2}}
\,,
\end{equation}
and therefore~\eqref{eq:characterdef} becomes~\eqref{eq:character}.

\section{Including a Finite Bridge Length}
\label{sec:withBridge}

It is straightforward to include a finite bridge-length in the octagon
strong-coupling derivation by simply adding an extra factor $e^{-a
\ell/\sqrt{u^2-4g^2}}$ to $T_a$ in~\eqref{eq:Ta}. This leads to a
slightly less pretty expression in the integration-by-parts
step~\eqref{eq:IBP}, and to an $i0$ prescription in the subsequent
$\theta$ shift. That is the only modification,
so we now have%
\footnote{Here we assumed $\varphi<0$ -- the
opposite of the main text -- to illustrate how the shift would go in
that case. For the relation between positive and negative
$\varphi$ and on the symmetry which flips the sign of $\varphi$, see
\appref{sec:EvenAp}.}
\begin{align}
\frac{1}{{\color{red}n}} \int_{-2g}^{2g} du
\, e^{i {\color{red}n} \varphi \frac{u}{\sqrt{4g^2-u^2}}
        - {\color{red}n} \brk*{\mathcal{L}\equiv{\ell}/{2g}} \frac{2g}{\sqrt{4g^2-u^2} } }
= \frac{2g}{{\color{red}n}} \int_{-\infty}^{\infty} d\theta
\, \frac{d \tanh(\theta)}{d\theta}
e^{i {\color{red}n} \varphi \sinh(\theta)- {\color{red}n} \mathcal{L}\cosh(\theta)}
\nn \\
=-2g i \int_{-\infty}^{\infty} d\theta
\, \brk*{\varphi \sinh(\theta)+i \mathcal{L} \frac{\sinh^2(\theta)}{\cosh(\theta)} }
e^{i {\color{red}n} \varphi \sinh(\theta)- {\color{red}n} \mathcal{L}\cosh(\theta)}
\nn  \\
=2g \int_{-\infty}^{\infty} d\theta
\, \brk*{\varphi\cosh(\theta) +i  \mathcal{L} \frac{\cosh^2(\theta)}{\sinh(\theta+i0)} }
e^{{\color{red}n} \varphi \cosh(\theta)+i {\color{red}n} \mathcal{L}\sinh(\theta)}
\end{align}
when getting rid of the $1/n$ obstruction to factorization, and
thus find
\begin{equation}
\log\mathbb{O}_{{l}} \simeq
\frac{\sqrt{\lambda}}{2\pi }
\int\limits_{-\infty}^\infty \frac{d\theta}{2\pi}
\,\brk*{\varphi\cosh(\theta) + i\mathcal{L} \frac{\cosh^2(\theta)}{\sinh(\theta+i 0)} }
\log\brk{1+Y_\ell(\theta)}
\,,
\label{predictionWithL}
\end{equation}
where $Y_\ell(\theta)$ is given by~\eqref{eq:Yfunction} with
$\varphi\cosh(\theta)\to \varphi\cosh(\theta)+i \mathcal{L}
\sinh(\theta)$, with $\mathcal{L} \equiv \ell/2g$. If the bridge
length scales with $g\sim \sqrt{\lambda}$, then the bridge presence
significantly affects the final result, and it would be
interesting to reproduce this more general result from a string
sigma-model minimal-area computation. If ${\ell} = \order{1}$, then
$\mathcal{L}\to 0$, and the bridge has no effect at strong coupling, as expected.

\section{Minimal Areas Ending on Geodesics}
\label{sec:minimal-areas-ending}

\begin{figure}
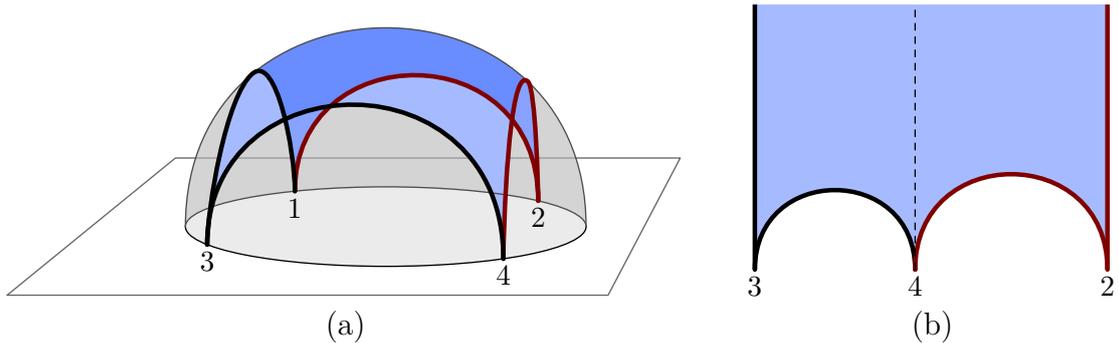

\centering
\begin{tabular}{c@{\qquad}c}
\includegraphics{FigAdSDome} &
\includegraphics{FigAdSDomeFlattened} \\
(a) & (b)
\end{tabular}
\caption{(a) Several geodesics ending on the same circle are
conformally equivalent to (b) Geodesics ending on the same straight
line. In the latter picture, we used conformal symmetry to put one of
the operators at infinity. We see very clearly in this frame that the
area becomes the sum of two pieces, separated by the dashed line. More
general, had we started with~$n$ points on a line, we would have
ended with~$n-2$ such world-sheet patches. In the text, we show that
the area of each patch is $\pi$. In the left figure removing the area
below the geodesics amounts to removing the gray patches of the
spherical dome, leaving only the blue cap.}
\label{fig:dome}
\end{figure}

This appendix followed from the following observation by Martin
Kruczenski: \emph{If we have some concatenation of geodesics on
$AdS$, whose endpoints lie on a common circle, then the minimal surface
which ends on those geodesics is nothing but the part of a spherical
dome ending on the circle that is enclosed by the geodesics},%
\footnote{This is in fact a general property of minimal surfaces: The
condition for minimality (vanishing of the mean curvature) is a local
condition. Hence cutting off arbitrary parts of any given minimal
surface (in our case, a half-sphere in the Poincar\'e plane) again
yields a minimal surface, with the boundary conditions given by the
chosen cut contours.}
see~\figref{fig:dome}a.
That circle configuration can be mapped to the
straight line, where the area is even simpler and given in~\figref{fig:dome}b.

It then becomes a straightforward exercise to compute this area. Of
course, this problem is \emph{not} the actual problem we want to solve, as
here we are totally ignoring the sphere. Indeed, instead of obtaining
the rich result~\eqref{eq:prediction} in the circle limit $z\to \bar z$
(or $\phi \to 0$), this simpler minimal area computation yields
a simple constant, an integer multiple of $\pi$. As
explained in the figure, for an $n$-point function we would simply need
to consider the area of $n-2$ world-sheet patches, each of which lies above
its own geodesic circle over the straight line. Each of these thus gives
\begin{equation}
\int\limits_{x_i}^{x_{i+1}} dx \int\limits_{b}^\infty dz\, \frac{1}{z^2} = \pi
\,,\qquad
b=\sqrt{\brk*{\frac{x_i-x_{i+1}}{2}}^2-\brk*{\frac{x_i+x_{i+1}}{2}-x}^2}
\,,
\end{equation}
so that the area for $n$ such geodesics simply is $(n-2)\pi$
in this sphere-free toy model. Note that each area piece is a pure
number, independent of the geodesic end-points~$x_i$; this is because
of conformal invariance. Relatedly, note that this
area bounded by geodesics is manifestly finite, without any need of subtractions, as
anticipated in~\cite{Bargheer:2019kxb}. This is in contrast with other, more
conventional minimal surface problems in $AdS/CFT$, where the surfaces
go all the way to the boundary, thus picking up a divergent piece which
one should re-normalize.

Of course, our actual result for the area is not as simple, although it does simplify a
little bit once we put all operators on a common line/circle: It becomes a
simple function of $\varphi$:
\begin{equation}
\log\mathbb{O}
\simeq
\frac{\sqrt{\lambda}}{2\pi}
\int\limits_{-\infty}^\infty
\frac{d\theta}{2\pi}
\,\varphi\cosh(\theta)
\log\brk[s]*{1-\frac{
\sinh^2\big(\frac{\varphi}{2}\big)
}{
\sinh^2\big(\frac{\varphi}{2}\cosh(\theta)\big)
}}
\,.
\end{equation}

\section{Analytic Structure}
\label{sec:AppendixContinuations}

After partial integration, the area~\eqref{eq:predictionA}
becomes (boundary terms vanish)
\begin{equation}
\log\mathbb{O}
\simeq
\frac{\sqrt{\lambda}}{2\pi}
\int\limits_{-\infty}^\infty
\frac{d\theta}{2\pi}
\frac{
\varphi^2
\sinh(\theta)^2
\sinh(\varphi\cosh\theta)
\brk*{\cos\phi-\cosh\varphi}
}{
\brk*{\cosh\varphi-\cosh(\varphi\cosh\theta)}
\brk*{\cos\phi-\cosh(\varphi\cosh\theta)}
}
\;\;\quad
(\varphi<0)
\,.
\label{eq:logoctpi}
\end{equation}
The integrand is now a meromorphic function of $\theta$, and most of
the analytic structure of $\log\oct$ can be inferred from the behavior
of its poles and their residues. The poles of the integrand are the
zeros of the two factors in the denominator, which are located at
\begin{equation}
\theta=\pm\arccosh\brk*{1+\frac{2\pi i\Integers_{\neq 0}}{\varphi}}+\pi i\Integers
\,,\qquad
\theta=\pm\arccosh\brk*{\frac{i\phi+2\pi i\Integers}{\varphi}}+\pi i\Integers
\,.
\label{eq:integrandpoles}
\end{equation}
At the points $\theta=\pi i\Integers$, the numerator factor
$\sinh^2(\theta)$ vanishes, canceling the zero of the denominator; we
hence excluded those points above.
In Euclidean kinematics, both $\varphi$ and $\phi$ are real
($\bar z$ is the complex conjugate of $z$). For all real values of
$\varphi$ and $\phi$, all poles remain away
from the real axis.
Moreover, the locations of the poles~\eqref{eq:integrandpoles} as well es
their residues are invariant under $\phi\to\phi\pm2\pi i$. Hence,
the function $\log\oct$ is a single-valued smooth function in Euclidean
kinematics, as it should be.

In complexified kinematics, both $z$ and $\bar z$ are complex and
independent of each other. Euclidean kinematics are located on the
real section $\bar z=z^*$, where $z$ and $\bar z$ are complex
conjugates. In contrast, $z$ and $\bar z$ are real and
independent in Lorentzian kinematics.

\subsection{The Area is Even}
\label{sec:EvenAp}

\begin{figure}[t]
\centering
\includegraphics{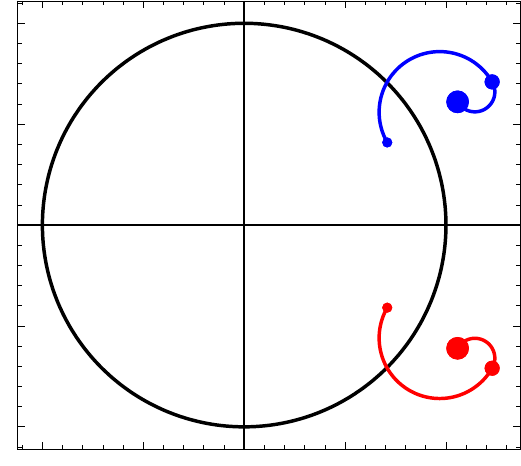}
\caption{Inversion $\varphi\to-\varphi$ in the $z,\bar z$ plane: The
figure shows the continuation path for $z$
(blue) and $\bar z$ (red). The continuation proceeds in four steps:
(1) Blue segment from small to medium dot, (2) red segment from small
to medium dot, (3) blue segment from medium to large dot, and (4) red
segment from medium to large dot. This path of continuation avoids
singular points where infinitely many residues accumulate at
$\theta=\infty$. In this example, the three dots have coordinates
$\phi=\pi/6$, $\varphi_1=-1/5$, $\varphi_2=7/20$, $\varphi_3=1/5$.}
\label{fig:inversionpath}
\end{figure}
\begin{figure}[t]
\includegraphics{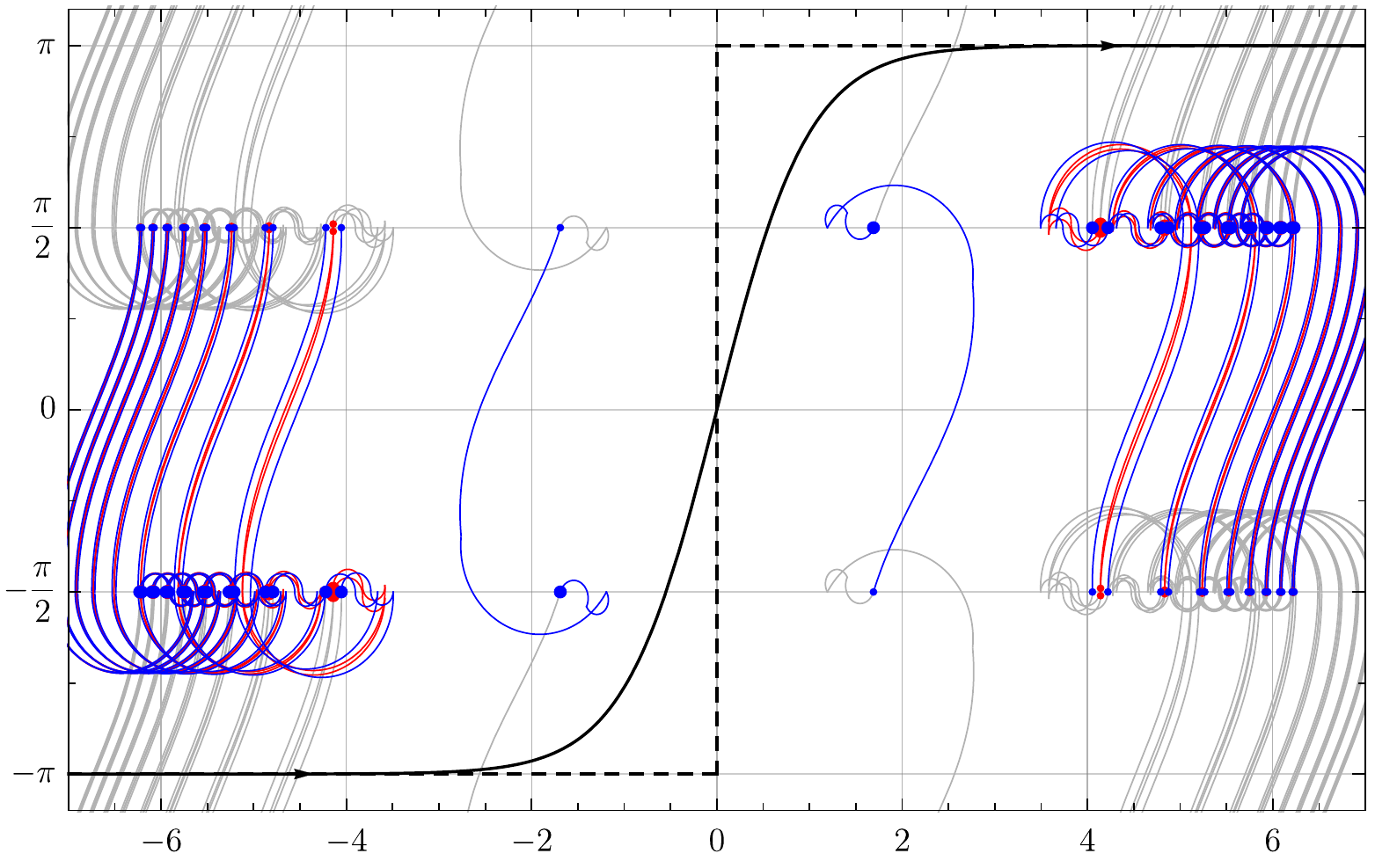}
\caption{Inversion $\varphi\to-\varphi$ in the $\theta$ plane: During
the continuation explained in~\figref{fig:inversionpath}, the poles in
the left/right half plane shift down/up along the contours shown (from
small to large dots). This pushes the contour of integration to the
black solid line. Deforming the contour further to the black dashed
line, it can be easily related back to the original function,
producing an overall minus sign.}
\label{fig:inversioncont}
\end{figure}

In particular, the area should be invariant
under inversions $z\to1/\bar z$, $\bar z\to1/z$, that is
$\varphi\to-\varphi$. In contrast, the integrand
of~\eqref{eq:logoctpi} behaves non-trivially as one passes from
$\varphi<0$ to $0<\varphi$. Performing this continuation within the
Euclidean section, almost all poles diverge to $\theta=\infty$ as
$\varphi$ approaches $\varphi=0$, which makes it difficult to see what
happens during the continuation. We can circumvent this problem by
deforming the continuation into complex kinematics. A convenient path
of continuation is shown in~\figref{fig:inversionpath}. Along such a
path, all towers of poles stay at finite $\theta$, and two such towers
cross the real line as shown in~\figref{fig:inversioncont}. This
deforms the contour of integration to the solid black line in the
figure. Deforming it further to the dashed black line, we can relate
the integral back to the original expression. To that end, first note
that the integrand in~\eqref{eq:logoctpi} acquires a minus sign under
a shift $\theta\to\theta\pm i\pi$. Hence the integrations along the
horizontal dashed contours in~\figref{fig:inversioncont} equal
\emph{minus} the original integration along the real line. Moreover,
the integrand is invariant under $\theta\to-\theta$; hence the
integration along the vertical dashed contour from $-i\pi$ to $i\pi$
gives zero. Since the integrand in~\eqref{eq:logoctpi} is odd under
$\varphi\to-\varphi$, one finds indeed that
\begin{equation}
\log\oct(-\varphi)=\log\oct(\varphi)
\,,
\label{eq:logoctinversion}
\end{equation}
as required by invariance under conformal inversions.

\subsection{Analytic Continuations}

The complexified function
$\log\oct$ is locally holomorphic in $z$ and $\bar z$ independently. Globally,
the function has branch points: For fixed~$\bar z$ ($z$), there are
branch points at $z=0$ and $z=1$ ($\bar z=0$ and $\bar z=1$). Because
$\log\oct$ is single-valued in Euclidean kinematics, the monodromies
in $z$ and $\bar z$ are related:
\begin{equation}
\disc_{z \rotpos 0}\log\oct(z,\bar z)
+\disc_{\bar z \rotneg 0}\log\oct(z,\bar z)
=0
\,,\qquad
\disc_{z \rotpos 1}\log\oct(z,\bar z)
+\disc_{\bar z \rotneg 1}\log\oct(z,\bar z)
=0
\,,
\label{eq:zzbdiscrel}
\end{equation}
where \eg $\disc_{z \rotpos z_0}f(z)$ is the discontinuity that is picked up by
$f(z)$ as $z$ follows a closed path encircling $z_0$ once
counterclockwise, with all other variables held fixed.

\paragraph{The Space of Analytic Continuations.}

We want to explore analytic continuations of the area $\log\oct$ as a
complexified function of the two variables $z$ and $\bar z$. We will
focus on continuations where $\bar z$ is held fixed, while $z$ follows
some non-trivial cycles around $z=0$ and/or $z=1$. The extension to
continuations that also involve cycles of $\bar z$ is not difficult,
using relations such as~\eqref{eq:zzbdiscrel}. Without loss of
generality, we will have our paths of analytic continuation start and
end in the Lorentzian region $0<z,\bar z<1$. The space of analytic
continuations of $\log\oct$ is then a representation of the
fundamental group of the sphere (compactified complex plane) with
three marked points $z=\brk[c]{0,1,\infty}$. The fundamental group is a
free group with two generators. As our two generators, we choose
$\gen_0$ and $\gen_1$ that wind $z$ counterclockwise
around $z=0$ and $z=1$, respectively:
\begin{equation*}
\includegraphics[align=c]{FigGenerators}
\end{equation*}
%

\paragraph{The Branch Point at \texorpdfstring{$z=0$}{z=0}.}

\begin{figure}[t]
\includegraphics{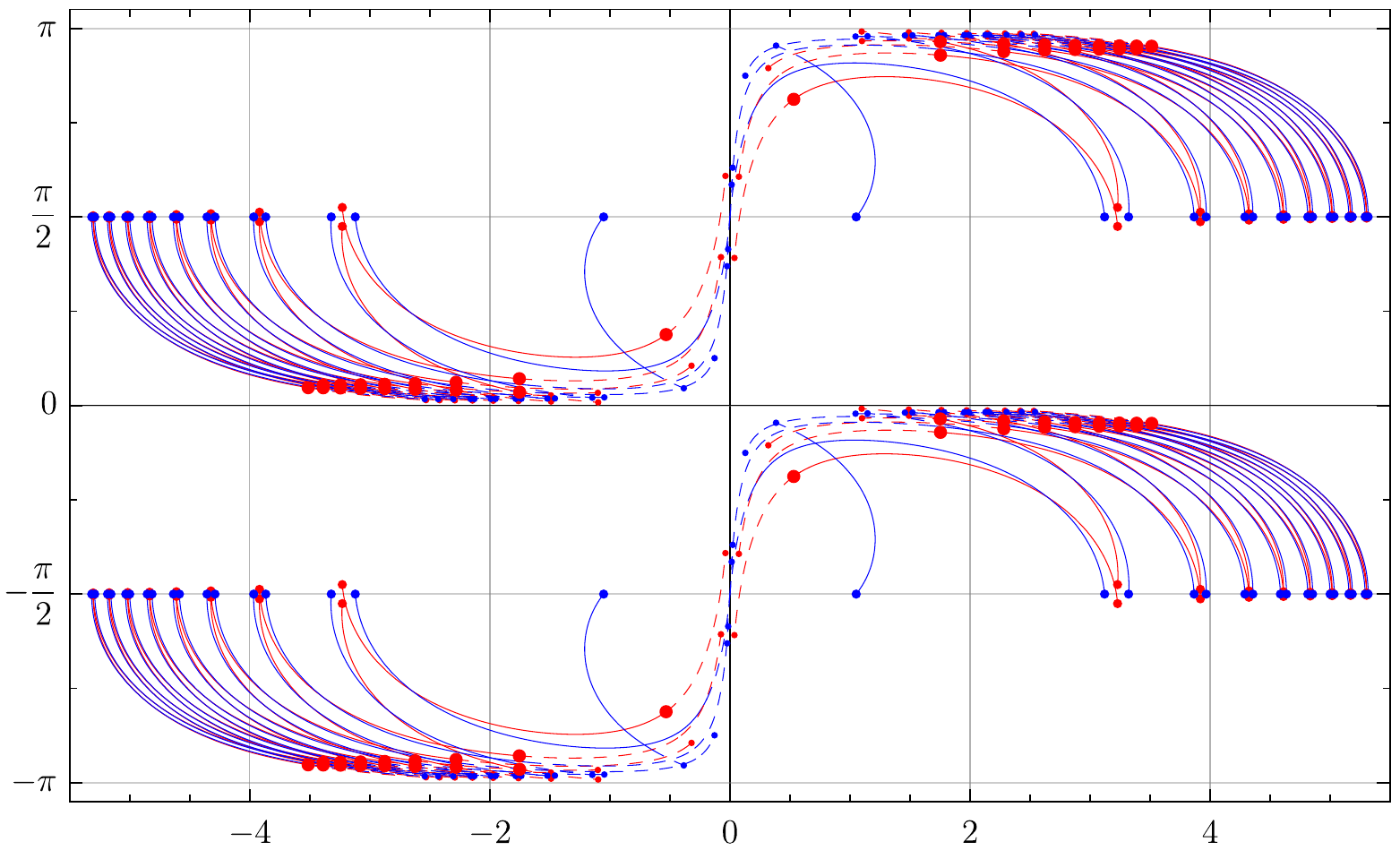}
\caption{Continuation of $z$ around $z=0$, with $\bar z$ held fixed.
The figure shows the relevant poles of the integrand of $\log\oct$ in the $\theta$
plane. The first sequence in~\protect\eqref{eq:integrandpoles} is shown
in red, the second in blue. The path of continuation is
$\varphi=1/2-i\alpha/2$, $\phi=\pi/5+\alpha/2$; it starts and
ends in the Euclidean section. Before the
continuation, almost all poles lie near the lines $\im(z)=\pm\pi/2$
(medium points). After one full rotation of $z$ ($\alpha=2\pi$, large
points), many poles have moved close to the real axis. After two more
rotations of $z$ ($\alpha=6\pi$, dashed lines, small points), some
poles approach the imaginary axis. Going further, more and more poles
accumulate near the imaginary axis. At no point does any pole cross
the real line (contour of integration).}
\label{fig:zcont0}
\end{figure}

In the analytic continuation of $z$ around $0$, the complete
discontinuity comes from the map $z\mapsto\log z$ contained in the
inversion of~\eqref{eq:crossRatios}. In other words, $\varphi$ and
$\phi$ provide uniformizing coordinates that resolve the branch point
at $z=0$ (and hence equally the branch point at $\bar z=0$). In order
to evaluate the continued function, we simply have to evaluate the
integral~\eqref{eq:logoctpi} at appropriately shifted values of
$\varphi$ and $\phi$. For example, a rotation of $z$ around the
origin, with $\bar z$ held fixed, is realized by
$\varphi\mapsto\varphi-i\pi$, $\phi\mapsto\phi+\pi$. As shown
in~\figref{fig:zcont0}, no poles cross the real line during such
continuations, and hence the $\theta$ contour of integration can be
maintained without picking up any residues. Hence we find
\begin{equation}
\log\oct(\varphi,\phi)
\xrightarrow{\,\brk{z\rotpos 0}^n\,}
\gen_0^n\log\oct(\varphi,\phi)
=\log\oct(\varphi-i\pi n,\phi+\pi n)
\end{equation}
%

\paragraph{The Branch Point at \texorpdfstring{$z=1$}{z=1}.}

\begin{figure}[t]
\includegraphics{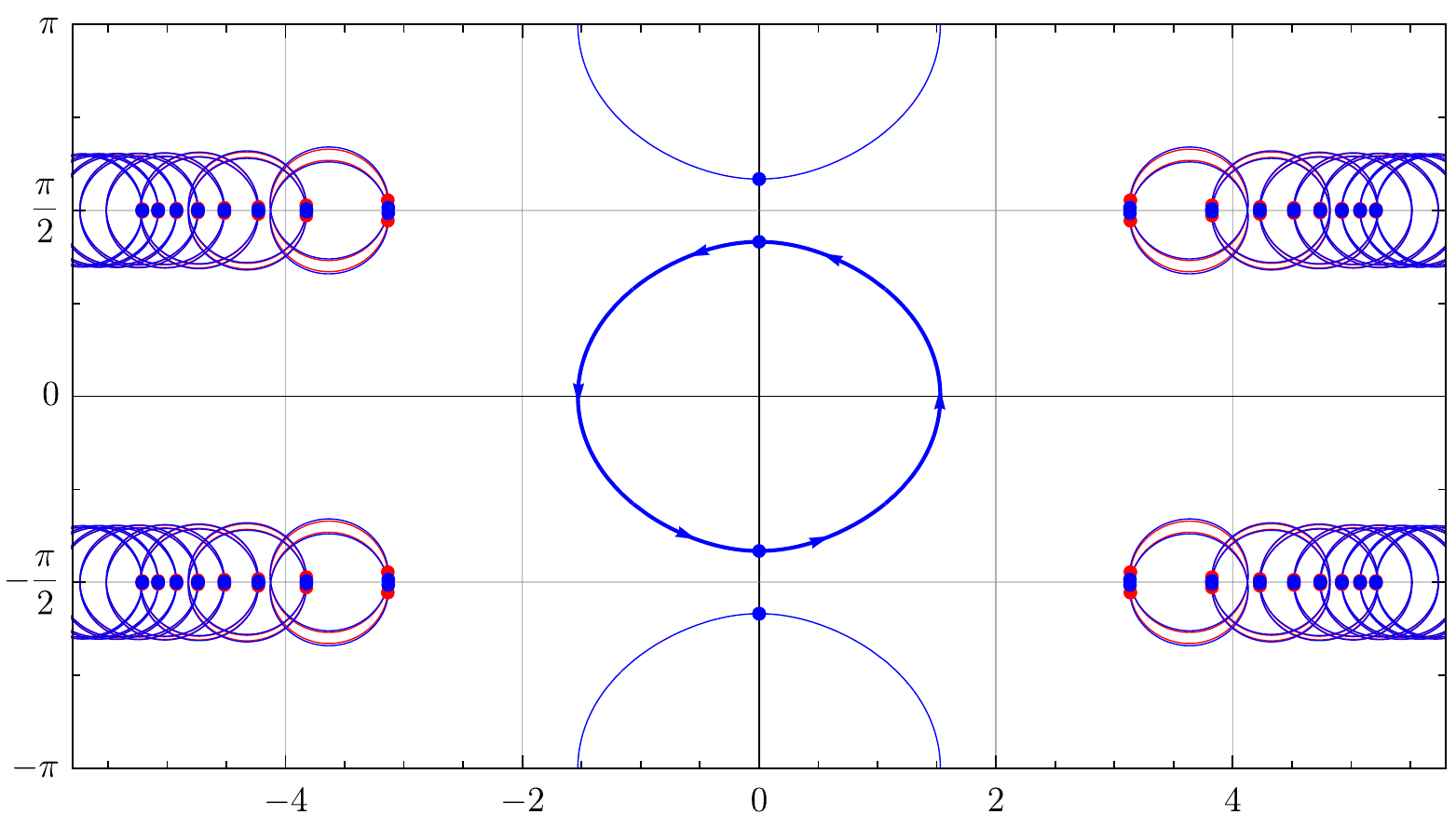}
\caption{Continuation of $z$ around $z=1$, with $\bar z$ held fixed.
The figure shows the relevant poles of the integrand of $\log\oct$ in the theta
plane. The path of continuation in this example is
$z=1-1/3\mathinner{e^{i\alpha}}$, $\bar z=1/2$; it starts and ends
in Lorentzian kinematics. Almost no poles come
close to the real line during the continuation. The configurations
before and after the continuation (at $\alpha=0$ and $\alpha=2\pi$)
are identical, but two poles have crossed the real line and
interchanged (shown in red and blue). Hence the function $\log\oct$
picks up the residues of these poles.}
\label{fig:zcont1}
\end{figure}

When we analytically continue in $z$ around $z=1$, with $\bar z$
sufficiently far away from $\bar z=1$, two poles cross the real axis,
and hence the integral~\eqref{eq:logoctpi} picks up the residues of
those poles, see~\figref{fig:zcont1}.%
\footnote{Only these two poles cross as long as either $z\bar z<0$ or
$z\bar z>0$ throughout the complete continuation. Otherwise, the
end result~\eqref{eq:z1monodromy} remains correct, but infinite towers of
poles cross the real axis during the continuation, rendering the
analysis slightly more complicated.}
Starting (and ending) the
continuation at $0<\bar z<z<1$ with $\varphi$ real and $\phi$ purely
imaginary, the poles that cross the real
axis are located at
\begin{equation}
\theta_\pm
=\mp\arccosh\brk*{-\frac{\log z-\log\bar z}{\log z+\log\bar z}}
=\mp\arccosh\brk*{\frac{i\phi}{\varphi}}
\,.
\label{eq:C1poles}
\end{equation}
They start on the imaginary axis, and rotate counterclockwise by
$\ang{180}$, ending up at their initial, but now interchanged,
locations. At the beginning of the continuation, $\theta_+$ lies in
the upper half plane, and $\theta_-$ lies in the lower half plane.
Hence $\theta_+$ crosses the real axis from above, while
$\theta_-$ crosses it from below. Before the continuation, the residues of the integrand
in~\eqref{eq:logoctpi} at these poles are
\begin{equation}
\pm\frac{i}{2\pi}
\sqrt{\log(z)\log(\bar z)}
\,.
\label{eq:C1residues}
\end{equation}
But the continuation rotates $\log(z)$ around $0$, hence the sign of the
square root switches. Combining all signs and contour orientations, we
therefore find
\begin{equation}
\log\oct
\xrightarrow{\,z\rotpos 1\,}
\gen_1\log\oct
=\log\oct-\frac{\sqrt{\lambda}}{\pi}\sqrt{\log(z)\log(\bar z)}
\label{eq:z1monodromy}
\end{equation}
under the counterclockwise rotation of $z$ around $z=1$ with $\bar z$
fixed.
We find the same result when the start and end points of the
continuation lie in the region $1<z<\bar z$.
At the end of the continuation, the poles~\eqref{eq:C1poles} have
interchanged, but their residues~\eqref{eq:C1residues} have also
swapped, so that again the residue with the plus sign is in the
upper half plane, while the residue with the minus sign is in the lower
half plane. When we now rotate back, \ie apply $\gen_1^{-1}$, the area
$\log\oct$ thus obtains the same discontinuity:
\begin{equation}
\log\oct
\xrightarrow{\,z\rotneg 1\,}
\gen_1^{-1}\log\oct
=\log\oct-\frac{\sqrt{\lambda}}{\pi}\sqrt{\log(z)\log(\bar z)}
\,.
\label{eq:z1monodromyneg}
\end{equation}
This is consistent, as the discontinuity term is a square root that
itself changes sign upon continuation:
\begin{equation}
\gen_1^{-1}\frac{\sqrt{\lambda}}{\pi}\sqrt{\log(z)\log(\bar{z})}
=\gen_1\frac{\sqrt{\lambda}}{\pi}\sqrt{\log(z)\log(\bar{z})}
=-\frac{\sqrt{\lambda}}{\pi}\sqrt{\log(z)\log(\bar{z})}
\,,
\end{equation}
such that consistently
\begin{equation}
\gen_1^{-1}\gen_1\log\oct
=\gen_1^{-1}\log\oct-\gen_1^{-1}\frac{\sqrt{\lambda}}{\pi}\sqrt{\log(z)\log(\bar{z})}
=\log\oct
\,.
\end{equation}
From the above discussion, it is clear what happens when $z$ winds
around $z=1$ any number of times: For each winding (in either
direction), $\log\oct$ picks up a term as in~\eqref{eq:z1monodromy}. In
addition, an already present term of this type will change sign, thus
canceling the new term. Hence we conclude that
\begin{equation}
\log\oct
\xrightarrow{\,\brk{z\rotpos 1}^n\,}
\gen_1^n\log\oct
=\log\oct-\delta_{1,(n\,\text{mod}\,2)}\frac{\sqrt{\lambda}}{\pi}\sqrt{\log(z)\log(\bar z)}
\,.
\end{equation}
%

\paragraph{Combined Continuations.}

We can now consider more complicated continuations that combine
windings around $z=0$ and $z=1$. Suppose that we
start in the Lorentzian section with $0<(z,\bar z)<1$, $\varphi$ real
and $\phi$ imaginary, keep $\bar z$ fixed, and let $z$ undergo a sequence of windings around $z=0$
and/or $z=1$. This amounts to applying a sequence of continuations
$\gen=\dots\gen_1^{p_4}\gen_0^{p_3}\gen_1^{p_2}\gen_0^{p_1}$ to $\log\oct$. We saw above that $\gen_0$ merely shifts
$\varphi$ and $\phi$ in the formula~\eqref{eq:logoctpi} for
$\log\oct$, not affecting the contour of integration. In contrast,
$\gen_1$ picks up residues from two specific poles of the
integrand, as shown in~\figref{fig:zcont1}. Now it turns out that when
we first apply a number of continuations $\gen_0^n$ around zero, and
then continue around $z=1$, a
\emph{different} pair of poles will cross the integration contour.
Namely, applying $\gen_1\gen_0^n$, the two
poles that cross the integration contour are
\begin{equation}
\theta_{n,\pm}=\mp\arccosh\brk*{\frac{i\phi-2\pi in}{\varphi}}
\,.
\end{equation}
Again, $\theta_{n,+}$ crosses from the upper half plane, and $\theta_{n,-}$
crosses from the lower half plane. At the end of the continuation,
the residues of the integrand in~\eqref{eq:logoctpi} at these poles are
\begin{equation}
\mp\frac{i}{2\pi}\sqrt{\brk!{\varphi+i\phi-2\pi i n}\brk!{\varphi-i\phi+2\pi in}}
=\mp\frac{i}{2\pi}\sqrt{\log(z)\brk!{\log(\bar z)+2\pi in}}
\,,
\end{equation}
where the logarithms are evaluated on the standard branch, that is
$-i\pi<\im\brk{\log(\alpha)}\leq i\pi$.
We therefore find
\begin{equation}
\gen_0^n\log\oct\xrightarrow{\,z\rotpos 1\,}
\gen_1\gen_0^n\log\oct=
\gen_0^n\log\oct-\frac{\sqrt{\lambda}}{\pi}\sqrt{\log(z)\brk!{\log(\bar z)+2\pi in}}
\,.
\end{equation}
The total effect of $n$ windings around $z=0$ followed by one winding
around $z=1$ (all counterclockwise) hence is
\begin{equation}
\gen_1\gen_0^n\log\oct(\varphi,\phi)
=\log\oct(\varphi-i\pi n,\phi+\pi n)
-\frac{\sqrt{\lambda}}{\pi}\sqrt{\log(z)\brk!{\log(\bar z)+2\pi in}}
\,.
\label{eq:monodromyz0nz1}
\end{equation}
As in the simple case~\eqref{eq:z1monodromyneg},
performing a reverse rotation around $z=1$ yields the same result:
\begin{equation}
\gen_1^{-1}\gen_0^n\log\oct(\varphi,\phi)
=\log\oct(\varphi-i\pi n,\phi+\pi n)
-\frac{\sqrt{\lambda}}{\pi}\sqrt{\log(z)\brk!{\log(\bar z)+2\pi in}}
\,.
\label{eq:monodromyz0nz1neg}
\end{equation}
In contrast, letting $z$ first wind once around $z=1$ and then $n$
times around $z=0$ yields
\begin{equation}
\gen_0^n\gen_1\log\oct(\varphi,\phi)
=\log\oct(\varphi-i\pi n,\phi+\pi n)
-\frac{\sqrt{\lambda}}{\pi}\sqrt{\brk!{\log(z)+2\pi i n}{\log(\bar z)}}
\,.
\end{equation}
This clearly shows that the two generators $\gen_0$ and $\gen_1$ do
not commute.

\medskip

\noindent
With these results, we can compute all analytic continuations in $z$:
Any path of analytic continuation can be written as a product of
factors $\gen_1\gen_0^n$ and $\gen_1^{-1}\gen_0^n$. The action of
these factors on $\log\oct$ is given above
in~\eqref{eq:monodromyz0nz1} and~\eqref{eq:monodromyz0nz1neg}, and the
action on the extra terms
\begin{equation}
\Delta_{p,q}:=\frac{\sqrt{\lambda}}{\pi}\sqrt{\brk!{\log(z)+2\pi ip}\brk!{\log(\bar{z})+2\pi iq}}
\end{equation}
produced by preceding factors is simple:
\begin{equation}
\gen_0^n\Delta_{p,q}
=\Delta_{p+n,q}
\,,\qquad
\gen_1^{\pm1}\Delta_{p,q}
=\brk{1-2\delta_{p,0}}\Delta_{p,q}
=\begin{cases}
-\Delta_{p,q} & p=0 \,, \\
+\Delta_{p,q} & p\neq0 \,.
\end{cases}
\end{equation}
With these continuations, we can evaluate the area
$\log\oct$ in all Euclidean and Lorentzian sections in all kinematics.
For example,
with the shorthand
\begin{equation}
\log\oct_n:=
\log\oct(\varphi-i\pi n,\phi+\pi n)
\,,
\end{equation}
we find
\begin{align}
\gen_1^{\pm1}\gen_0^n\gen_1^{\pm1}\gen_0^m\log\oct
&=\gen_1^{\pm1}\gen_0^n\brk!{\log\oct_{m}-\Delta_{0,m}}
\nn\\
&=\log\oct_{m+n}-\Delta_{0,m+n}-\gen_1^{\pm1}\gen_0^n\Delta_{0,m}
\nn\\
&=\log\oct_{m+n}-\Delta_{0,m+n}-\brk{1-2\delta_{n,0}}\Delta_{n,m}
\,.
\end{align}
Combining these formulae, we arrive at the
expressions~\eqref{eq:alloctcond} for the continuation of $\log\oct$
from the Euclidean into any Lorentzian region.

\section{Weak Coupling Comparison}
\label{sec:Weak}

\tabref{tab:limitsbig} summarizes the main similarities and
differences between weak and strong coupling. See the main text for all
details and precise pre-factors.

\begin{landscape}
\begin{table}[p]
\centering
\begin{tabular}{L{3.3cm} L{2.5cm} L{5.7cm} L{10.9cm}}
\toprule
  \rule{0pt}{13pt}{\bf Common Name}
& \rule{0pt}{13pt}{\bf Kinematics}
& \rule{0pt}{13pt}{\bf Strong Coupling $\mathbb{O}\sim e^{-\sqrt{\lambda} \mathbb{A}}$}
& \rule{0pt}{13pt}{\bf Perturbative Result}
\\ \midrule
Euclidean neighbor OPE
& $\bar z, z \to 0$
& $\mathbb{A} \sim \sqrt{\log (z\bar z)}  $
& $\mathbb{O} \simeq \sum \lambda^{n} \sum\limits_{k=0}^{{\color{blue}
n}} \log(z\bar z)^k\,\sum_{r,p}z^r \bar z^p c_{k,n,r,p}  $
\\ \midrule
Euclidean diagonal OPE
& $\bar z, z \to 1$
& $\mathbb{A} \sim  \sqrt{(1-z)(1-\bar z)}  $
& $\mathbb{O} \simeq \sum \lambda^{n} \sum\limits_{k=0}^{{\color{red}
1}} \log(y\bar y)^k\,\sum_{r,p}y^r \bar y^p d_{k,n,r,p}   $
\\ \midrule
Double light-like neighbor OPE
& $z \to 0$ and $\bar z\to \infty$
& $\mathbb{A} \sim \log(z/\bar z)^2$
& $\log \mathbb{O} \simeq
-\frac{(\log(-z)+\log(-1/\bar z))^2}{8\pi^2} \Gamma(\lambda) +\frac{1}{8} C(\lambda) +\frac{\lambda}{16\pi^2} \log(z\bar z)^2
$
\\ \midrule
Double light-like diagonal OPE
& $z \to 0$, $ \bar z\to 1$
& $\mathbb{A} \sim  \sqrt{\log(\tfrac{1}{z})} \sqrt{1-\bar z}$
& $\mathbb{O} \simeq \sum \lambda^{n} \sum\limits_{k=0}^{{\color{blue} n}} \sum\limits_{k'=0}^{{\color{red} 1}} \log(z)^k \log(\bar y)^{k'}\,\sum_{r,p}z^r \bar y^p e_{k,k',n,r,p}
$
\\ \midrule
Single light-like neighbor OPE
& $z \to 0$
& $\mathbb{A} \sim \sqrt{\log(\tfrac{1}{z})} (\text{Li}_{\tfrac{3}{2}}(1)-\text{Li}_{\tfrac{3}{2}}(\bar z)) $
& $ \mathbb{O} \simeq 1  -g^{2} \frac{(1-\bar{z})\left(\log(z\bar{z})
\log(1-\bar{z})+2\text{Li}_{2}(\bar{z})\right)}{\bar{z}}+\cdots $
\\ \midrule
Single light-like diagonal OPE
& $\bar z\to 1$
& $\mathbb{A} \sim \sqrt{\log(\tfrac{1}{z})} \sqrt{1-\bar z}$
& $ \mathbb{O} \simeq 1+ g^{2}\bar{y}\left(\frac{-\pi^{2}}{3}-\log
\bar{y}\,\log z +\log(1-z)\log z + 2 \text{Li}_{2}(z)\right)+\cdots
$
\\ \midrule
Diagonal equal neighboring length
& $z\bar z \to 1$
& $\mathbb{A}\sim  \log(\tfrac{z}{\bar z})(2\pi -\tfrac{1}{2i}\log(\tfrac{z}{\bar z}))$
& $ \mathbb{O} = \text{no particular simplification}$
\\ \midrule
Regge (\figref{fig:Regge}b)
& $z, \bar z\to 0$ after $z\rotpos 1$
& $\mathbb{A} \sim \sqrt{\log(z)\log(\bar z)}$
& $\mathbb{O} \simeq 1+ g^{2}\,2\pi i \frac{\log z-\log
\bar{z}}{z-\bar{z}} +\cdots $
\\ \midrule
Regge (\figref{fig:Regge}a)
& $z, \bar z\to 1$ after $z\rotpos 0$
& $\mathbb{A} \sim \sqrt{y}+\sqrt{\bar{y}}-\sqrt{y+\bar{y}}$
& $\mathbb{O} \simeq 1+ g^{2}\,2\pi i\,y\bar{y}\,\frac{\log y-\log
\bar{y}}{y-\bar{y}} +\cdots $
\\ \midrule
Bulk point
& $z\to \bar z$ after $z\rotpos 0$ \& $z\rotpos 1$
& $\mathbb{A}\sim \text{regular}$
& $ \mathbb{O} \simeq 1+g^{2}\,4\pi^{2}\frac{(1-\bar{z})^{2}}{z-\bar{z}}+\dots$
\\
\bottomrule
\end{tabular}
\caption{Telegraphic summary of the differences and similarities
between strong coupling and perturbation theory. We expect the strong
coupling results to be most representative of the full finite coupling
behavior. On the weak coupling column we use the
short-hand notation: $y\equiv1-z,\bar{y}\equiv1-\bar{z}$ and
$\bar{w}\equiv\frac{1}{\bar{z}}$}
\label{tab:limitsbig}
\end{table}
\end{landscape}

\section{Expansions of the Strongly-Coupled Octagon}

This appendix is devoted to the derivation of the OPE limits of the
octagon at strong coupling. We map this problem to the study of
various limits of the integral:
\begin{equation}
\label{eq:Int}
\mathbb{I}(\varphi,\mu)
\,=\,
-\varphi\,\int_{0}^{\infty} d\theta \cosh\theta
\log\brk*{1-e^{\mu}\,e^{-\varphi\,\cosh\theta}}
\,=\,
-\varphi\,\int_{1}^{\infty} dt
\,\frac{\log\left(1-e^{\mu}\,e^{-\varphi\,t}\right)}{\sqrt{1-1/t^2}}
\end{equation}
which serves as a building block for the octagon upon identifying the
integrand~\eqref{eq:Yfunction} of this latter:
\begin{align}
\log\left(1+Y\right)
&=\log\frac{
\brk!{1-e^{-{\log(z\bar{z})}/{2}}\,e^{{\log(z\bar{z})}/{2}\,\cosh(\theta)}}
\brk!{1-e^{ {\log(z\bar{z})}/{2}}\,e^{{\log(z\bar{z})}/{2}\,\cosh(\theta)}}
}{
\brk!{1-e^{-{\log(z/\bar{z})}/{2}}\,e^{{\log(z\bar{z})}/{2}\,\cosh(\theta)}}
\brk!{1-e^{ {\log(z/\bar{z})}/{2}}\,e^{{\log(z\bar{z})}/{2}\,\cosh(\theta)}}
}
\nn\\
&=\log\frac{
\brk!{1-e^{ \varphi}\,e^{-\varphi\,\cosh(\theta)}}
\brk!{1-e^{-\varphi}\,e^{-\varphi\,\cosh(\theta)}}
}{
\brk!{1-e^{ i\phi}\,e^{-\varphi\,\cosh(\theta)}}
\brk!{1-e^{-i\phi}\,e^{-\varphi\,\cosh(\theta)}}
}
\end{align}
can be decomposed into four pieces which give the following
representation:
\begin{equation}
\log\mathbb{O}\,=\,
-\frac{\sqrt{\lambda}}{2\pi^{2}}
\,\brk*{
     \mathbb{I}(\varphi,\varphi)
\,+\,\mathbb{I}(\varphi,-\varphi)
\,-\,\mathbb{I}(\varphi,i\phi)
\,-\,\mathbb{I}(\varphi,-i\phi)
}
\label{eq:FourInt}
\end{equation}
with cross ratios:
\begin{equation}
\varphi = -\frac{1}{2}\,\log\left(z\bar{z}\right)
\quad\text{and}\quad
i \phi = \frac{1}{2} \log\left(\frac{z}{\bar{z}}\right)
\end{equation}
The closed form of the integral \eqref{eq:Int} is not known for arbitrary values of
$\varphi$ and $\mu$, nevertheless we can obtain closed-form
expressions in the limits where these parameters are very large or
very small. For instance the case $\mu=0$ has been well studied, see
\cite{Gromov:2011de}, and the limit for large argument is
known in terms of modified Bessel functions of the second kind:
\begin{equation}
\mathbb{I}(\varphi,0)
\overset{\varphi\to\infty}{=}
\varphi\sum_{n=1}^{\infty}\frac{1}{n}\,\mathbf{K}_{1}(n\varphi)
\label{eq:Large}
\end{equation}
This series is convergent and can be well approximated by the first
few terms thanks to the exponential suppression
$\mathbf{K}_{1}(n\varphi) \simeq
\sqrt{\pi/2}\,{e^{-n\varphi}}/{\sqrt{n\varphi}}$ in the
large limit $\varphi \to \infty$.

The small limit $\varphi\to0$ is also known as:
\begin{equation}
\mathbb{I}(\varphi,0)
\overset{\varphi\to 0}{=}
\frac{\pi^{2}}{6}{\color{red}-\frac{\pi}{2}\varphi}
+\varphi^{2}
\brk*{
\frac{1}{8}
-\frac{\gamma_{E}}{4}
-\frac{1}{4}\log\brk*{\frac{\varphi}{4\pi}}
}
+\sum_{k=1}^{\infty}\mathcal{C}_{k}\,\varphi^{2k+2}
\label{eq:Small}
\end{equation}
but unlike the former case, this series has a finite radius of
convergence as dictated by the coefficients:
\begin{equation}
\mathcal{C}_{k}=
\frac{(-1)^{k+1}\zeta(2k+1)\Gamma(2k+1)}{2^{4k+2}\pi^{2k}\Gamma(k+1)\Gamma(k+2)}
\text{ with }
k\geq 1
\,.
\end{equation}
In the context of this paper the series in \eqref{eq:Large} and
\eqref{eq:Small} give us access to OPE limits of our four-point
function in the restricted kinematics $\phi=0$ or $z=\bar{z}$. In
order to address more generic OPE limits we need to find out how to
incorporate the chemical potential $\mu$ in these series. The rest of
this appendix is devoted to this task.

In \secref{sec:LargeExpansionSection} we revisit the large $\varphi$
series \eqref{eq:Large}, now including the chemical potential $\mu\neq
0$. This series allows us to obtain the Euclidean OPE limits in
\tabref{tab:OPElimits}. In particular we obtain the full series of the
OPE limits $z\to 0$ and/or $\bar{z}\to 0$. Furthermore, thanks to the
competition between the four terms in \eqref{eq:FourInt}, we also get
access to the leading term of the OPE series $z\to 1$ or $\bar{z} \to
1$ (not both limits together).

The remaining sections concern the limit of the integrals necessary to
obtain the double light-cone limit of the octagon $-i\phi\to\infty$ in
the restricted kinematics $\varphi\to0$. We start warming up
in \secref{app:SmallExpansionWithout} showing how to obtain the small
$\varphi$ series in \eqref{eq:Small} without chemical potential $\mu=0$. This
we achieved by starting with the large $\varphi$ series in
\eqref{eq:Large}, consider the small $\varphi$ series expansion of the
Bessel function $\mathbf{K}_{1}(n\varphi)$ and then finally exchange
the order of sums to perform the re-summation over $n$ in
\eqref{eq:Large}. This latter step requires the use of a Zeta
regularization and reproduces the result in
\eqref{eq:Small}\footnote{Except for the linear term in $\varphi$, see discussion below \eqref{eq:DeriveSmall}}. In the following sections we incorporate large and
small chemical potential $\mu$ in the regime of small $\varphi$. In
\secref{app:LargeChemical} we consider a large chemical potential
$\mu$ which gives us access to the leading term of the double light-cone limit of our
four-point function under the identification $\mu = -i\phi \to \infty$.
Finally in \secref{app:LargeChemical} we consider the limit
$\mu=\pm\varphi\to0$ and obtain a series representations for the
contributions of $\mathbb{I}(\varphi,\pm\varphi)$ showing how they
modify the sub-leading term of the double light-cone limit.

In all these derivations for series in small $\varphi$ we perform
dangerous steps such as exchanging order of sums and regularizing
infinite sums. We do not fully justify them but have verified
our results numerically in their corresponding regimes of validity.

\subsection{Series
\texorpdfstring{$\varphi\to\infty$}{phi to infinity}
Including a Chemical Potential \texorpdfstring{$\mu$}{mu}}
\label{sec:LargeExpansionSection}

The first representation in \eqref{eq:Large} can be obtained by
expanding the $\log$ in the integrand considering large $\varphi$ and
then performing the integral for each term in the series. Each term
evaluates to the modified Bessel function of the second kind
$\mathbf{K}_{1}$. Following this recipe we can easily incorporate the
chemical potential as
\begin{equation}
\mathbb{I}(\varphi,\mu)
=\varphi\sum_{n=1}^{\infty}\frac{e^{n \mu}}{n}\,\int_{1}^{\infty}dt\,\frac{e^{-n\varphi t}}{\sqrt{1-1/t^2}}
=\varphi\sum_{n=1}^{\infty}\frac{e^{n \mu}}{n}\,\mathbf{K}_{1}(n\varphi)
\label{eq:NewLarge}
\end{equation}
This representation is suitable for large $\varphi$ since
$\mathbf{K}_{1}(n\varphi)\simeq
\frac{e^{-n\varphi}}{\sqrt{\pi\varphi}}+\cdots$.

In order to write an explicit series in $\frac{1}{\varphi}$ we first
introduce the large $\varphi$ expansion of the Bessel function:
\begin{equation}
\mathbf{K}_{1}(n\varphi)
\,=\,
\sqrt{\varphi}\sqrt{\frac{\pi}{2}}\frac{e^{-n\varphi}}{\sqrt{n}}
\,\sum_{k=1}^{\infty}c_{k}
\brk*{\frac{1}{2n\varphi}}^{k-1}
\qquad \text{with}\quad
c_{k}=(-1)^{k}\frac{(2k-1)!!(2k-5)!!}{2^{2k-2}(k-1)!}
\end{equation}
Plugging this latter series into \eqref{eq:NewLarge} and exchanging
the sums we have:
\begin{align}
\mathbb{I}(\varphi,\mu)
&=\sqrt{\varphi}\sqrt{\frac{\pi}{2}}
\sum_{n=1}^{\infty}\frac{e^{\mu}}{n}
\,\frac{e^{-n\varphi}}{\sqrt{n}}
\,\sum_{k=1}^{\infty}c_{k}\left(\frac{1}{2n\varphi}\right)^{k-1}
\nn\\
\mathbb{I}(\varphi,\mu)
&=\sqrt{\varphi}\,\sqrt{\frac{\pi}{2}}
\,\sum_{k=1}^{\infty}\frac{c_{k}}{(2\varphi)^{k-1}}
\,\Li_{k+\frac{1}{2}}(e^{\mu-\varphi})
\label{eq:LargeInt}
\end{align}
Using this latter representation we find the $\varphi\to\infty$ or
$z\bar{z}\to 0$ limit of our correlator as:
\begin{multline}
\log \mathbb{O}=
-\frac{\sqrt{\lambda}}{4\pi^{3/2}}\,\sqrt{-\log z\bar{z}}
\cdot{}\\\cdot
\,\sum_{k=1}^{\infty}
\,\frac{c_{k}}{
\brk*{-\log z\bar{z}}^{k-1}}
\brk*{
 \Li_{k+\frac{1}{2}}(1)
+\Li_{k+\frac{1}{2}}(z\bar{z})
-\Li_{k+\frac{1}{2}}(z)
-\Li_{k+\frac{1}{2}}(\bar{z})
}
\label{eq:LargeExpansion}
\end{multline}
This is a good representation for the limits $z\to0$ or $\bar{z}\to 0$
or both $z,\bar{z}\to 0$. In particular its leading terms reproduces
the corresponding Euclidean OPE limits in \tabref{tab:OPElimits}.

Furthermore, thanks to the competition between the four functions in
\eqref{eq:LargeExpansion}, the first term of the series ($k=1$) provides the correct
leading term of the limits $z\to1$ or $\bar{z}\to 1$ or both
$z,\bar{z}\to1$. For instance for $z\to 1$ the leading term
reproduces the result in \tabref{tab:OPElimits}:
\beq\label{eq:LeadingZ1}
\log\mathbb{O}
\overset{z\to 1}{=}-\frac{\sqrt{\lambda}}{2\pi}\sqrt{1-z}\sqrt{-\log\bar{z}} \qquad\text{and}\qquad \log\mathbb{O}\overset{z,\bar{z}\to 1}{=}-\frac{\sqrt{\lambda}}{2\pi}\sqrt{1-z}\sqrt{1-\bar{z}}
\eeq
However the rest of the series does not provide a good expansion in
this limit.  This can be noticed when trying to compute the first
sub-leading term in the limit $z\to1$, which requires a re-summation of all the series for finite $\bar{z}$. While for $z,\bar{z}\to 1$ the series is no longer convergent.
\begin{multline}
\log\mathbb{O}
\overset{z\to 1}{=}
-\frac{\sqrt{\lambda}}{2\pi}\sqrt{1-z}\sqrt{-\log\bar{z}}
-\frac{\sqrt{\lambda}}{4\pi^{3/2}}(1-z)\sqrt{-\log\bar{z}}
\cdot{}\\\cdot
\sum_{n=1}^{\infty}\frac{c_{n}}{\brk*{\log\bar{z}}^{n-1}}
\,\brk*{\Li_{n-\frac{1}{2}}(1)-\Li_{n-\frac{1}{2}}(\bar{z})}
+ \mathcal{O}(1-z)^{\frac{3}{2}}
\end{multline}
%

\subsection{The Double Light-Cone Limit
\texorpdfstring{$-i\phi\to \infty$}{-i phi to infinity} with
\texorpdfstring{$\varphi\to 0$}{varphi to 0}}
\label{app:DoubleLight}

In this appendix, we consider the light-cone limit $z\to 0^{-},
\bar{z}\to-\infty$ in a restricted kinematics with small $\varphi$ \ie
$z\bar{z}\simeq 1$. To address this limit, we find it more convenient to
make the arguments of the logarithms explicitly positive:
\begin{equation}
\log\left(1+Y\right)
=\log\frac{
\brk*{1-e^{-\frac{\log(z\bar{z})}{2}}e^{\frac{\log(z\bar{z})}{2}\cosh(\theta)}}
\brk*{1-e^{\frac{\log(z\bar{z})}{2}}e^{\frac{\log(z\bar{z})}{2}\cosh(\theta)}}
}{
\brk*{
1{\color{blue}+e^{\frac{1}{2}\brk*{\log(-z)+\log(-\frac{1}{\bar{z}})}}}
\,e^{\frac{\log(z\bar{z})}{2}\cosh(\theta)}
}
\brk*{
1{\color{blue}+e^{-\frac{1}{2}\brk*{\log(-z)+\log(-\frac{1}{\bar{z}})}}}
\,e^{\frac{\log(z\bar{z})}{2}\cosh(\theta)}}
}
\end{equation}
and in order to avoid dealing with $i\pi$ shifts on the chemical
potential we introduce a slightly modified integral:
\begin{equation}
\mathbb{I}_{+}(\varphi,\mu)
=-\varphi\,\int_{0}^{\infty} d\theta \cosh\theta
\log\brk*{1\,{\color{blue}+}\,e^{\mu}\,e^{-\varphi\,\cosh\theta}}
=-\varphi\,\int_{1}^{\infty} dt
\,\frac{\log\brk*{1\,{\color{blue}+}\,e^{\mu}\,e^{-\varphi\,t}}}{\sqrt{1-1/t^2}}
\label{eq:Intplus}
\end{equation}
Using this new building block we rewrite the octagon as:
\begin{equation}
\log\mathbb{O}
=\frac{\sqrt{\lambda}}{2\pi^{2}}
\brk*{
 \mathbb{I}_{+}(\varphi,\mu_{\text{light}})
+\mathbb{I}_{+}(\varphi,-\mu_{\text{light}})
-\mathbb{I}(\varphi,\varphi)
-\mathbb{I}(\varphi,-\varphi)
}
\label{eq:lightOnew}
\end{equation}
and the double-light cone limit is obtained with
$\mu_{\text{light}}\equiv-\frac{1}{2}\left(\log(-z)+\log(-\frac{1}{\bar{z}})\right)\to
+\infty$.

The first integral $\mathbb{I}_{+}(\varphi,\mu_{\text{light}})$ gives
the leading contribution and is obtained in section
\ref{app:LargeChemical} in a $\mu\to \infty$ and $\varphi\to0$
expansion. The second integral
$\mathbb{I}_{+}(\varphi,-\mu_{\text{light}})$ is exponentially
suppressed $\mathcal{O}(e^{-\mu})$ and can be neglected. The tail
$-\mathbb{I}(\varphi,\varphi)\,-\,\mathbb{I}(\varphi,-\varphi)$
corrects the sub-leading term in the double light-cone limit and is
obtained in section \ref{app:SmallChemical} in a $\varphi\to 0$
expansion. Finally we put these results together in section
\ref{app:doublelight} reproducing  \eqref{perturbation2}.

Before addressing the relevant integrals for this light-cone limit, we
start by deriving the small $\varphi\to 0$ expansion in \eqref{eq:Small} with $\mu=0$.
This exercise teaches us the steps such as sum exchanges and
regularization necessary to find the series expansions of the
integrals in \eqref{eq:lightOnew}.

\subsubsection{Series \texorpdfstring{$\varphi\to 0$}{varphi to 0}
\texorpdfstring{{\color{blue}without}}{without} a Chemical Potential}
\label{app:SmallExpansionWithout}

Here we derive the series expansion in small $\varphi$ for $\mu=0$ given in \eqref{eq:Small}.
For this we start with the representation \eqref{eq:Large} and
consider the small $\varphi$ expansion of the Bessel function:
\begin{align}
\frac{\varphi \mathbf{K}_{1}(n\varphi)}{n}
&\overset{\varphi\to 0}{=}
{\color{blue} \frac{1}{n^{2}} }
+\sum_{k=0}^{\infty} \varphi^{2k+2}\left(c_{k}\,\,{\color{blue}n^{2k}\log n}
+ \left(c_{k}\log\left(\frac{\varphi}{2}\right)-d_{k}\right) {\color{blue}n^{2k}}\right)
\label{eq:KSmall}
\end{align}
with:
\begin{equation}
c_{k}=\frac{1}{2^{2k+1}}\frac{1}{\Gamma(k+1)\Gamma(k+2)}
\qquad\text{and}\qquad
d_{k}=\frac{1}{2^{2k+2}}\frac{\psi(k+1)+\psi(k+2)}{\Gamma(k+1)\Gamma(k+2)}
\,.
\end{equation}
Combining this latter representation \eqref{eq:KSmall} for
$\mathbf{K}_{1}$ with the series representation in \eqref{eq:Large} we
want to derive the series representation in \eqref{eq:Small}:
\begin{align}
\lim_{\varphi \to 0}\,\mathbb{I}(\varphi,0)
&=\sum_{n=1}^{\infty}\left(\lim_{\varphi\to 0}\frac{\varphi\mathbf{K}_{1}(n\varphi)}{n}\right)
\nn\\
&=\sum_{n=1}^{\infty}\left(\sum_{k=0}^{\infty}\cdots\text{ in \eqref{eq:KSmall}}\right)
\end{align}
This requires exchanging the sums in $n$ and $k$. The leading term
$\varphi^{0}$ comes from the sum:
\begin{equation}
\sum_{n=1}^{\infty} \frac{1}{n^{2}} = \zeta(2) = \frac{\pi^{2}}{6}
\end{equation}
The coefficients of the subleading terms in $\varphi$ seem to diverge
but performing a zeta-regularization we obtain the finite
contributions:
\begin{align}
\sum_{n=1}^{\infty} n^{2k}\log n &= -\zeta'(-2k) = \begin{cases}
\frac{\log 2\pi}{2}\quad \text{if $k=0$} \\
\frac{(-1)^{k+1} \zeta(2k+1) \Gamma(2k+1)}{2(2\pi)^{2k}} \quad \text{if $k\in\mathbb{Z}^{+}$} \end{cases}\\
\sum_{n=1}^{\infty} n^{2k} &= \zeta(-2k) = \begin{cases}
-\frac{1}{2}\quad \text{if $k=0$} \\
 0\qquad \text{if $k\in\mathbb{Z}^{+}$} \end{cases}
\end{align}
Combining these regularized sums we obtain the series expansion in \eqref{eq:Small} as:
\begin{align}\label{eq:DeriveSmall}
\sum_{n=1}^{\infty}\frac{\varphi \mathbf{K}_{1}(n\varphi)}{n} &
\overset{\varphi\to 0}{=}\;
{\color{blue}\frac{\pi^{2}}{6}}
+\sum_{k=0}^{\infty}\varphi^{2k+2}\brk*{
    c_{k}\,({\color{blue} -\zeta'(-2k)})
    +\brk*{c_{k}\log\brk*{\frac{\varphi}{2}}-d_{k}}{\color{blue}\zeta(-2k)}
}
\\ &
={\color{blue}\frac{\pi^{2}}{6}}
+\varphi^{2}\brk*{
    c_{0} {\color{blue} \frac{\log 2\pi}{2}}
    +\brk*{c_{0}\log\left(\frac{\varphi}{2}\right)-d_{0}}\brk*{{\color{blue}-\frac{1}{2}}}
}
+\sum_{k=1}^{\infty}\brk*{{\color{blue}-\zeta'(-2k)}\,c_{k}}\,\varphi^{2k+2}
\nn\\ &\qquad\qquad
+\sum_{k=1}^{\infty}\varphi^{2k+2}\brk*{c_{k}\log\brk*{\frac{\varphi}{2}}-d_{k}}{\color{blue}\zeta(-2k)}
\\ &
=\frac{\pi^{2}}{6}{\color{red}-\frac{\pi}{2}\varphi}
+\varphi^{2}\brk*{\frac{1}{8}-\frac{\gamma_{E}}{4}-\frac{1}{4}\log\brk*{\frac{\varphi}{4\pi}}}
+\sum_{k=1}^{\infty}\mathcal{C}_{k}\,\varphi^{2k+2}
\label{eq:redterm}
\end{align}
where we have used $\mathcal{C}_{k} = -\zeta'(-2k)\,c_{k}$ and $\zeta(-2k)=0$ for positive integer $k$.

\paragraph{Remark.}

The red term in~\eqref{eq:redterm} was added by hand. It does not come naturally out of this
illegal derivation with the dangerous summation manipulations
performed above. All other terms come out perfectly and beautifully
agree with the expansion quoted in~\cite{Gromov:2011de}. In this case
we could easily fix the linear term since it is the sub-leading term.
Together with the logarithmic term, it is a kind of anomaly in the sense that
it is the only term which does not respect the $\varphi\to -\varphi$
symmetry of the result. Since it is the subleading term in the small
$\varphi$ expansion, it is trivial to restore it by an independent
small $\varphi$ analysis. Now, in what follows we will carry on
similar derivations for the case of interest with chemical potentials.
We will again be missing the analogue of these linear red terms which
we should add back at the end. However, we are in better shape here,
since we know that the full result is actually $\varphi \to -\varphi$
invariant, as this is an exact symmetry of the octagon. As such, these
linear terms must cancel when adding up all four free energies.
Indeed we checked that they do. With this hindsight, we will thus
ignore them completely in the discussion that follows. Let us also
stress again that in the end all final expansions are carefully
checked numerically anyway.

\subsubsection{Including a Large Chemical Potential
\texorpdfstring{$\mu\to\infty$}{mu to infinity}}
\label{app:LargeChemical}

Now we study the slightly modified integral
$\mathbb{I}_{+}(\varphi,\mu)$ of \eqref{eq:Intplus} in the combined
limits $\mu\to\infty$ and $\varphi\to0$. For this we start with the
series expansion:
\begin{equation}
\mathbb{I}_{+}(\varphi,\mu)
=\sum_{n=1}^{\infty}\frac{\varphi \mathbf{K}_{1}(n\varphi)\,(-e^{\mu})^{n}}{n}
\end{equation}
After taking the small $\varphi$ limit of the summand in \eqref{eq:NewLarge} we
obtain:
\begin{multline}
\frac{\varphi\mathbf{K}_{1}(n\varphi)\,(-e^{\mu})^{n}}{n}
\overset{\varphi\to 0}{=} \\
{\color{blue} \frac{(-e^{\mu})^{n}}{n^{2}}}
+\sum_{k=0}^{\infty}\varphi^{2k+2}
\brk*{
c_{k}\,\,{\color{blue}(-e^{\mu})^{n} n^{2k}\log n}
+\brk*{c_{k}\log\left(\frac{\varphi}{2}\right)-d_{k}}
{\color{blue}(-e^{\mu})^{n}\,n^{2k}}
}
\,
\end{multline}
The sums on $n$ can be performed using a generalization of the
Zeta-regularization which now gives polylogarithms $\text{Li}_{k}(x)$
and their derivatives\footnote{Notice the derivative is over the index
of the Polylogarithm and not on its argument.} $\text{Di}_{k}(x)
\equiv -\partial_{m}\text{Li}_{m}(x)\big{|}_{m\to k}$,
\begin{align}
\sum_{n=1}^{\infty}\frac{\varphi \mathbf{K}_{1}(n\varphi)\,(-e^{\mu})^{n}}{n}&\overset{\varphi\to 0}{=}{\color{blue} \text{Li}_{2}(-e^{\mu}) }\,+\, \varphi^{2}\left( {\color{blue}\text{Di}_{0}(-e^{\mu})}\,c_{0} +{\color{blue}\text{Li}_{0}(-e^{\mu})}\left(c_{0}\log\left(\frac{\varphi}{2}\right)-d_{0}\right)   \right) \nonumber\\
&\qquad+\, \sum_{k=1}^{\infty} \varphi^{2k+2} \left({\color{blue}\text{Di}_{-2k}(-e^{\mu})}\,c_{k}+ {\color{blue}\text{Li}_{-2k}(-e^{\mu})}\left(c_{k}\log\left(\frac{\varphi}{2}\right)-d_{k}\right) \right)
\end{align}
In the large $\mu$ limit these regularized sums behave as:
\begin{align}
\sum_{n=1}^{\infty} \frac{(-e^{\mu})^{n}}{n^{2}} &= \text{Li}_{2}(-e^{\mu})\,\overset{\mu\to\infty}{=}\,-\frac{\mu^{2}}{2}-\frac{\pi^{2}}{6}  \, +\mathcal{O}(e^{-\mu})  \\
\sum_{n=1}^{\infty} (-e^{\mu})^{n}\,n^{2k}\log n &= \text{Di}_{-2k}(-e^{\mu})\,\overset{\mu\to\infty}{=}\,
\begin{cases}
\log\mu +\gamma_{E} +\mathcal{O}(\frac{1}{\mu^{2}})\quad\text{if }k=0 \\
-\frac{\Gamma(2k)}{\mu^{2k}}+\mathcal{O}(\frac{1}{\mu^{2k+2}})\quad\text{if }k\in\mathbb{Z}^{+}
\end{cases}\\
\sum_{n=1}^{\infty} (-e^{\mu})^{n}\,n^{2k} &=\text{Li}_{-2k}(-e^{\mu})\,=\, \partial_{\mu}^{2k}\frac{-e^{\mu}}{1+e^{\mu}}\,\overset{\mu\to\infty}{=}\,\begin{cases}  -1+e^{-\mu} + \mathcal{O}(e^{-2\mu}) \qquad \text{if } k=0\\
 e^{-\mu}+ \mathcal{O}(e^{-2\mu}) \qquad \text{if } k  \in \mathbb{Z}^{+} \end{cases}
\end{align}
The derivative of the polylogarithm $\text{Di}_{0}$ contains power law
corrections given by:
\begin{equation}
\lim_{\mu \to \infty} \text{Di}_{0}(-e^{\mu}) \,= \, \log\mu + \gamma_{E} - \sum_{m=1}^{\infty}  \frac{\left(2-\frac{1}{2^{2m-2}}\right)\Gamma(2m)\,\zeta(2m)}{\mu^{2m}} \,+\cdots
\end{equation}
where the ellipsis stands for terms further suppressed exponentially
$e^{-\mu}$. This function can be used as a seed to obtain the
derivatives with negative indices from the recursion relation:
$\partial_{\mu}^{2k}\text{Di}_{0}(-e^{\mu}) =
\text{Di}_{-2k}(-e^{\mu})$. This relation makes obvious the
logarithmic divergence $\log\mu$ disappear and we only have
power-suppressed corrections from these terms:
\beq
\lim_{\mu \to \infty} \text{Di}_{-2k}(-e^{\mu}) = \frac{-\Gamma(2k)}{\mu^{2k}}  - \sum_{m=1}^{\infty}  \frac{\left(2-\frac{1}{2^{2m-2}}\right)\Gamma(2m+2k)\,\zeta(2m)}{\mu^{2m+2k}} \,+\cdots
\qquad\text{with }k\in\mathbb{Z}^{+}
\eeq
Finally, neglecting exponentially suppressed terms
$\mathcal{O}(e^{-\mu})$ we have the series:
\begin{align}
\lim_{\mu\to\infty}\mathbb{I}_{+}(\varphi,\mu)\,&\simeq
\,{\color{blue}-\frac{\mu^{2}}{2}-\frac{\pi^{2}}{6}  } \,+\, \varphi^{2}\left(  {\color{blue}\text{Di}_{0}(-e^{\mu})}\,c_{0} +{\color{blue}(-1)}\left(c_{0}\log\left(\frac{\varphi}{2}\right)-d_{0}\right)   \right) \nonumber\\
&\qquad\qquad\qquad \, + \sum_{k=1}^{\infty} \varphi^{2k+2} {\color{blue}\text{Di}_{-2k}(-e^{\mu})}\,c_{k}
\label{eq:CompleteLargeMu}
\end{align}
More simply, neglecting $\mathcal{O}(1/\mu^{2})$ corrections, we obtain:
\begin{align}
\lim_{\mu\to\infty}\mathbb{I}_{+}(\varphi,\mu)\,
&\simeq \, -\frac{\mu^{2}}{2}-\frac{\pi^{2}}{6}  \,+\, \varphi^{2}\left(\frac{\log\mu}{2}+\frac{1}{4} -\frac{\log(\varphi/2)}{2}\right)
\label{eq:LargeChemicalFinal}
\end{align}
We thank Nikolay Gromov for providing us with a clean alternative
derivation of these large chemical potential results which was
extremely useful in cross-checking these manipulations.

\subsubsection{Series \texorpdfstring{$\varphi\to 0$}{varphi to 0}
\texorpdfstring{{\color{blue}with}}{with} a Chemical Potential
\texorpdfstring{$\mu=\pm\varphi$}{mu=+-varphi}}
\label{app:SmallChemical}

In this section we consider the contribution of
$\mathbb{I}(\varphi,\varphi)+\mathbb{I}(\varphi,-\varphi)$ as a series
expansion in small $\varphi$. For this we introduce the chemical
potentials as a series expansion:
\begin{equation}
e^{n\varphi}+e^{-n\varphi}=\sum_{l=0}^{\infty}f_{l} \varphi^{2l}
\quad\text{with }
f_{l} =2 \Gamma(2l)
\,.
\end{equation}
inside:
\begin{align}
\mathbb{I}(\varphi,\varphi)+\mathbb{I}(\varphi,-\varphi) = \sum_{n=1}^{\infty}\frac{\varphi(e^{n\varphi}+e^{-n\varphi}) \mathbf{K}_{1}(n\varphi)}{n}&\overset{\varphi\to 0}{=}.  \sum_{n=1}^{\infty}\left({\color{blue} \frac{\sum_{l=0}^{\infty}f_{l} \varphi^{2l}}{n^{2}} } +  S^{\text{extra}}_{\color{blue} n } \right)
\end{align}
The leading term can be obtained after regularizing the sum over $n$ as:
\beq\label{eq:sum1}
\sum_{n=1}^{\infty} \sum_{l=0}^{\infty} f_{l} \varphi^{2l} n^{2l-2} = \sum_{l=0}^{\infty} \varphi^{2l} \,f_{l}\,\zeta(2-2l) \,=\, \frac{\pi^{2}}{3} - \frac{\varphi^{2}}{2}
\eeq
The subleading terms come from $S^{extra}_{n}$ which,
 after rearranging the sums on $k$ and $l$, is given by:
\begin{align}
S^{\text{extra}}_{\color{blue} n }
& = \sum_{k=0}^{\infty} \varphi^{2k+2}\left(\sum_{l=0}^{\infty}f_{l}\,{\color{blue}n^{2l}}\varphi^{2l}\right) \left(c_{k}\,\,{\color{blue}n^{2k}\log n}+ \left(c_{k}\log\left(\frac{\varphi}{2}\right)-d_{k}\right) {\color{blue}n^{2k}}\right) \\
& = \sum_{k=0}^{\infty} \sum_{l=0}^{\infty}  \varphi^{2k+2l+2} \left( {\color{blue} n^{2k+2l} \log n} \,f_{l}\,c_{k}  + {\color{blue} n^{2k+2l}} \left(f_{l} \,c_{k}\log\left(\frac{\varphi}{2}\right)-f_{l} \,d_{k}\right)\right)\nonumber\\
& = \sum_{M=0}^{\infty} \varphi^{2M+2} \left({\color{blue} n^{2M}\log n} \sum_{k=0}^{M}f_{M-k}\,c_{k} + {\color{blue}n^{2M}}\sum_{k=0}^{M}\left(f_{M-k} \,c_{k}\log\left(\frac{\varphi}{2}\right)-f_{M-k} \,d_{k}\right)\right)
\nn
\end{align}
and performing the sum over $n$ using zeta-regularization we obtain:
\begin{align}\label{eq:sum2}
\sum_{n=1}^{\infty} S^{extra}_{n}  &= \sum_{M=0}^{\infty} \varphi^{2M+2} \left({\color{blue} -\zeta'(-2M)} \sum_{k=0}^{M}f_{M-k}\,c_{k} + {\color{blue}\zeta(-2M)}\sum_{k=0}^{M}f_{M-k}\left(c_{k}\log\left(\frac{\varphi}{2}\right)-\,d_{k}\right)\right)\nonumber\\
& = \varphi^{2}\,f_{0}\,\left({\color{blue}\frac{\log 2\pi}{2}}\,c_{0}\,{\color{blue}-\frac{1}{2}}\left(\,c_{0}\log\left(\frac{\varphi}{2}\right)-d_{0}\right)\right) + \sum_{M=1}^{\infty}\varphi^{2k+2} \left({\color{blue} -\zeta'(-2M)} \sum_{k=0}^{M}f_{M-k}\,c_{k} \right)
\end{align}
Adding up \eqref{eq:sum1} and \eqref{eq:sum2} we get
\begin{align}
\mathbb{I}(\varphi,\varphi)+\mathbb{I}(\varphi,-\varphi)&=\sum_{n=1}^{\infty}\frac{\varphi(e^{n\varphi}+e^{-n\varphi}) \mathbf{K}_{1}(n\varphi)}{n}\nonumber\\
&\overset{\varphi\to0}{=} \frac{\pi^{2}}{3} \,+\,  \varphi^{2}\left(-\frac{1}{2}+\frac{\log 2\pi}{2} f_{0}\,c_{0}\,+\frac{1}{2}\,f_{0}\,d_{0}-\frac{1}{2}\,f_{0} \,c_{0}\log\left(\frac{\varphi}{2}\right) \right)\nonumber\\
&\qquad\quad + \sum_{M=1}^{\infty}\varphi^{2k+2} \left( -\zeta'(-2M) \sum_{k=0}^{M}f_{M-k}\,c_{k} \right)
\label{eq:CompleteSmallMu}
\end{align}
Finally neglecting $\mathcal{O}(\varphi^{4})$ terms we obtain:
\begin{equation}\label{eq:SmallChemicalFinal}
\mathbb{I}(\varphi,\varphi)+\mathbb{I}(\varphi,-\varphi)\overset{\varphi\to0}{=}  \frac{\pi^{2}}{3} + \varphi^{2}\left(-\frac{1}{2}\log\frac{\varphi}{4\pi}-\frac{1}{4}-\frac{\gamma_{E}}{2}\right)
\end{equation}
%

\subsubsection{The Double Light-Cone Limit}
\label{app:doublelight}

The double light-cone limit corresponds to the limit $i\phi\to\infty$.
Considering also the limit $\varphi\to 0$, we can provide an
analytic expression for the octagon~\eqref{eq:lightOnew} by
combining formulas~\eqref{eq:SmallChemicalFinal} and~\eqref{eq:LargeChemicalFinal}
with $\mu \to \mu_{\text{light}} \equiv
-\frac{1}{2}\left(\log(-z)+\log(-\frac{1}{\bar{z}})\right)$:
\begin{align}\label{eq:FinalDoubleApp}
\log \mathbb{O}
&=\frac{\sqrt{\lambda}}{2\pi^{2}}
\brk!{
\mathbb{I}_{+}(\varphi,\mu_{\text{light}})
+\mathbb{I}_{+}(\varphi,-\mu_{\text{light}})
-\mathbb{I}(\varphi,\varphi)
-\mathbb{I}(\varphi,-\varphi)
}
\nn\\
&\simeq
\frac{\sqrt{\lambda}}{2\pi^{2}}
\Bigg(
-\frac{\mu_{\text{light}}^{2}}{2}-\frac{\pi^{2}}{6}
+\varphi^{2}
\brk*{\frac{\log\mu_{\text{light}}}{2}+\frac{1}{4}-\frac{\log(\varphi/2)}{2}}
\nn\\ & \mspace{300mu}
-\brk*{
\frac{\pi^{2}}{3}
+\varphi^{2}\brk*{-\frac{1}{2}\log\frac{\varphi}{4\pi}-\frac{1}{4}-\frac{\gamma_{E}}{2}}
}
\Bigg)
\nn\\
&=
\frac{\sqrt{\lambda}}{2\pi^{2}}
\brk*{
-\frac{\mu_{\text{light}}^{2}}{2} -\frac{\pi^{2}}{2}
+ \varphi^{2} \brk*{\frac{\log\mu_{\text{light}}}{2}\,+\,\frac{1+\log 2 +\gamma_{E}-\log 4\pi}{2}}
} \nn\\
& = -\frac{(-\log(-z)-\log(-1/\bar{z}))^{2}}{8\pi^{2}}\left(\frac{\sqrt{\lambda}}{2}\right)  \,+\, \frac{1}{8}\left(-2\sqrt{\lambda}\right)
\\ \nn & \mspace{100mu}
+\frac{\sqrt{\lambda}}{16\pi^{2}}(\log z\bar{z})^{2}
\brk2{\log\brk!{-\log(-z)-\log(-1/\bar{z})} +1+\gamma_{E}-\log 4\pi}
\,.
\end{align}
We have neglected terms suppressed as
$\mathcal{O}(1/\mu_{\text{light}}^{2},e^{-\mu_{\text{light}}},\varphi^{4})$
in order to compare with \eqref{perturbation2} and
\eqref{eq:OurLightCs} in the main text. The power-like corrections on
$\mathcal{O}(1/\mu^{2}_{\text{light}})$ and $\mathcal{O}(\varphi^{2})$
can be recovered from the series in \eqref{eq:CompleteLargeMu} and
\eqref{eq:CompleteSmallMu}.

\pdfbookmark[1]{\refname}{references}
\addcontentsline{toc}{section}{References}
\bibliographystyle{nb}
\bibliography{references}

\end{document}